\documentclass[traditabstract]{aa}  
\usepackage[varg]{txfonts}
\usepackage{graphicx}
\usepackage{subfigure}
\usepackage{rotating}
\usepackage{color}
\usepackage{natbib}
\usepackage{ragged2e}
\usepackage{soul} 
\usepackage{float}
\usepackage{wrapfig,lipsum}
\usepackage[flushleft]{threeparttable} 
\bibpunct{(}{)}{;}{a}{}{,} 
\definecolor{darkgreen}{rgb}{0.0,0.5,0.0}
\definecolor{darkred}{rgb}{0.5,0.0,0.0}
\definecolor{grey}{rgb}{0.4,0.5,0.6}
\definecolor{bluepigment}{rgb}{0.2, 0.2, 0.6}
\definecolor{darkslateblue}{rgb}{0.28, 0.24, 0.55}
\definecolor{midnightblue}{rgb}{0.1, 0.1, 0.44}
\definecolor{navyblue}{rgb}{0.0, 0.0, 0.5}

\begin{document} 
\title{Double-peak emission line galaxies in the SDSS catalogue. A minor merger sequence.}
\titlerunning{Double-peak emission line galaxies in SDSS}
\authorrunning{Maschmann et al.}

\author{Daniel Maschmann\inst{1, 2}, Anne-Laure Melchior\inst{1}, Gary A. Mamon\inst{3}, Igor V. Chilingarian\inst{4,5}, Ivan Yu. Katkov\inst{5,6}\\ 
}
\institute{Sorbonne {Universit\'e},
LERMA, Observatoire de Paris, PSL research university, CNRS, F-75014, Paris, France\\
\email{Daniel.Maschmann@rwth-aachen.de, A.L.Melchior@obspm.fr}
\and
RWTH Aachen University, Institute for Theory of Science and Technology, Aachen, Germany
\and 
Institut d'Astrophysique de Paris (UMR 7095: CNRS \& Sorbonne Universit\'e), 98 bis bd Arago, F-75014 Paris, France
\and
Center for Astrophysics -- Harvard and Smithsonian, 60 Garden St. MS09, Cambridge, MA, 02138, USA 
\and
Sternberg Astronomical Institute, M.V. Lomonosov Moscow State University, 13 Universitetsky prospect, Moscow, 119991, Russia
\and 
New York University Abu Dhabi, Saadiyat Island, PO Box 129188, Abu Dhabi, UAE
}
\date{Received ?; accepted ?}
\abstract {


Double-peak narrow emission line galaxies have been studied extensively in the past years, in the hope of discovering late stages of mergers. 
It is difficult to disentangle this phenomenon from disc rotations and gas outflows with the sole spectroscopic measurement of the central $3^{\prime \prime}$. 
We aim to properly detect such galaxies and distinguish the underlying mechanism with a detailed analysis of the host-galaxy properties and their kinematics.
Relying on RCSED, we developed an automated selection procedure and found $5\,663$ double-peak emission line galaxies at $z < 0.34$ corresponding to $0.8\%$ of the parent database. To characterise these galaxies, we built a single-peak no-bias control sample (NBCS) with the same redshift and stellar mass distributions as the double-peak sample (DPS).
These two samples are indeed very similar in terms of absolute magnitude, [OIII] luminosity, colour-colour diagrams, age and specific star formation rate, metallicity, and environment. 
We find an important excess of S0 galaxies in the DPS, not observed in the NBCS, and which cannot be accounted for by the environment, as most of these galaxies are isolated or in poor groups.
Similarly, we find a relative deficit of pure discs in the DPS late-type galaxies, that are preferentially of Sa type. In parallel, we observe a systematic central excess of star formation and extinction for DP galaxies.
Finally, there are noticeable differences in the kinematics: 
the gas velocity dispersion is correlated with the galaxy inclination in the NBCS, whereas this relation does not hold for the DPS. Furthermore, the DP galaxies show larger stellar velocity dispersions and they deviate from the Tully-Fisher relation for both late-type and S0 galaxies. These discrepancies can be reconciled if one considers the two peaks as two different components.
Considering the morphological biases in favour bulge dominated galaxies and the star-formation central enhancement, we suggest a scenario of multiple sequential minor mergers driving the increase of the bulge size, leading to larger fractions of S0 galaxies and a deficit of pure disc galaxies.
}

\maketitle
\section{Introduction}\label{sect:introduction}

The evolution of galaxies over cosmic time is largely determined by their mass growth and is thus connected to their environment and their merger rate. It is well observed that the mix of morphological types of galaxies depends on the environment \citep{1980ApJ...236..351D,1993ApJ...407..489W}. The star-formation rate of galaxies is a well-suited diagnostic to characterise their evolutionary state. Galaxies can, on the one hand, enhance their star formation rate through interaction with their environment \citep{1986ApJ...301...57B, 2002MNRAS.331..333P}, but on the other hand, they can be also be quenched by the environment \citep{1998ApJ...504L..75B}. Isolated galaxies are thought to refuel their discs with gas from extended halos and from cosmic filaments, while galaxies located in massive clusters will evolve passively \citep{1998ApJ...504L..75B}.
The assembly and growth of galactic discs and galaxies in general are some of the key issues of galaxy simulations \citep[e.g.][]{1998MNRAS.295..319M}. 
Accretion from filaments is motivated by numerical simulations \citep[e.g.][]{1996Natur.380..603B}, while observational detection are based on filaments of galaxies in cluster environments \citep[e.g.][]{2018MNRAS.474.5437L,2019A&A...632A..49S} and the Ly$\alpha$ forest tomography \citep[e.g.][]{2018ApJS..237...31L}. The latter approach is the only one that directly detects so-called gas accretion. 

The identification of merging galaxies is usually based on morphology \citep[e.g.][]{2004AJ....128..163L} or detection of dynamically close pairs \citep[e.g.][]{2005AJ....130.1516D}. Relying on the latter technique, \citet{2008AJ....135.1877E} identified 1716 galaxies with companions in the Sloan Digital Sky Survey (SDSS) Data Release (DR) 4 with stellar mass ratios between $0.1 < {\rm M}_1/{\rm M}_2 < 10$. Further studies of this sample found that star formation due to galaxy interactions can be triggered in low-to-intermediate density environments \citep{2010MNRAS.407.1514E}. By extending their search to SDSS DR7, they increased their sample to $21\,347$ galaxy pairs and found evidence for a central starburst induced by galaxy interactions \citep{2011MNRAS.412..591P}. By including quasi stellar objects (QSO) in their search, \citet{2011MNRAS.418.2043E} found that AGN activity can occur well before the final merging of a galaxy pair and is accompanied by ongoing star formation.

The original prediction that merging should go up to the black hole coalescence \citep[e.g.][]{1980Natur.287..307B} has not been observed yet.
But earlier steps have been explored, and several dual AGNs, which is a late stage of a galaxy merger, \citep{2001ApJ...563..527G,2016ApJ...824L...4K,2018Natur.563..214K,2019ApJ...879L..21G} or even triple nucleus \citep{2014Natur.511...57D,2019ApJ...883..167P} have been detected. While about $40\%$ of ultra-luminous infrared galaxies exhibit a double nucleus \citep{2001AJ....122...63C}, \citet{2018Natur.563..214K} discuss that gas-rich luminous AGNs are often hidden mergers. \citet{2010ApJ...710.1578G} were first to identify a galaxy merger resulting in a binary quasar with a projected separation of 21\,kpc and a radial velocity difference of 215\,km\,s$^{-1}$. 
Mergers with a binary quasar have also been associated with an offset and/or double-peak [OIII]$\lambda$5008 emission line \citep[e.g.][]{2009ApJ...698..956C,2013ApJ...777...64C}. 
Many systematic searches for dual AGN have been conducted at different wavelengths \citep{2011ApJ...737..101L,2013ApJ...762..110L,2012ApJ...746L..22K,2015ApJ...799...72F} to discuss the nature of dual AGN.

Using the direct detection of double-peak narrow emission lines, \citet{2009ApJ...705L..76W}, \citet{2010ApJ...708..427L}, \citet{2010ApJ...716..866S} and \citet{2012ApJS..201...31G} have selected large galaxy samples from several galaxy surveys. In most of these works, the search for double-peak emission lines are motivated by the search of dual AGNs or dual galactic cores. 
Starting from such samples, \citet{2012ApJ...753...42C} conducted long slit observations on double-peak emission line galaxies to find kiloparsec-scale spatial offsets and to constrain the selection of dual AGNs. Using the Hubble Space Telescope and the space based X-ray telescope Chandra, \citet{2015ApJ...806..219C} confirmed a dual AGN, with a separation of $2.2\,{\rm kpc}$, resulting from an extreme minor merger (460:1) creating a double-peak [OIII]$\lambda$5008 emission line. Follow-up observations with the Very Large Array enabled the detection of 3 dual AGNs, AGN wind-driven outflows, radio-jet driven outflows and one rotating narrow-line region producing double-peak narrow emission lines \citep{2015ApJ...813..103M}.
Long-slit observations of double-peak galaxies enabled to distinguish between AGN driven outflows and a rotating disc \citep{2016ApJ...832...67N} as further supported by Monte Carlo simulations in \citet{2018MNRAS.473.2160N}. Furthermore, \citet{2018ApJ...867...66C} associated double-peak emission-line galaxies with galaxy mergers, concluding that at least $3\%$ galaxies with double-peak narrow AGN emission lines found in the SDSS spectra are galaxy mergers identified in SDSS snapshots. 

In this article, we build up an objective selection procedure for double-peak narrow emission line galaxies, to test whether we can identify different merger stages. We do not constrain our search to dual AGN candidates and we do not include a visual selection in contrast to previous galaxy samples. We base our work on the value-added Reference catalogue of Spectral Energy Distributions (RCSED) \citep{2017ApJS..228...14C}.  

This work is organised as follows. In Sect. \ref{sect:data}, we describe the pipeline developed to automatically select galaxies with spectra exhibiting double-peak emission lines and describe the selection of a no-bias control sample (NBCS). Sect. \ref{sect:classification} classifies this double-peak sample (DPS) relying on ionisation diagrams and on morphology and compare it with previous works. In Sect. \ref{sect:analysis}, we analyse the properties of the DPS and compare them with the NBCS. In Sect. \ref{sect:discussion}, we discuss our results followed by a conclusion in Sect. \ref{sect:conclusion}.

A cosmology of $\Omega_{m} = 0.3$, $\Omega_{\Lambda} = 0.7$ and $h = 0.7$ is assumed in this work.

\section{Detection of double-peak emission-line galaxies in RCSED catalogue }\label{sect:data}

\subsection{Spectroscopic Data}\label{ssect:spec}
\begin{figure*}
    \centering
    \begin{subfigure}
        \centering
        \includegraphics[width=0.98\linewidth]{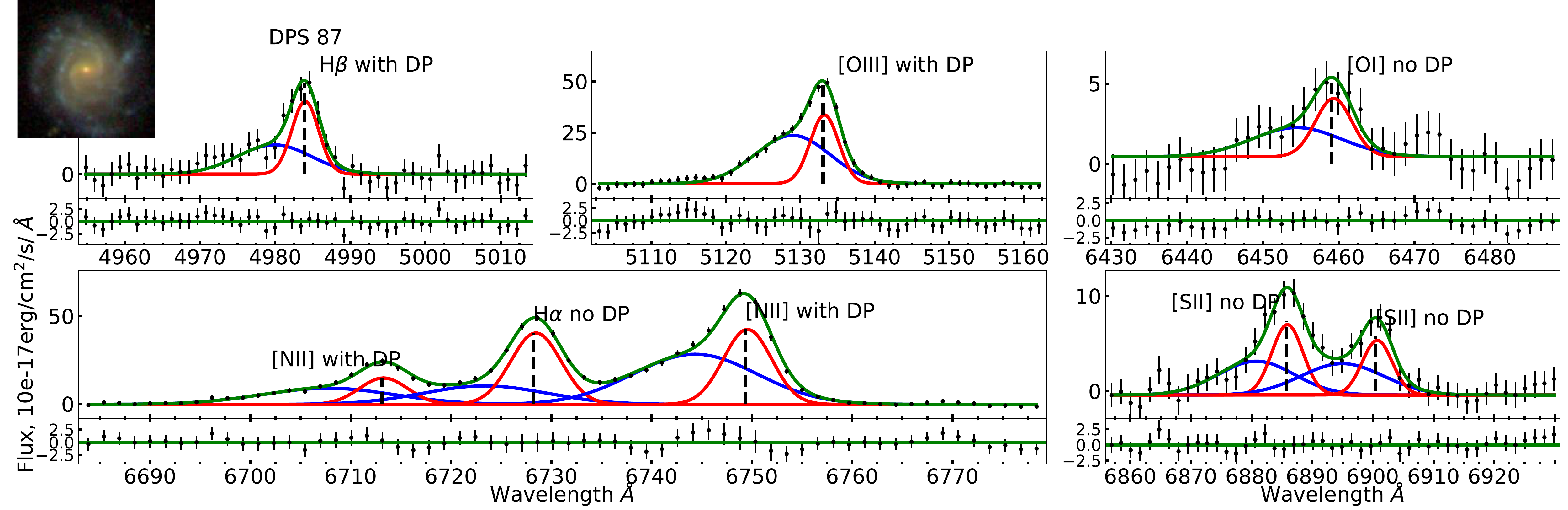}
    \end{subfigure}
    \begin{subfigure}
        \centering
        \includegraphics[width=0.98\linewidth]{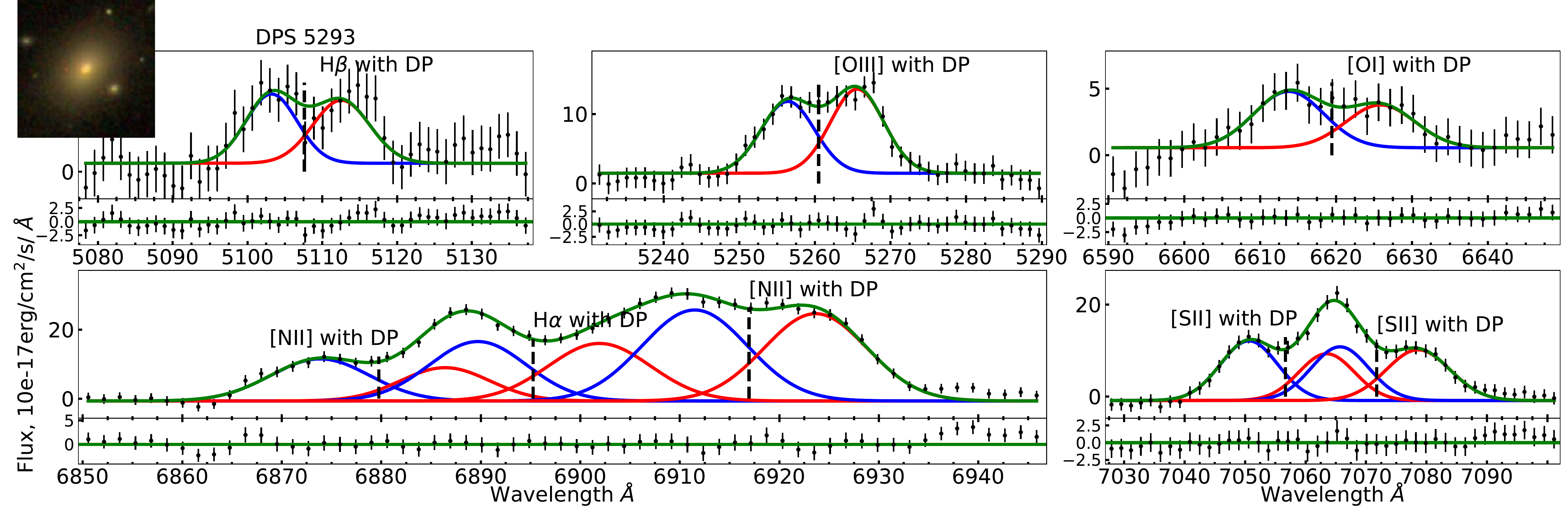}
    \end{subfigure}
    \begin{subfigure}
        \centering
        \includegraphics[width=0.98\linewidth]{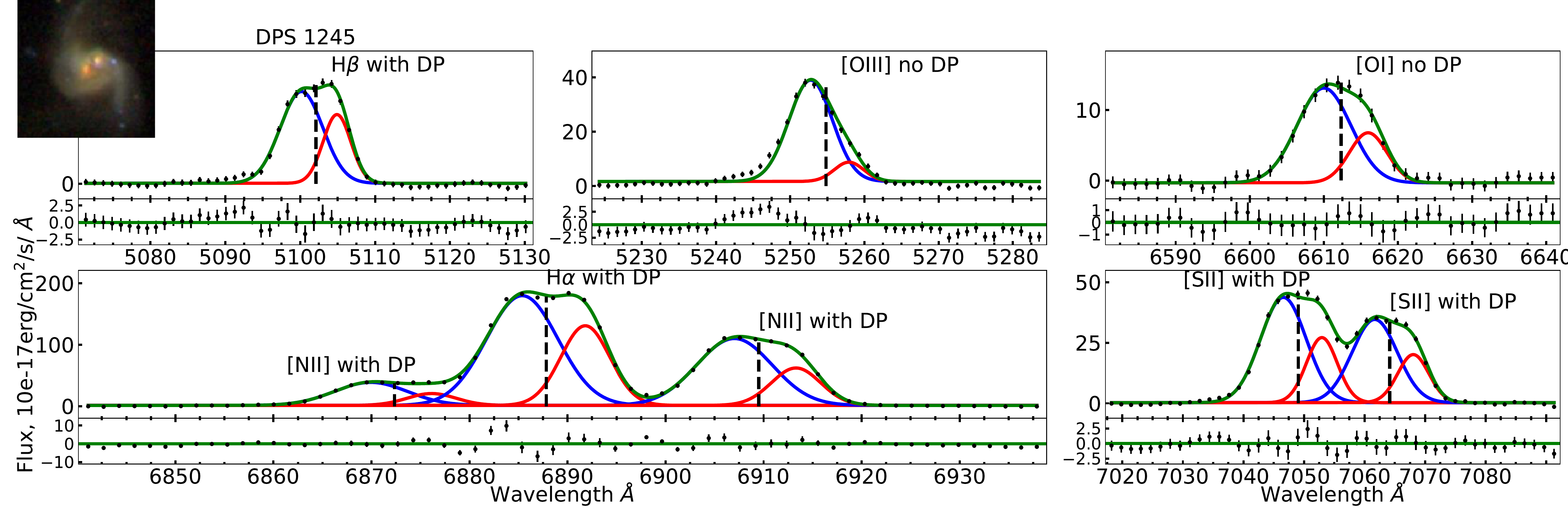}
    \end{subfigure}    
    \caption{Emission lines of three DP galaxies, namely ${\rm H}{\beta}\lambda$4863, {\rm [OIII]}$\lambda$5008, $\rm [OI]\lambda$6302, ${\rm [NII]}\lambda$ 6550, ${\rm H}{\alpha}\lambda$6565, ${\rm [NII]}\lambda$ 6585, ${\rm [SII]}\lambda$6718 and ${\rm [SII]}\lambda$ 6733. For each panel, we display on the top left, the $62^{\prime\prime}\times 62^{\prime\prime}$ SDSS snapshot. Each displayed line is fitted with a double Gaussian function as explained in Sect.\,\ref{ssect:automated:selection:procedure}. We show the blueshifted Gaussian component as blue lines and the redshifted component as red lines. The green line is a superposition of the two Gaussian components. Below each emission line we show residuals. We flag lines with a confirmed DP selected with criteria described in Sect.\,\ref{ssect:automated:selection:procedure} as "with DP" and flag the others as "no DP" for failing the criteria. The black dashed vertical line indicates the position of the stellar velocity of the host galaxy, computed by \citet{2017ApJS..228...14C}.
    The top spectra are for a face-on spiral galaxy at a redshift of ${\rm z}=0.02$ with a $\Delta {\rm v_{DP}} = 232\, {\rm km\,s^{-1}}$, which resembles an outflow (discussed in Sect.\,\ref{ssect:alternatives}). The middle spectra are for an elliptical galaxy at ${\rm z}=0.05$ showing a $\Delta {\rm v_{DP}} = 495\, {\rm km\,s^{-1}}$. The bottom spectra are for a galaxy merger at ${\rm z}=0.05$ and $\Delta {\rm v_{DP}} = 269\, {\rm km\,s^{-1}}$. This is one of 58 galaxies discussed in \citet{2019A&A...627L...3M} showing a DP structure probably associated with recent galaxy merger.}
    \label{fig:Spec}
\end{figure*}
The Reference catalogue of Spectral Energy Distribution (RCSED) contains $800\,299$ galaxies selected from the SDSS DR7 spectroscopic sample (with a spectral resolving power $R=1500\dots2500$) in the redshift-range $0.007 < {\rm z} < 0.6$ \citep{2017ApJS..228...14C}. This catalogue provides $k$-corrected photometric data in the ultraviolet, optical and near-infrared bands observed by the Galaxy Evolution Explorer (GALEX), SDSS and the UK Infrared Telescope Deep Sky Survey (UKIDSS).

The RCSED catalogue also provides optical SDSS spectra in 3-arcsec circular apertures up to a magnitude limit of ${\rm r} = 17.77$ mag \citep{2009ApJS..182..543A} and a best-fitting template. The template assumes either a simple stellar population model (SSP) or an exponentially declining star formation history (EXP-SFH) \citep{2017ApJS..228...14C}. The best-fitting template subtracted from an original spectrum provides an emission-line spectrum in the observed wavelength range [3600 \AA, 6790 \AA]. In Fig.\,\ref{fig:Spec}, we show major emission lines extracted from stellar continuum subtracted spectra of three different galaxies, studied in this article as they exhibit DP emission lines as described in Sect. \ref{ssect:automated:selection:procedure}.

In the RCSED catalogue, each emission-line spectrum is fitted with two different functions: (1) a Gaussian function and (2) a non-parametric distribution. In case (1), two Gaussian functions are adjusted to all allowed and all forbidden transitions. The case (2) is based on an algorithm, which adapts an arbitrary shape to all emission lines simultaneously, again grouped by the transition type. The non-parametric fit is able to fit complex line shapes such as a DP and AGN-driven outflows, and can also reveal low-luminosity AGN broad line components \citep{2018ApJ...863....1C}. The catalogue provides the fluxes resulting from these two procedures for several emission lines, $\chi^2$ per number degree of freedom (Ndof), hereafter $\chi^2_{\nu}$, the equivalent width (EW) and other parameters, as specified in \citet{2017ApJS..228...14C}.

\subsection{Automated Selection Procedure} 
\label{ssect:automated:selection:procedure}
\begin{figure*}
  \centering 
 \includegraphics[width=0.65\textwidth]{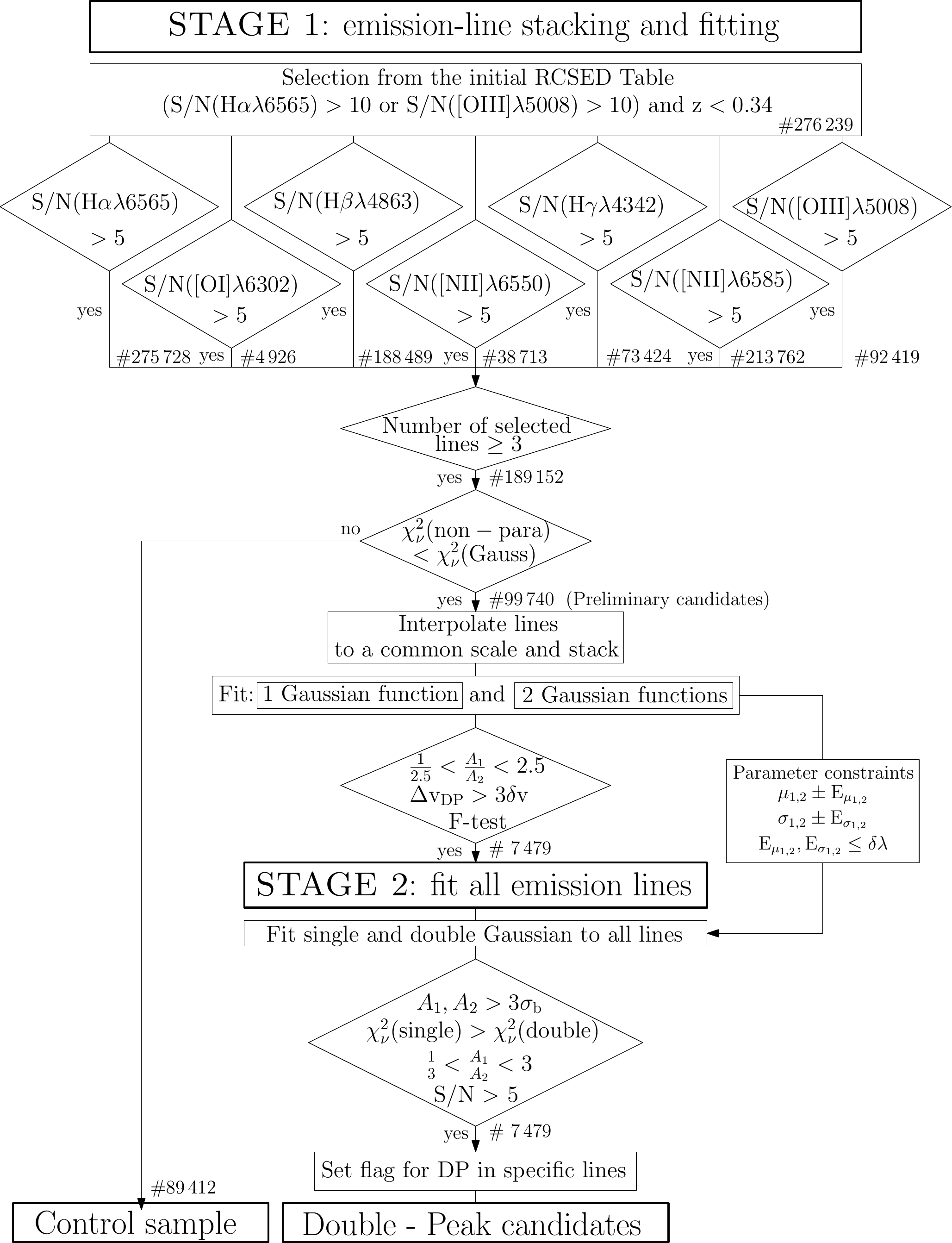}
  \caption{Flowchart describing the first two stages of the automated selection procedure, detailed in Section \ref{ssect:automated:selection:procedure}. We list up all selection criteria and note the number of galaxies at each selection step. In stage 1 we select preliminary candidates and a control sample (CS) using emission line properties and $\chi^2$ ratios from emission-line fitting computed by \cite{2017ApJS..228...14C}. We visualise the stacking and fitting procedure and all selection criteria described in \ref{ssect:stack:and:fit}. In stage 2 we describe the individual fitting of each emission line and list up all criteria for the DP flag detailed in \ref{ssect:individual:line:fit}. We finally select $7\,479$ DP candidates and $89\,412$ galaxies for the CS.}
  \label{fig:Flowchart_1}%
\end{figure*}

We developed an automated three-stage selection procedure to find DP galaxies. The first stage pre-selects galaxies with a threshold on the S/N rates, and performs successively the emission-line stacking, line adjustments and empirical selection criteria. Some emission lines are individually fitted at the second stage to select first DP candidates. We also select candidates showing no DP properties to be the control sample (CS). Stages 1 and 2 are summarised in Fig.\,\ref{fig:Flowchart_1}. At the third stage, we obtain the final DPS using the fit parameter of each line, as shown in Fig.\,\ref{fig:Flowchart_2}. Hereafter, we explain in detail each step of the selection procedure.
\subsubsection{Preliminary Candidates}\label{ssect:preliminary:candidates}
We first restrict the analysis to galaxies with detectable emission lines, in order to define a preliminary sample on which a subsequent emission-line fitting can be applied. Hence, we select objects with a ${\rm S/N} > 10$ in the [OIII]$\lambda$5008 or ${\rm H}{\alpha}\lambda$6565 lines. To secure the detection of the S[II]$\lambda$6718 and S[II]$\lambda$6733 lines within the spectra bandwidth, we add the condition $z<0.34$. We thus keep a sample containing $276\,239$ objects from the RCSED catalogue.

We then select $189\,152$ galaxies, that have a ${\rm S/N} > 5$ in at least 3 emission lines among ${\rm H}{\alpha}\lambda$6565, ${\rm H}{\beta}\lambda$4863, ${\rm H}{\gamma}\lambda$4342, [OIII]$\lambda$5008, [OI]$\lambda$6302, [NII]$\lambda$6550 and [NII]$\lambda$6585.

As described above, the non-parametric fit adapts to the line shape and is thus able to fit a DP structure. It is hence possible to disentangle single Gaussian profiles from  non-Gaussian profiles. With the reduced $\chi^2_{\nu}$ value of the single Gaussian and non-parametric RCSED fit, we can compare the spectra whose emission lines resemble a Gaussian shape with those spectra showing an unlikely Gaussian shape, e.g. a DP. We select $99\,740$ galaxies with a larger $\chi^2_{\nu}$ value for the Gaussian fit than for the non-parametric fit. They are classified as preliminary candidates. Spectra with a larger $\chi^2_{\nu}$ value for the non-parametric fit are selected as the control sample (CS), since they have more likely a Gaussian shape (see detailed description in Sect.\,\ref{ssect:CS:selection}).

\subsubsection{Emission Line Stacking and Fitting Procedure}\label{ssect:stack:and:fit}
The emission-line fittings performed in the RCSED catalogue are separated for Balmer and forbidden emission lines since they can originate from different parts of the galaxy \citep{2017ApJS..228...14C}. For the preliminary selected galaxies, we only find $\sim1\%$ showing a deviation greater than the SDSS spectral resolution between the two fits regarding the emission line position or dispersion. The RCSED catalogue has also excluded SDSS objects classified as quasars or Seyfert 1 \citep{2010AJ....139.2360S} since the stellar population analysis or the k-correction \citep[e.g.][]{2010MNRAS.405.1409C} is not supporting these kind of objects \citep{2017ApJS..228...14C}. Hence, most broad-line galaxies are not included and we are not investigating such line shapes in this study.

We can thus assume {\em a priori} the same or a similar shape in all emission lines. This motivates a stacking procedure of the different emission lines of each spectra. While the shape of a single emission line can be distorted by noise, genuine signals will be enhanced in the stacked spectra characterised by a reduced noise. To guarantee the significance of this procedure, we require ${\rm S/N}>5$ for all stacked emission lines. To stack the emission lines, we select each emission line in the range of $\pm 30 \AA$ with respect to the emission line position and transform it into velocities.
We calculate the velocity of each wavelength bin $i$ as $v_i = {\rm c} \, (\Delta \lambda_{i})/\lambda_{\rm rest}$, where $\Delta \lambda_{i}$ is the difference between the wavelength of the bin $i$ and the observed emission-line position and $\lambda_{\rm rest}$ its rest-frame wavelength. We calculate the stacked spectra by summing the flux of each selected emission line in each velocity bin $v_i$ and add the uncertainties quadratically.

We isolate the ${\rm H}{\alpha}\lambda$6565, [NII]$\lambda$6550 and [NII]$\lambda$6585 emission lines for the stacking procedure. They can overlap and cause artefacts in the stacked spectra. The [NII]$\lambda$6550, $\lambda$6585  doublet is characterised by a fixed flux ratio of ${\rm [NII]}\lambda6585/{\rm [NII]}\lambda6550 = 2.92 \pm 0.32$ \citep{1989Msngr..58...44A}. Using the non-parametric emission-line fit from  RCSED we can extrapolate and subtract the [NII] doublet, as done in \citet{2013ApJ...763...60S}. 

We fit a single Gaussian function (${\rm g_{single}}$) and a double Gaussian function (${\rm g_{double}}$) against each stacked spectrum. We use the following functions for the adjustments: 
\begin{equation} \label{eq:single:gauss}
{\rm g_{single}}(v) = A \, \exp\left({\frac{ (v - \mu)^2}{2\sigma^2}}\right) + B \, ,
\end{equation}
where $A$ is the amplitude, $\mu$ the mean and $\sigma$ the standard deviation of the Gaussian function, $v$ the velocity and $B$ a constant accounting for the background noise level.
\begin{equation} \label{eq:double:gauss}
{\rm g_{double}}(v) = A_{1} \, \exp\left({\frac{(v - \mu_{1})^2}{2\sigma_{1}^2}}\right) + A_{2} \, \exp\left({\frac{(v - \mu_{2})^2}{2\sigma_{2}^2}}\right) + B
\end{equation}
In Eq. (\ref{eq:double:gauss}), we use the same notation as Eq. (\ref{eq:single:gauss}) with subscripts (1,2) defining the first and second Gaussian components. All fitting procedures are performed using the data analysis framework ROOT\footnote{\copyright\,\,Copyright CERN 2014-18 (http://root.cern.ch/).}

We then apply criteria to select DP candidates with the fit procedure, as follows:
\begin{enumerate}
  \item   $1/2.5 < {\rm A}_{1}/{\rm A}_{2} < 2.5$ 
  \item   $\Delta v_{\rm DP} = |\mu_2 - \mu_2| > 3\,\delta v$ 
  \item   F-test
\end{enumerate}
Criteria (1) ensures that one of the two possible peaks is not suppressed or does not represent only noise. Criteria (2) demands the separation of the two peaks to be three times greater than $\delta v$, the bin-width of the spectroscopic observation, transformed into a velocity, which is $3\delta v = 207 \,{\rm km\,s^{-1}}$. The F-test of criteria (3) directly compares the two fitted models and demands a significant decrease in $\chi^2$ relative to the increase in the number degree of freedom (Ndof) for the double Gaussian fit (subset "d") in comparison to the single Gaussian fit (subset "s"). Following \citet{mendenhall2011second}, we calculate the F-statistic, as follows:
\begin{equation}
    {\rm f}_{\rm stat} = \frac{(\chi^2_{\rm s} - \chi^2_{\rm d})/ ({\rm Ndof}_{\rm s} - {\rm Ndof}_{\rm d})}{\chi^2_{\rm d} / {\rm Ndof}_{\rm d}}.
\end{equation}
and demand the Fisher-distribution F to reject the single Gaussian hypothesis with a probability of less than $5\%$ by using the cumulative distribution function:
\begin{equation}
{\rm F_{cdf}}({\rm f}_{\rm stat}| {\rm Ndof}_{\rm s} - {\rm Ndof}_{\rm d}, {\rm Ndof}_{\rm d}) > 0.95
\end{equation}
With these criteria, we select $7\,479$ galaxies.

The Gaussian velocity dispersions $\sigma_{\rm i}$, that we measure directly from the spectra, need to be corrected for the instrumental broadening $\sigma_{\rm inst}$  \citep[as e.g. discussed in][]{2004ApJ...617..903W}. We calculate the corrected dispersion as
$\sigma_{\rm i,\,corr} = \sqrt{\sigma_{\rm i}^2 - \sigma_{\rm inst}^2}$, where $\sigma_{\rm i}$ corresponds to $\sigma$, $\sigma_{\rm 1}$ and $\sigma_{\rm 2}$.
The resolution of the SDSS spectra is not constant for the covered wavelength range and decreases towards higher wavelengths. To correct the stacked-spectra velocity dispersions, we use the mean $\sigma_{\rm inst}$ computed over the selected emission lines. We find a mean $\sigma_{\rm inst} = 61 \pm 4 {\rm km\,s^{-1}}$. In the subsequent analysis, we only discuss corrected velocity dispersions.

\subsubsection{Control Sample Selection}\label{ssect:CS:selection}
For later analysis, we select a Control Sample (CS) to compare with our DPS. This sample is selected during the first stage and corresponds to galaxies showing no evidence of any DP feature. The preliminary sample, selected in Sect.\,\ref{ssect:preliminary:candidates}, contains $189\,152$ galaxies, and is then divided into two subsamples using the Gaussian and the non-parametric fits provided in the RCSED. We keep spectra exhibiting a larger $\chi^2_{\nu}$ value for the non-parametric than for the Gaussian fit to select galaxies showing Gaussian shaped emission lines. With this criterion, we select $89\,412$ galaxies, building up the CS. Since we consider the same S/N thresholds for emission lines and the same maximal redshift as for the DP candidates, this is a representative control sample. Nevertheless, as further discussed in Sect.\,\ref{ssect:bias}, this CS still shows a selection bias in the redshift and stellar mass distributions, and a no-bias control sample (NBCS) will be selected.
\subsubsection{Individual emission line fitting}\label{ssect:individual:line:fit}
In the second stage of the selection procedure, we examine the following emission lines separately: ${\rm H}{\gamma}\lambda$4342, ${\rm H}{\beta}\lambda$4863, [OIII]$\lambda$5008, $\rm [OI]\lambda$6302, ${\rm [NII]}\lambda$ 6550, ${\rm H}{\alpha}\lambda$6565, ${\rm [NII]}\lambda$6585, ${\rm [SII]}\lambda$6718 and ${\rm [SII]}\lambda$6733. We fit a single and a double Gaussian function to each line. 
For the double Gaussian function, we set the parameters $\mu_{1,2}$ and $\sigma_{1,2}$ provided by the best fit of the stacked emission lines (see Sect.\,\ref{ssect:stack:and:fit}) and let them vary only inside their uncertainty range $\pm{\rm E_{\mu_{1,2}}}$ and $\pm{\rm E_{\sigma_{1,2}}}$. The uncertainties are usually smaller than the spectral bin size $\delta \lambda$ of the SDSS which means that these values are quasi fixed. In the case of uncertainties larger than $0.4\times \delta \lambda$, we fix them to $0.4\times \delta \lambda$.

Using the best fit results of the single and double Gaussian fit functions, we apply the following criteria to flag each line if we detect a DP:
\begin{enumerate}
	\item $A_{1}, A_{2} > 3\,\sigma_{b}$
	\item $\chi^2_{\nu}({\rm single}) > \chi^2_{\nu}({\rm double})$\
	\item ${1}/{3} < A_{1}/A_{2} < 3$
    \item $\rm S/N > 5$
\end{enumerate}
where $\sigma_{b}$ is the root mean square (RMS) of the background noise level measured on both sides of the emission line. The first criterion ensures that the amplitude of each DP component is significantly larger than the background noise. The second criterion constrains that the double Gaussian function is better fitting the data than the single Gaussian one. With the third criterion, we exclude emission lines where one DP component is suppressed. In this case, it is likely that the weak component represents noise or an artefact of a clumpy line shape. \citet{2019A&A...627L...3M} discussed genuine cases where one of the components is weak or suppressed despite a high S/N ratio. Criteria four ensures that the fitted lines are detectable and are not just noise.

For those lines showing a DP according to the selection criteria above, we set a flag to highlight the specific line as a DP line.

\subsubsection{Final selection of double-peak galaxies}\label{ssect:dp:selection}
\begin{figure}
  \centering 
 \includegraphics[width=0.48\textwidth]{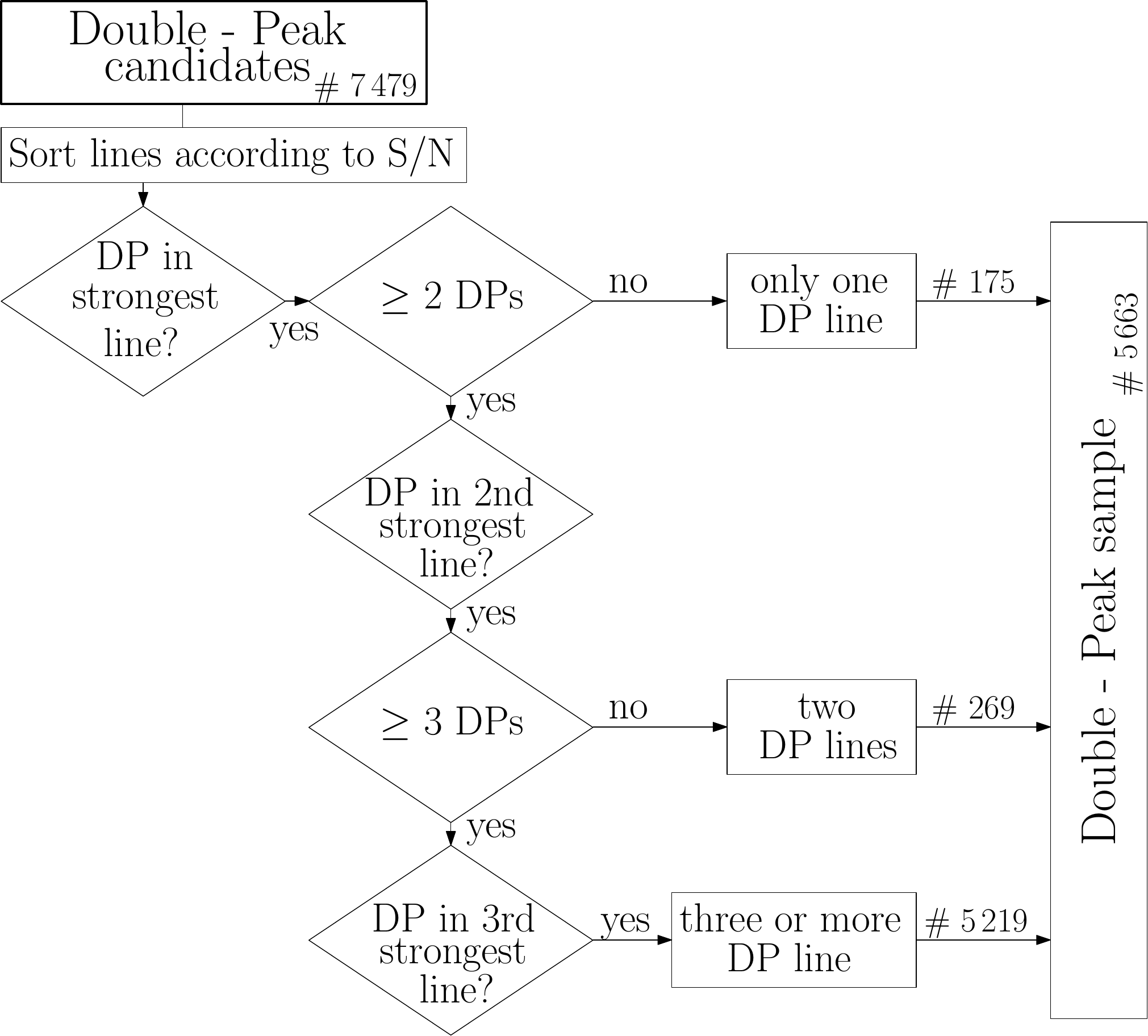}
  \caption{Stage 3 of the selection procedure to find the final DPS. This algorithm sorts the fitted emission lines according to their S/N and uses the DP flags indicating confirmed DP in the specific line. All selected galaxies must have a confirmed DP in the strongest line. If the object has more than 1 (resp. 2) confirmed line(s), we must also find a DP in the second (resp. second and third) strongest line. This procedure excludes objects which have a falsely confirmed DP in some weaker lines. We finally get the DPS counting $5\,663$ galaxies. In Table\,\ref{table:number:dp}, we present the distribution of DP galaxies for different number of confirmed DP.}
  \label{fig:Flowchart_2}%
\end{figure}
In the third stage of the selection procedure, we exclude galaxies which do not show any DP in the strongest emission lines which are mostly misclassified due to an artificial DP structure created by the stacking procedure. This third-stage selection is illustrated in Fig.\,\ref{fig:Flowchart_2}.

We keep spectra with the strongest line flagged as DP. If we have two (resp. more than two) emission lines flagged as DP, we also demand the second (resp. second and third) strongest line to be flagged as DP.
With these criteria, we exclude $1\,816$ galaxies, which do not show a DP in their strong lines. This can occur for different reasons: (1) the fitting procedure can fail at specific lines because of a noisy line shape, (2) the spectra show only a DP structure in the stacked spectra but not in any individual line or (3) DP occur only in weak lines, which are dominated by noise.

\begin{table*}
\centering      
\caption{The DPS sorted into groups of number of confirmed DP lines}\label{table:number:dp}
\vspace{-0.2cm}
\begin{tabular}{c c c c c c c c c c}   
\hline\hline
Number of lines with confirmed DP & 1 & 2 & 3 & 4 & 5 & 6 & 7 & 8 & 9 \\
\hline
Number of Objects & 175  & 269 & 1149 & 1634 & 897 & 800 & 543 & 169 & 27 \\
\hline
\end{tabular}
\end{table*}
\begin{table*}
\caption{The double-peak galaxy sample}\label{table:fitdata}
\vspace{-0.6cm}
\begin{center}
\begin{tabular}{c c c c c c c c}        
\hline\hline                       
Identifier  & \multicolumn{1}{c}{$\mu_{1}$} & \multicolumn{1}{c}{$\mu_{2}$} &\multicolumn{1}{c}{ $\sigma_{1}$} & \multicolumn{1}{c}{$\sigma_{2}$} & ${\rm DP}_{\rm H_{\beta}}$ & \multicolumn{1}{c}{${\rm H}{\beta}$ $\rm flux_{1}$} & \multicolumn{1}{c}{${\rm H}{\beta}$ $\rm flux_{2}$}  \\ 
 & \multicolumn{1}{c}{$\rm km\,s^{-1}$} & \multicolumn{1}{c}{$\rm km\,s^{-1}$} & \multicolumn{1}{c}{$\rm km\,s^{-1}$} & \multicolumn{1}{c}{$\rm km\,s^{-1}$} & &\multicolumn{1}{c}{ $\rm 10^{-17}erg\,cm^{2}\,s^{-1}$} & \multicolumn{1}{c}{$10^{-17}\,\rm erg\,cm^{2}\,s^{-1}$ } \\
\multicolumn{1}{c}{(1)} & \multicolumn{1}{c}{(2)} & \multicolumn{1}{c}{(3)} & \multicolumn{1}{c}{(4)} & \multicolumn{1}{c}{(5)} & \multicolumn{1}{c}{(6)} & \multicolumn{1}{c}{(7)} & \multicolumn{1}{c}{(8)}  \\
\hline  
DPS 1 & $ -81 \pm 9$& $ 219 \pm 10$& $ 150 \pm 6$ & $ 119 \pm 6$& $ 1 $& $ 57 \pm 5$ &$ 33 \pm 4$ \\
DPS 2 & $ -194 \pm 7$& $ 43 \pm 11$& $ 91 \pm 5$ & $ 108 \pm 8$& $ 0 $& $ 25 \pm 4$ &$ 34 \pm 5$ \\
DPS 3 & $ -187 \pm 11$& $ 163 \pm 8$& $ 158 \pm 10$ & $ 109 \pm 7$& $ 1 $& $ 39 \pm 4$ &$ 27 \pm 3$ \\
DPS 4 & $ -89 \pm 28$& $ 125 \pm 27$& $ 110 \pm 15$ & $ 99 \pm 13$& $ 1 $& $ 47 \pm 7$ &$ 42 \pm 8$ \\
DPS 5 & $ -99 \pm 6$& $ 148 \pm 5$& $ 110 \pm 4$ & $ 106 \pm 3$& $ 1 $& $ 71 \pm 5$ &$ 58 \pm 5$ \\
\hline 
\end{tabular}
\end{center}
\vspace{0.1cm}
{\justifying \small
{\noindent {\bf Notes:} We provide the parameters obtained by the two fitting procedures described in sec. \ref{ssect:stack:and:fit} and \ref{ssect:individual:line:fit}. The first column (1) provides the object identifier. We provide the fit-parameters and their uncertainties for the stacked spectra in columns (2-5). In column (6) we provide the DP flag exemplary for the  ${\rm H}{\beta}\lambda$4863 line and the flux of the two DP components in column (7) and (8). This table is available in its entirety in digital form and provides parameters for: ${\rm H}{\gamma}\lambda$4342, ${\rm H}{\beta}\lambda$4863, [OIII]$\lambda$5008, $\rm [OI]\lambda$6302, ${\rm [NII]}\lambda$ 6550, ${\rm H}{\alpha}\lambda$6565, ${\rm [NII]}\lambda$ 6585, ${\rm [SII]}\lambda$6718 and ${\rm [SII]}\lambda$ 6733.}}
\end{table*}
\begin{figure*}
\centering 
\includegraphics[width=0.98\textwidth]{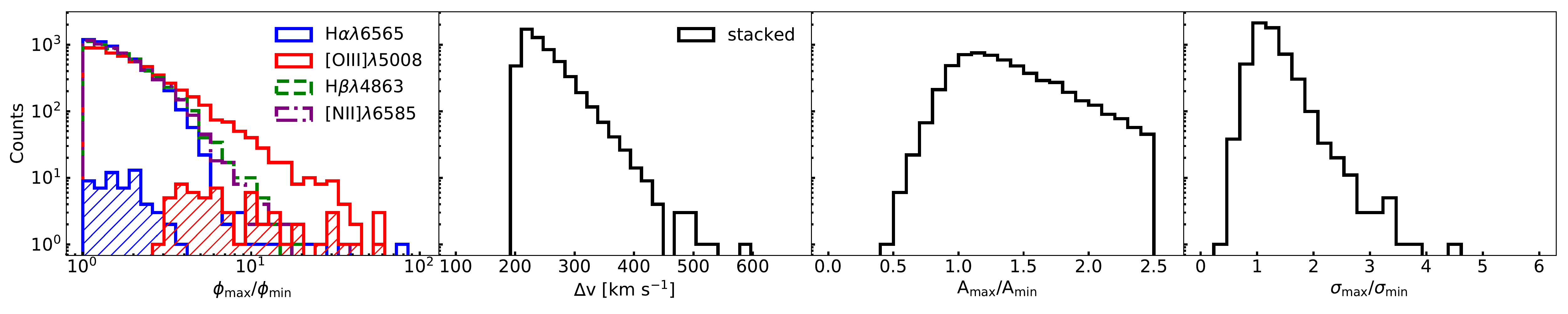}
\caption{Characteristics of the galaxies selected with a DP feature in their emission lines. From left to right: The first panel shows the flux ratio $\phi_{\rm max} / \phi_{\rm min}$ of the stronger line divided by the weaker line for the individual emission lines ${\rm H}{\alpha}\lambda$6565, [OIII]$\lambda$5008, ${\rm H}{\beta}\lambda$4863 and ${\rm [NII]}\lambda$6585. We display the ${\rm H}{\alpha}\lambda$6565 and [OIII]$\lambda$5008 emission line ratios of the objects found in \citet{2019A&A...627L...3M} with blue and re hatched surfaces. The other panels display parameters computed on the stacked spectra  (see Sect.\,\ref{ssect:stack:and:fit}).  The second panel shows the velocity difference $\Delta {\rm v}$ between the two peak components taken from the stacking procedure. The third (resp. fourth) panel displays the amplitude (resp. velocity dispersion) ratio of the stronger and weaker line components of the stacked spectra.}\label{fig:peak:properties}
\end{figure*}
Our final DPS contains $5\,663$ galaxies with $\Delta v = |\mu_2 - \mu_1|$ between 211 and 582 km/s. In Fig.\,\ref{fig:peak:properties}, we show distributions of the flux ratios between the two fit components of the ${\rm H}{\alpha}\lambda$6565, [OIII]$\lambda$5008, ${\rm H}{\beta}\lambda$4863 and ${\rm [NII]}\lambda$6585 emission lines. There is a noticeable difference between the flux ratio $\phi_{\rm max}/\phi_{\rm min}$ of ${\rm H}{\alpha}$ and ${\rm [OIII]}$. We also present the distribution of $\Delta {\rm v_{DP}}$ and the measured ratio of the amplitudes and velocity dispersions $\sigma_{\rm max}/\sigma_{\rm min}$ of the two components in the stacked spectra. 

The number of confirmed DP lines varies between 1 and 9 and is presented in Table\,\ref{table:number:dp}. $92\%$ (resp. $72\%$) of the selected galaxies exhibit 3 (resp. 4) or more DP emission lines. 

The automated selection procedure selects DP galaxies with an objective algorithm. This means that we do not need any visual inspection, which would be a subjective factor in the sample selection. We show in Table\,\ref{table:fitdata} some fitting parameters of the first five galaxies of our DPS.

\subsection{No-bias control sample.}\label{ssect:bias}
\begin{figure*}
  \centering 
 \includegraphics[width=0.98\textwidth]{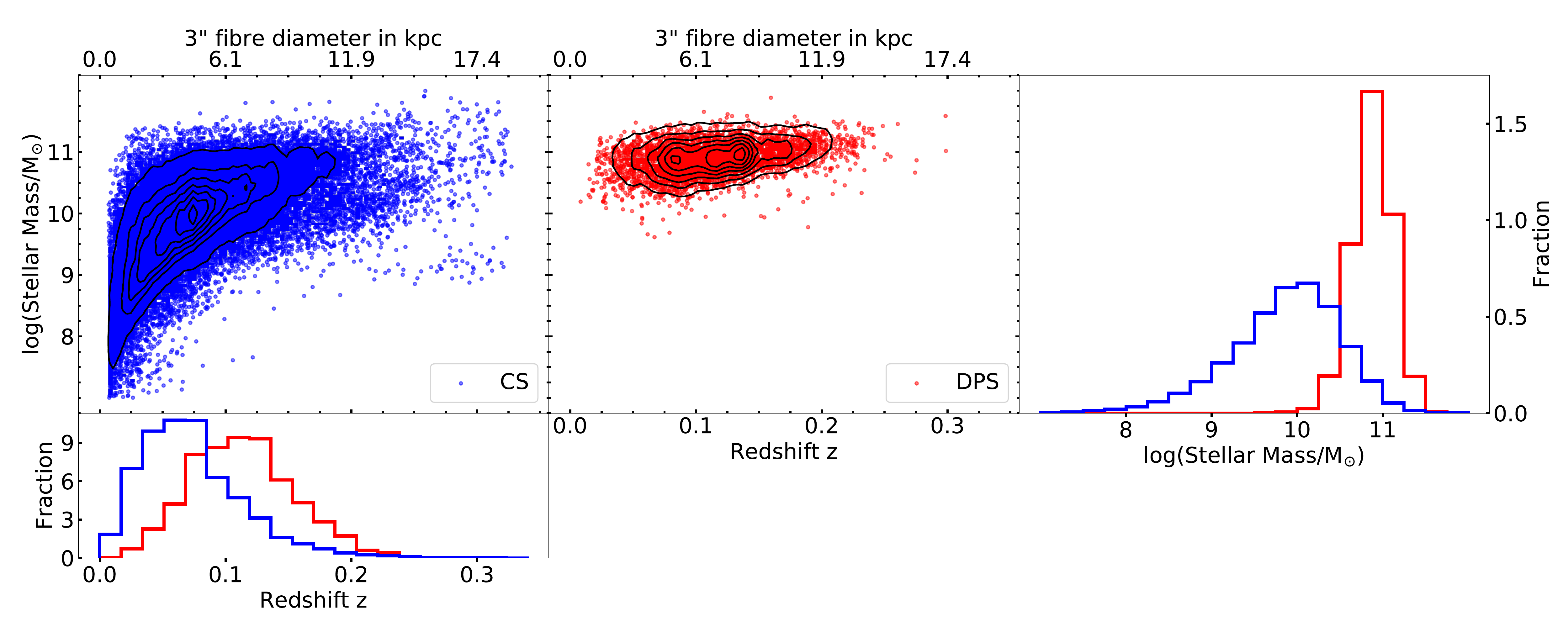}
  \caption{Stellar mass-redshift distribution for the CS (blue, top left panel) and the DPS (red, top middle panel). The contours indicate the density level. In the top right panel, we show the histogram of stellar masses and, on the bottom left panel, the histogram of redshifts. We also display the fiber diameter corresponding to $3"$ on top of the upper left and middle panels.}
  \label{fig:stellar:mass:z:dist:cs}
\end{figure*}

\begin{figure}
  \centering 
 \includegraphics[width=0.48\textwidth]{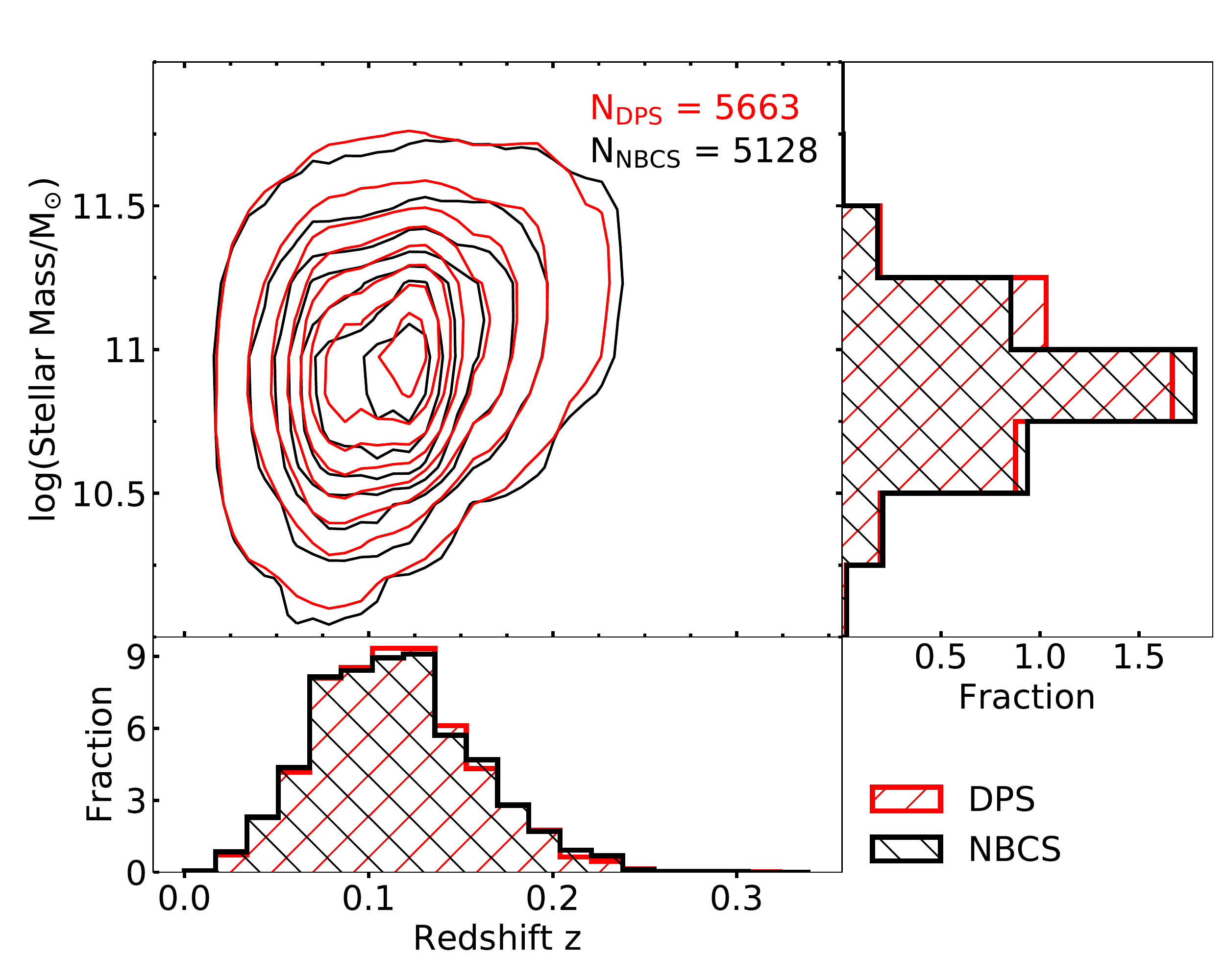}
  \caption{Stellar mass-redshift distribution (see Fig.\,\ref{fig:stellar:mass:z:dist:cs}) for the DPS (red) and the NBCS (black). The selection of the NBCS is explained in Sect.\,\ref{ssect:bias}. We also compute histograms for the redshift and the stellar mass for both samples for a better comparison. The NBCS follows the same distribution as the DPS and contains $5\,128$ galaxies.}
  \label{fig:stellar:mass:z:dist:nbcs}
\end{figure}

Figure \ref{fig:stellar:mass:z:dist:cs} displays stellar mass \citep{2003MNRAS.341...33K} as a function of redshift for the CS and for the DPS. We can observe that very few low-redshift objects are present in the DPS. This is obviously a selection bias: our method ends up excluding small systems with ${\rm M}_* < 10^{10}\,{\rm M}_\odot$. This is due to the fact that we cut ${\rm \Delta v_{DP}} < 211{\rm km\,s^{-1}}$: those two quantities are related through the Tully Fisher relation \citep{1977A&A....54..661T}, as  further discussed in Sect.\,\ref{ssect:TF:FJ}. We thus indirectly cut on the redshift and the fiber size. Hence, most double-peak detections correspond to a fibre size larger than 3\,kpc and massive galaxies which have ${\rm M}_* > 10^{10.4}\,{\rm M}_\odot$. The maximal redshift observed in the DPS is ${\rm z} = 0.32$ (but this particular galaxy has no stellar mass approximation and is thus not represented in Fig.\,\ref{fig:stellar:mass:z:dist:cs}). Only $< 1\%$ of the DP galaxies have a redshift $z > 0.25$. This can be explained by the S/N cut of emission lines in the selection procedure (see Sect.\,\ref{ssect:preliminary:candidates}).
For comparison purposes, we define a sample of ordinary galaxies showing no DP emission lines but following the same redshift and stellar mass distribution as the DPS. We randomly select galaxies from the CS (defined in Sect.\,\ref{ssect:CS:selection}) following the same redshift-stellar mass distribution as the DPS. Therefore, we divide the redshift-stellar mass space into a grid of 20$\times$20 boxes and randomly draw galaxies from the CS in each box, following the probability of finding a DP galaxy in the specific box. We are thus able to select $5128$ galaxies from the CS as shown in Fig.\,\ref{fig:stellar:mass:z:dist:nbcs}. In order to keep the same redshift-stellar mass distribution for the NBCS as the DPS, it is not possible to exceed $5128$ galaxies in the NBCS. This new sample has approximately the same redshift and stellar mass properties as the DPS but shows single Gaussian shaped emission lines and is hereafter called the no-bias control sample (NBCS). We present the redshift and stellar mass distributions for both samples in Fig.\,\ref{fig:stellar:mass:z:dist:nbcs}.

\subsection{Comparison with other works on DP}\label{ssect:comparison}
\begin{table*}
\caption{Comparison of our DPS to other similar works with respect to the selection procedure of this work.}\label{table:comparison}
\vspace{-0.6cm}
\begin{center}
\begin{tabular}{l r r r r r r}
\hline\hline
\multicolumn{2}{l}{Previous work} & \multicolumn{4}{l}{Cross matched samples}\\
\hline
Sample & Size & RCSED & Preliminary Candidates & DP Candidates & DPS \\
\hline
\citet{2009ApJ...705L..76W} & (87) & 83 & 82 & 75 & 58 \\ 
\citet{2010ApJ...708..427L} & (167) & 135 & 129 & 114 & 73 \\
\citet{2010ApJ...716..866S} & (148) & 1 & 1 & 1 & 0 \\
\citet{2012ApJS..201...31G} G12-DP & ($3\,030$) & $2\,794$ & $1\,255$ & $1\,179$ & 947 \\
\citet{2012ApJS..201...31G} G12-TFAS & ($12\,582$) & $11\,475$ & $6\,277$ & $3\,585$ & $2\,967$ \\
\hline
\end{tabular}
\end{center}
\vspace{0.1cm}
{\justifying \small
{\noindent {\bf Notes:} We present the comparison of our DPS to other similar works with respect to the selection procedure of this work. We show in the first column the reference of the sample and the size in the second. In the third column, we show a cross-match with the RCSED catalogue \citep{2017ApJS..228...14C}. We display the number of galaxies found in the preliminary candidate sample, the DP candidates and the final DP galaxies in the last three columns, respectively (see Sect.\,\ref{ssect:preliminary:candidates}, \ref{ssect:stack:and:fit} and \ref{ssect:dp:selection}).}}
\end{table*}
In this Section, we compare our DPS with previous samples. In Table\,\ref{table:comparison}, we cross-identify our samples at different selection steps with 4 other works, which released galaxy samples defined with DP galaxies or asymmetric features. Namely, we present a comparison with the samples found by \citet{2009ApJ...705L..76W} and \citet{2010ApJ...708..427L}, which comprise 87 and 167 Type 2 DP AGNs detected with the [OIII]$\lambda$5008 line. The sample of \citet{2010ApJ...716..866S} comprises 148 quasars classified as type 1 and type 2 AGNs. \citet{2012ApJS..201...31G} conducted a much broader selection procedure and provide two samples: one composed of 3030 galaxies with DP emission lines (hereafter G12-DP) and a second one gathering $12\,582$ galaxies with top flat or asymmetric line shapes (hereafter G12-TFAS).

Our DP detection algorithm is based on a stacking procedure, which enables to study galaxies with at least three significant emission lines including a strong [OIII]$\lambda$5008 or ${\rm H}{\alpha}\lambda$6565 line. These requirements are encoded among others in the preliminary selection (see Sect.\,\ref{ssect:preliminary:candidates}). By comparing the sample of our preliminary candidates of $99\,740$ galaxies (see Sect.\,\ref{ssect:preliminary:candidates}), we detect $71\%$ (resp. $57\%$) of the sample found by \citet{2009ApJ...705L..76W} \citep[resp.][]{2010ApJ...708..427L}. 
Those galaxies, which fail our DP detection algorithm, show mostly different emission-line shapes between the [OIII]$\lambda$5008 and ${\rm H}{\alpha}\lambda$6565 lines. Those galaxies are then systematically sorted out in stage three of our selection procedure (see Sect.\,\ref{ssect:dp:selection}).
We find only one DP galaxy in the RCSED catalogue selected by \citet{2010ApJ...716..866S}, which also fails our final selection procedure due to an irregular ${\rm H}{\alpha}\lambda$6565 shape where the algorithm is not able to adjust a double Gaussian function with fixed $\mu_{1,2}$ and $\sigma_{1,2}$ (see Sect.\,\ref{ssect:dp:selection}). These are studies based on AGN quasars \citep{2010AJ....139.2360S} which were excluded from both the RCSED \citep{2017ApJS..228...14C} and our present study.

By comparing our work to the catalogues provided by \citet{2012ApJS..201...31G}, we find a detection rate of $75\%$ (resp. $47\%$) for the G12-DP (resp. G12-TFAS) sample of those detected in our preliminary sample. We find some similarities between the two catalogues but we find only 947 galaxies on the G12-DP sample and 2967 from the G12-TFAS sample. In addition, most of the galaxies found by \citet{2012ApJS..201...31G} have been discarded in our algorithm since they do not show a ${\rm S/N} > 5$ in at least 3 emission-lines, or are better fitted by a single Gaussian function than a non-parametric function (see Sect.\,\ref{ssect:preliminary:candidates}).

In this work, we present a new DP sample with a very different selection procedure in comparison to previous works which all relied on visual inspection at some stage. Our sample is selected by an algorithm without any subjective post processing. 

\subsection{Summary}
We developed an automated selection procedure to find double-peak emission-line galaxies. We present 5663 such galaxies showing a wide range of possible emission-line shapes (different Flux, Amplitude or $\sigma$ ratios, see Fig.\,\ref{fig:peak:properties}). Due to a wide range of explanation for DP phenomena, we classify our sample in the following and present an analysis.  

\section{Sample Classification}\label{sect:classification}

\subsection{Classification based on ionisation diagrams}\label{ssect:bpt}
\begin{figure*}[h]
\centering 
\includegraphics[width=1\textwidth]{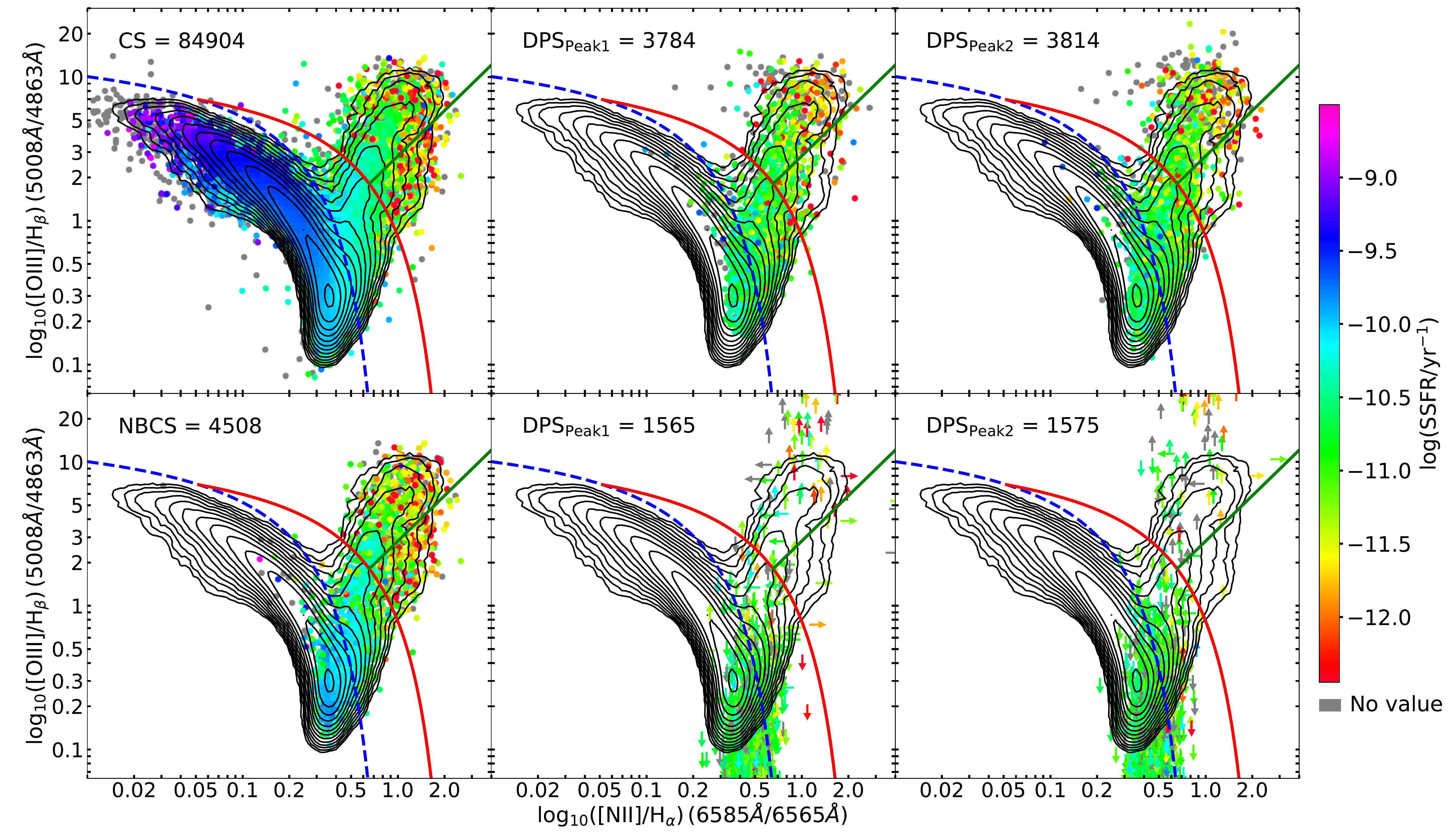}
\caption{BPT diagrams used to classify our samples into different galaxy types \citep{2006MNRAS.372..961K}. The red line separates active galactic nuclei (AGN) and composite galaxies (COMP) \citep{2001ApJ...556..121K}. AGNs and low ionisation narrow emission-line regions (LINER) are separated by the green line \citep{2007MNRAS.382.1415S} and the blue dashed lines separate star forming (SF) and COMP \citep{2006MNRAS.372..961K}. We present the CS in the upper left panel and the NBCS in the lower left panel. In the upper middle (resp. right) panel, we show the blueshifted (resp. redshifted) peaks of DP galaxies with a S/N$>3$ in the 4 emission-line fluxes needed for the BPT diagram. In the lower middle and right panels, we display those emission-line components of the DPS with three of the requested emission lines with S/N$>3$ and the fourth S/N$<3$. To classify them, we display arrows to indicate the limits derived by the uncertainties of weak emission lines. In all 6 panels, we apply the same colour-coding indicating the specific star formation rate (SSFR) computed by \citet{2016ApJS..227....2S}. We display in all panels the same contour lines corresponding to the density of the CS. The corresponding classifications are presented in Table \ref{table:bpt}.}
\label{fig:first:bpt}
\end{figure*}
To identify different galaxy types, BPT diagnostic diagrams provide an empirical classification based on optical emission-line flux ratios first introduced by \citet*{1981PASP...93....5B}, which have been further tuned by \citet{2006MNRAS.372..961K} and \citet{2007MNRAS.382.1415S}. 
We use three different types of BPT diagrams: the first type uses the emission-line flux ratio ${\rm [OIII]}$5008/${\rm H}{\beta}$4863 on the $y$-axis and ${\rm [NII]}$6585/${\rm H}{\alpha}$6565 on the $x$-axis as shown in Fig.\,\ref{fig:first:bpt}. 
The second and third types of BPT diagrams 
use (${\rm [SII]}$6718 + ${\rm [SII]}$6733)/${\rm H}{\alpha}$6565 (resp. $\rm [OI]$6302/${\rm H}{\alpha}$6565) on the $x$-axis, while the $y$-axis is the same as the first type. 
We classify galaxies with the first BPT diagram into 4 groups: star-forming (SF) galaxies, active galactic nuclei (AGN), composite (COMP) galaxies and low-ionisation narrow emission-line regions (LINER) \citep{2007MNRAS.382.1415S}. In the case of the DPS, we display each of the two emission-line components in the BPT-diagrams, as the two peaks are most probably emitted from two different regions.

We classify the two emission-line components according to their position on the BPT diagram if all four needed emission lines have a ${\rm S/N} > 3$. In the case where we only have three emission lines with a ${\rm S/N} > 3$, we compute the upper/lower limits derived from the uncertainties of the undetected emission line. If the constraints of these emission-line ratios are unambiguous, we can classify them. Since the COMP region in the first BPT diagram borders the AGN and the SF regions, it is not possible to classify galaxies as COMP unambiguously using the flux limits. In Fig.\,\ref{fig:first:bpt}, 
we show the classification of the single lines of the CS, the NBCS and each emission-line component of the DPS. We also show with arrows upper/lower limits of these DP galaxies, which only have three significant emission-line components. In all BPT diagrams, we colour-code the specific star formation ratio computed by \citet{2004MNRAS.351.1151B}.

\begin{table}[h]   
\caption{Classification based on BPT diagrams (see Fig.\,\ref{fig:first:bpt} 
)}\label{table:bpt}  
\vspace{-0.6cm}
\begin{center}
\begin{tabular}{l c c}          
\hline
Criteria & Counts & fraction  \\
\hline
\multicolumn{3}{c}{DPS} \\
\hline
double SF & 1617 & (28.6 \%) \\
double COMP & 939 & (16.6 \%) \\
double AGN & 385 & (6.8 \%) \\
double LINER & 29 & (0.5 \%) \\
\hline
SF + COMP & 784 & (13.8 \%) \\
SF + AGN & 17 & (0.3 \%) \\
SF + LINER & 9 & (0.2 \%) \\
COMP + AGN & 167 & (2.9 \%) \\
COMP + LINER & 64 & (1.1 \%) \\
AGN + LINER & 59 & (1.0 \%) \\
\hline
SF + uncertain & 570 & (10.1 \%) \\
COMP + uncertain & 592 & (10.5 \%) \\
AGN + uncertain & 110 & (1.9 \%) \\
LINER + uncertain & 63 & (1.1 \%) \\
\hline
Not classifiable & 258 & (4.6 \%) \\
\hline
\multicolumn{3}{c}{DPS with non-parametric fit} \\
\hline
SF & 2534 & (44.7 \%) \\
COMP & 2153 & (38.0 \%) \\
AGN & 630 & (11.1 \%) \\
LINER & 174 & (3.1 \%) \\
\hline
Not classifiable & 172 & (3.0 \%) \\
\hline
\multicolumn{3}{c}{No-Bias Control sample} \\
\hline
SF & 2811 & (54.8 \%) \\
COMP & 1226 & (23.9 \%) \\
AGN & 687 & (13.4 \%) \\
LINER & 308 & (6.0 \%) \\
\hline
Not classifiable & 96 & (1.9 \%) \\
\hline
\multicolumn{3}{c}{Control sample} \\
\hline
SF & 82065 & (91.8 \%) \\
COMP & 4721 & (5.3 \%) \\
AGN & 1623 & (1.8 \%) \\
LINER & 536 & (0.6 \%) \\
\hline
Not classifiable & 467 & (0.5 \%) \\
\hline
\end{tabular}
\end{center}                              
\vspace{0.1cm}
{\justifying \small
{\noindent {\bf Notes:} Following \citet{2006MNRAS.372..961K} and \citet{2007MNRAS.382.1415S}, we classify each two emission-line component of the DPS individually, as explained in Sect.\,\ref{ssect:bpt}. We show all combinations of classification such as double (e.g. SF + SF) and mixed (e.g. SF + COMP) classifications. We also display those galaxies which are not classifiable in only one component (e.g. SF + uncertain) or in both. The classifications for the CS and the NBCS are also displayed for comparison.}}
\end{table}
The classification of the first BPT diagram type using all four emission lines with a ${\rm S/N} > 3$ is used first. This classifies about $67\%$ of the bluer and the redder peaks. Galaxies with only three significant emission lines in the first diagram type are classified using the upper/lower limits, which comprises about $16\%$ of the blue and the redshifted peaks. With the second and the third types of BPT diagrams, we only classify $0.4\%$ of the two peak components with all four emission lines having a ${\rm S/N} > 3$. Numerous peak components cannot be classified using the upper/lower limits in the second or the third diagram types. This is because if the ${\rm [NII]}$6585 line is weak, the ${\rm [SII]}$6718, ${\rm [SII]}$6733 or the $\rm [OI]$6302 lines are usually not detected. To directly compare the DPS with the CS and the NBCS, we also perform the same BPT classification as for the CS and NBCS but with the non-parametric emission-line fit performed by \citet{2017ApJS..228...14C}. This classification shows $80$ to $90\%$ agreement with those DP, classified the same way in each component (e.g. double SF). We present the final BPT classification in Table\,\ref{table:bpt} for all the three samples, namely DPS, CS and NBCS.

We classify $88\,945\,(99.5\%)$ galaxies of the CS, $5032\,(98.1\%)$ of the NBCS and $5423\,(95.4\%)$ galaxies from the DPS with an individual classification and $4818\,(85.1\%)$ using the non-parametric fit. The majority of the CS galaxies are SF $92\%$, while only a small amount is classified as COMP ($5\%$) and only a few galaxies, about $2\%$ (resp $1\%$), are classified as AGN (resp. LINER). This is not surprising since we know from Fig.\,\ref{fig:stellar:mass:z:dist:cs} that the CS comprises many small galaxies. This classification is different for the NBCS: the majority ($55\%$) is still classified as SF but we find around $24\%$ to be classified as COMP, and $13\%$ (resp. $6\%$) as AGN (resp. LINER). In comparison, we classify $45\%$ of the DPS as SF using the non-parametric emission-line fit. About $38\%$ of the DPS are classified as COMP showing that we find less SF and more COMP galaxies in comparison to the NBCS. We find a similar AGN fraction of about $11\%$ ($13\%$). We find twice the fraction of galaxies classified as LINER in the NBCS ($6\%$) in comparison to the DPS ($3\%$). 

By classifying each emission-line component of the DPS, we find $29\%$ (resp. $17\%$) to be classified as SF (resp. COMP) in both emission-line components. We also find $14\%$ of the DPS galaxies with one component to be classified as SF and the other one as COMP. According to \citet{2006MNRAS.372..961K}, COMP galaxies are a combination of a SF and a AGN nucleus or a SF and LINER emission. Counting all galaxies which have at least one emission-line component classified as COMP, we find $45\%$. This is an excess of about a factor 2 in the DPS in comparison with the NBCS ($24\%$). Counting only all galaxies classified as "double SF" or "SF and uncertain" in the second peak, we find $39\%$ of the DPS which is significant less in comparison with the NBCS ($55\%$).

Many works published in the recent years were focused on double-peak emission-line AGNs with the motivation to find dual AGNs (see Sect.\,\ref{sect:introduction}). Here, we do not observe the DPS to be dominated by this type of galaxies. We observe about $7\%$ double AGN or LINER classification and $3\%$ with one AGN or LINER and one uncertain. We furthermore find $6\%$ to show a mixed classification with one AGN or LINER component. In comparison with the NBCS, which comprises about $19\%$ to be classified as AGN or LINER, we find less AGN and LINER in the DPS. Recall that QSO and Seyfert1 galaxies have been excluded from the RCSED \citep{2010AJ....139.2360S,2017ApJS..228...14C}, which can explain the lack of galaxies classified as AGN in this catalogue (see also Sect.\,\ref{ssect:comparison}).  \citet{2010ApJ...716..866S} concentrated on DP QSO and found 148 DP galaxies, which are not treated in this work (see Sect.\,\ref{ssect:comparison}).  

\subsection{Morphological Classification} \label{ssect:morph}
\begin{figure}
  \centering 
 \includegraphics[width=0.48\textwidth]{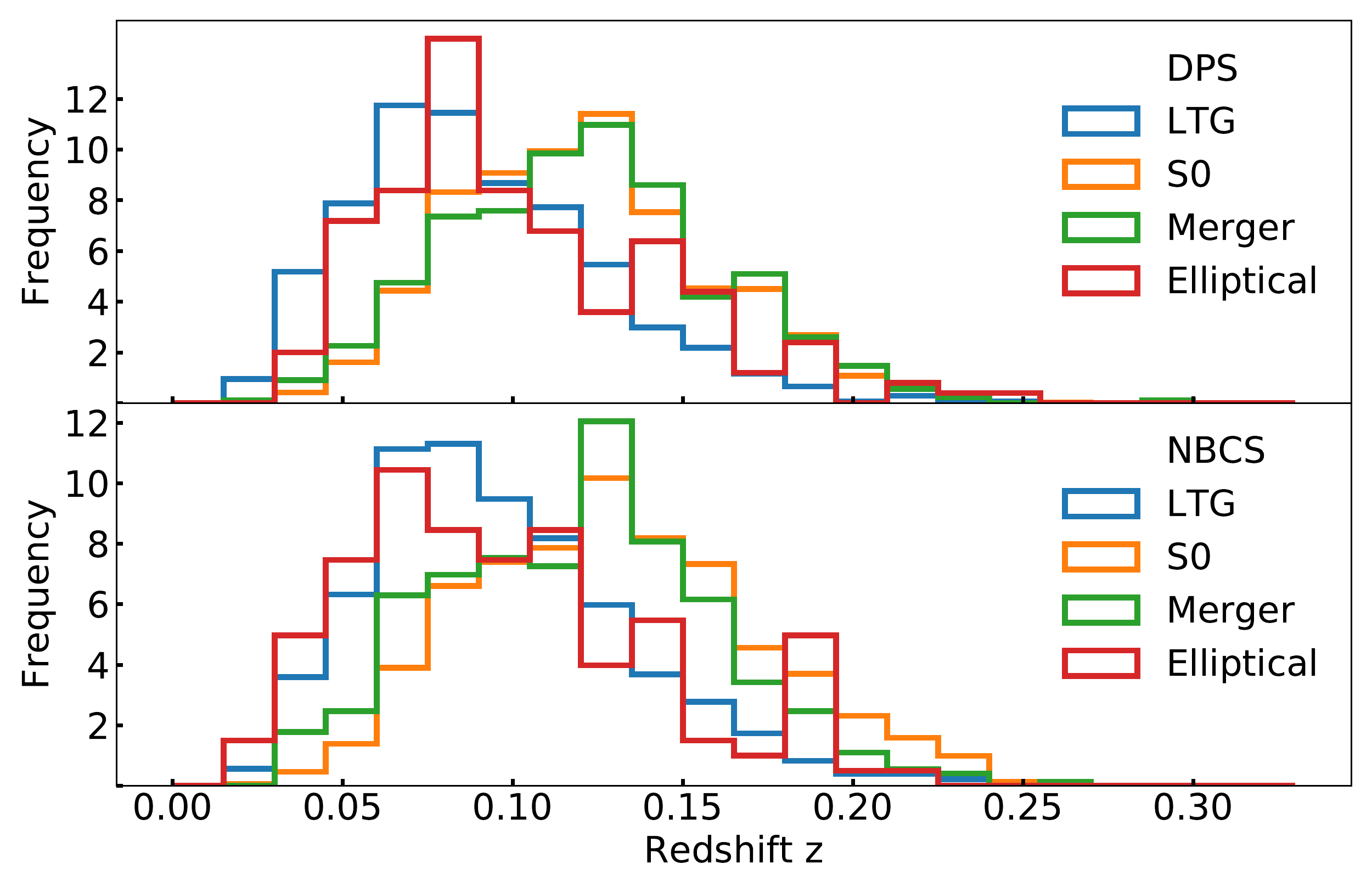}
  \caption{Redshift distribution of the DPS for LTG, S0, merger, and elliptical galaxies. We unified the histogram surface of all four samples to 1.}
  \label{fig:redshift_dist}
\end{figure}

We identify the morphological types of our galaxy samples using \citet{2018MNRAS.476.3661D}, which is a machine-learning-based algorithm to identify different types of galaxy morphologies. Different neuronal networks have been trained with SDSS {\em gri} colour composite image on various criteria to determine their morphology. To classify our samples, we use the following variables to define the probability of an observed feature: ${\rm P_{disc}}$ for disc features, ${\rm P_{S0}}$ for S0 galaxies, ${\rm P_{edge-on}}$ for edge-on orientation, ${\rm P_{merger}}$ for visual merger and ${\rm P_{bar}}$ for a bar structure. The T-type relates each galaxy to a type classification on the Hubble sequence. To classify spiral galaxies according to their inclination, we calculate the inclination angle $i$ as:
\begin{equation}\label{eq:inclination}
    \cos i  = \sqrt{\frac{(b/a)^2 - q_0^2}{1 - q_0^2}}
\end{equation}
where $b/a$ is the r-band minor-to-major axial ratio estimated by \citet{2015MNRAS.446.3943M}, while $q_0$ is the intrinsic axial ratio of galaxies observed edge-on and is set to $q_0 = 0.2$. The inclination for galaxies with $b/a < 0.2$ is set to $90^{\rm o}$  \citep{2012MNRAS.420.1959C, 2018MNRAS.479.2133A}. Relying on \citet{2018MNRAS.476.3661D} and through inspection of the selected SDSS images, we find the following arguments:
\begin{itemize}
    \item \textbf{Late-type galaxies (LTG)}, which are disc dominated\footnote{Please note that we refer throughout this article to LTG for all disc galaxies from Sa to Sd.}, are selected by T-type $>0$ and ${\rm P_{disc}} > 0.5$.
    \begin{itemize}
        \item \textbf{Face-on} are identified as LTG with an inclination angle smaller than $30^{\rm o}$. To further decrease the false-positive detection rate, we demand ${\rm P_{edge-on}} < 0.1$.
        \item \textbf{Edge-on} and strongly inclined galaxies are selected as LTGs with an inclination angle larger than $70^{\rm o}$ and ${\rm P_{edge-on}} > 0.4$.
        \item \textbf{Barred} galaxies are LTGs with ${\rm P_{bar}} > 0.9$ and ${\rm P_{edge-on}} < 0.5$
        \end{itemize}{}
    \item \textbf{S0} galaxies are selected by a T-type $\leq 0$, ${\rm P_{S0}} > 0.6$, ${\rm P_{disc}} < 0.3$ and ${\rm P_{edge-on}} < 0.4$
    \item \textbf{Elliptical} are characterised by a T-type $\leq0$ and ${\rm P_{S0}} < 0.3$
    \item \textbf{Merger} galaxies are selected by using ${\rm P_{merger}} > 0.9$. This high threshold provides a sub-sample of merger with high purity enabling to test if visible merging processes are at the origin of DP galaxies.
\end{itemize}
To compare our galaxy samples with respect to their morphological classification, we did not apply any subjective correction to the selection. We compute the fraction of all classified types for the DPS, the CS and the NBCS in Table\,\ref{table:morph}. Fig.\,\ref{fig:redshift_dist} displays the redshift distribution for the different morphological types of the DPS and the NBCS. We observe merger and S0 galaxies to be more distant in comparison to LTG and elliptical galaxies. This can be the result of a classification bias as discussed in the training sample \citep{2013MNRAS.435.2835W} used for the classification in \citet{2018MNRAS.476.3661D}. 
We thus might misclassified LTG as S0 galaxies because of their distance.
In Fig.\,\ref{fig:image:patern}, we illustrate the morphological types, by showing 20 randomly selected galaxies for each type (except for the face-on LTGs, for which we show all 10 cases).

\begin{figure}
  \centering 
 \includegraphics[width=0.48\textwidth]{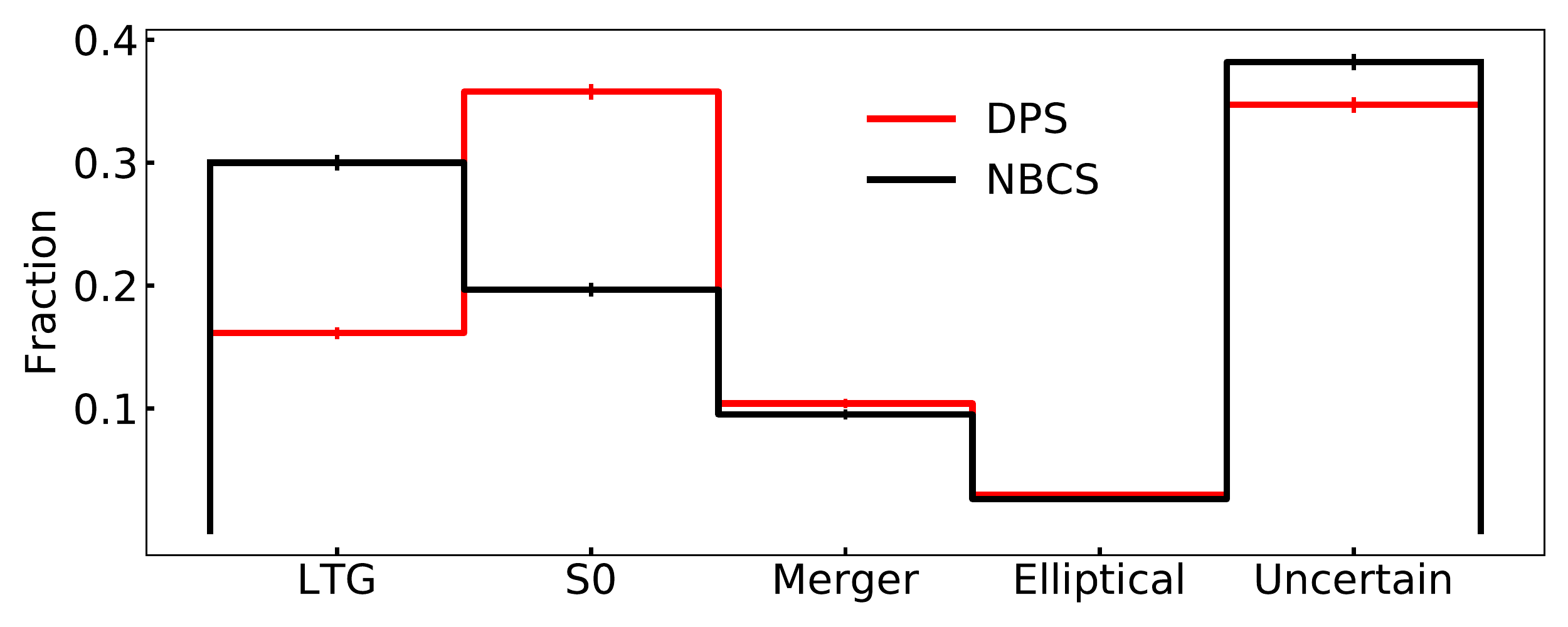}
  \caption{Fraction of different morphological types for the DPs and the NBCS. The error-bars represent the binomial errors.}
  \label{fig:morph_fraction}
\end{figure}
\begin{table}
\caption{The morphological classification of the DPS, the CS and the NBCS.}\label{table:morph}
\vspace{-0.6cm}
\begin{center}
\begin{tabular}{l r@{\hskip 0in}l r@{\hskip 0in}l r@{\hskip 0in}l}
\hline\hline
Type && DPS && NBCS && CS\\
\hline
LTG & 914&($16.1\%$) & 1539&($30.0\%$) & 20157&($22.5\%$) \\
\,\,\,\,Face-on & 10&($0.2\%$) & 149&($2.9\%$) & 1609&($1.8\%$) \\
\,\,\,\,Edge-on & 54&($1.0\%$) & 114&($2.2\%$) & 3844&($4.3\%$) \\
\,\,\,\,Bar & 105&($1.9\%$) & 147&($2.9\%$) & 956&($1.1\%$) \\
\hline
Elliptical & 167&($2.9\%$) & 134&($2.6\%$) & 302&($0.3\%$) \\
S0 & 2027&($35.8\%$) & 1009&($19.7\%$) & 7145&($8.0\%$) \\
\hline
Merger & 589&($10.4\%$) & 487&($9.5\%$) & 6524&($7.3\%$) \\
Uncertain & 1966&($34.7\%$) & 1959&($38.2\%$) & 55284&($61.8\%$) \\
\hline
\end{tabular}
\end{center}{}
\vspace{0.1cm}
{\justifying \small
{\noindent {\bf Notes:} We used parameters provided by \cite{2018MNRAS.476.3661D} and \citet{2015MNRAS.446.3943M}, described in detail in sec.\,\ref{ssect:morph}. The first column names the morphological type. The second (resp. third and fourth) column provides the number of galaxies and the percentage in parenthesis for the DPS (resp. CS and NBCS)}}
\end{table}

In Table\,\ref{table:morph}, we find that the majority of the CS galaxies are not classified ($62\%$). This is due to a classification bias against small and weak galaxies \citep{2018MNRAS.476.3661D}. The numbers of not classified galaxies in the DPS and the NBCS are comparable ($35\%$ and $38\%$). By comparing the morphological classification of the DPS and the NBCS, we find two noticeable effects. On the one hand, LTG are less numerous in the DPS ($16\%$) than in the NBCS ($30\%$). On the other hand, more S0 galaxies are present in the DP ($36\%$) than in the NBCS ($20\%$). These results are visualised in Fig.\,\ref{fig:morph_fraction}.

We find in both samples a similar merger rate of $\sim 10\%$. This observation is {\em a priori} contradictory to the idea of a DP structure due to a major galaxy merger but does not exclude the possibility that DP galaxies might be minor mergers, post-mergers or hidden merger. Furthermore, the same merger rate ($10\%$) was found in a recent double-peak sample in the Large Sky Area Multi-object Fiber Spectroscopic Telescope survey \citep{2019MNRAS.482.1889W}. As a cross-check, we performed a visual inspection of the first 1000 galaxies in the DPS and the NBCS. We found a merger rate of $11.4\,\%$ (resp. $11.7\,\%$) for the DPS (resp. NBCS), which is close to the merger rate we extracted from \citet{2018MNRAS.476.3661D}. By lowering the selection threshold for mergers (e.g. $\rm P_{merger} > 0.8$ or $\rm P_{merger} > 0.7$), we do not detect different merger rates in the DPS and the NBCS, but higher rates of misclassification.

We also classified LTGs in face-on ($0.2\%$), edge-on ($1.0\%$) and bar ($1.9\%$) galaxies. We chose strict selection criteria to avoid false-positive classification for face-on and edge-on galaxies. These galaxy types are interesting since edge-on galaxies may exhibit a double horn due to a rotating disc, while face-on galaxies should not exhibit rotation in form of a DP. We do not observe any excess of edge-on galaxies in comparison to the NBCS ($2.2\%$). In Sect.\,\ref{ssect:inclination} and \ref{ssect:TF:FJ}, we will discuss arguments using the inclination.

We also provide cross-matches between the morphological types and the BPT classifications in Table \ref{table:bpt:morph1} and \ref{table:bpt:morph2} and provide a statistical significance test in \ref{ssect:fishers:exact:test}.

\section{Analysis}\label{sect:analysis}
\begin{figure*}
\centering 
\includegraphics[width=0.98\textwidth]{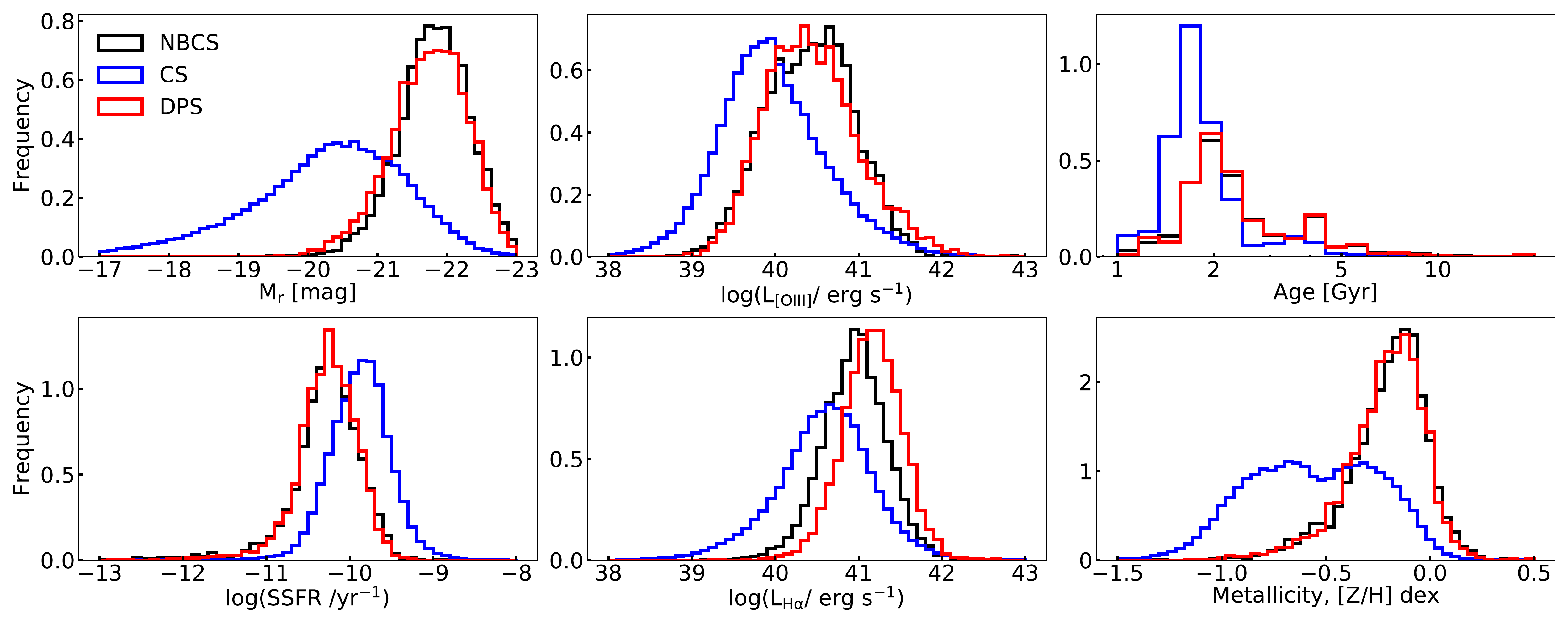}
\caption{Different parameters distributions of the DPS (red), the CS (blue) and the NBCS (black). The surfaces of all histograms are normalised to unity. In the upper left panel, we compute the absolute AB magnitude in the r-band with Galactic dust and K-corrections estimated by \citet{2017ApJS..228...14C}. The lower left panel shows the specific star formation ratio (SSFR) computed by \citet{2004MNRAS.351.1151B}. The middle upper (resp. lower) panel presents the absolute luminosity of the [OIII]$\lambda$5008 (resp. ${\rm H}{\alpha}\lambda$6565) emission line. The absolute luminosity is calculated using the flux of the non-parametric emission line fit provided by \citet{2017ApJS..228...14C}. The right upper panel computes the age assumed by a single stellar population fit by \citet{2017ApJS..228...14C}. The right lower panel shows the stellar metallicity computed by \citet{2017ApJS..228...14C}.}
\label{fig:characteristic:hist}%
\end{figure*}

\subsection{General characteristics}\label{ssect:general:characteristics}
To describe the characteristics of the DPS and compare them with the NBCS and the CS, we show in Fig.\,\ref{fig:characteristic:hist} the distribution of six parameters: the Galactic extinction and k-corrected absolute magnitude of the $r$-band $M_r$, the specific star formation rate ${\rm SSFR} = {\rm SFR / M_{\odot}}$, the luminosities of the [OIII]$\lambda$5008 and ${\rm H}{\alpha}\lambda$6565 emission lines, stellar population ages and stellar metallicities.

In every distribution, the CS differs from the DPS whereas the NBCS shows good agreement with the DPS, except for the ${\rm H}{\alpha}\lambda$6565 luminosity. We will discuss, in Sect. \ref{ssect:star:formation:agn}, that this is related to a significant difference in the star formation activity. As discussed in Sect.\,\ref{ssect:CS:selection} and \ref{ssect:bias}, the CS follows a different stellar mass-redshift distribution in comparison to the DPS and the NBCS. On average, the CS shows smaller $ M_r$, higher SSFR and smaller absolute luminosities in the [OIII]$\lambda$5008 and ${\rm H}{\alpha}\lambda$6565 line than the DPS and the NBCS. This different behaviour observed for galaxies of the CS is mainly due to a sensitivity bias: low-mass galaxies pass the signal-to-noise ratio cut on the emission lines while the DPS and NBCS are large-mass galaxies (see Sect.\,\ref{ssect:bias}). 

The differences in the DPS and NBCS ${\rm H}{\alpha}\lambda$6565 luminosities are small: ${\rm L_{H\alpha}(DPS)} =  41.16 \pm 0.36$ and ${\rm L_{H\alpha}(NBCS)} =  40.96 \pm 0.42$ (mean and standard deviation). 
However, using a the non-parametric Kolmogorov–Smirnov test to distinguish weather the two distributions are the same or not, we find a k-statistic of 0.235 marking the maximal distance of the cumulative distribution function and a p-value of $5 \times 10^{-130}$, strongly indicating that these two distributions are different.

We checked the extinction based on the Balmer decrement and found that DPS galaxies have slightly more extinction (0.25\,mag higher in V) than the NBCS galaxies, while this effect is stronger for LTG and S0 galaxies (see Fig.\,\ref{fig:appendix:sfr}).

To approximate stellar ages, we use the spectral continuum fit performed with a single stellar population \citep{2017ApJS..228...14C}. Due to the cut on the stellar masses (see Fig. \ref{fig:stellar:mass:z:dist:cs} and \ref{fig:stellar:mass:z:dist:nbcs}), the CS is dominated by young stellar populations with an SSP-based age between 1 and 2 Gyr in $52\%$ of the galaxies. Nearly all ($97\%$) of these young galaxies are classified as SF (see Sect.\,\ref{ssect:bpt}). The NBCS and DPS show comparable distributions but with on average an older stellar population than for the CS. The majority ($74\%$ and $71\%$ respectively) of the two samples is younger than 5 Gyr, while $47\%$ (DPS) and $66\%$ (NBCS) of these young galaxies are SF galaxies. We also find that $29\%$ (resp. $23\%$) of the young DP galaxies (resp. NBCS) are classified as COMP. Last, we find $15\%$ of the DPS to be classified with one peak as COMP and the second one as SF. As discussed in Sect.\,\ref{ssect:bpt}, we find an excess of the percentage of SF galaxies in the NBCS and CS whereas we do not find such a high excess in the DPS. 

We also show the stellar metallicity, computed by \citet{2017ApJS..228...14C}, and find that the DPS and the NBCS follow the same distribution, whereas the CS shows a different distribution with lower metallicity values. This is expected due to the difference in stellar mass of the different samples \citep{2004ApJ...613..898T}.

While we find many differences between the DPS and the CS, the NBCS and DPS distributions are similar. The NBCS is therefore well suited to identify the peculiarities of the DPS. 

\subsection{Star formation vs. AGN excitation}\label{ssect:star:formation:agn}
\begin{figure}
\centering 
\includegraphics[width=0.48\textwidth]{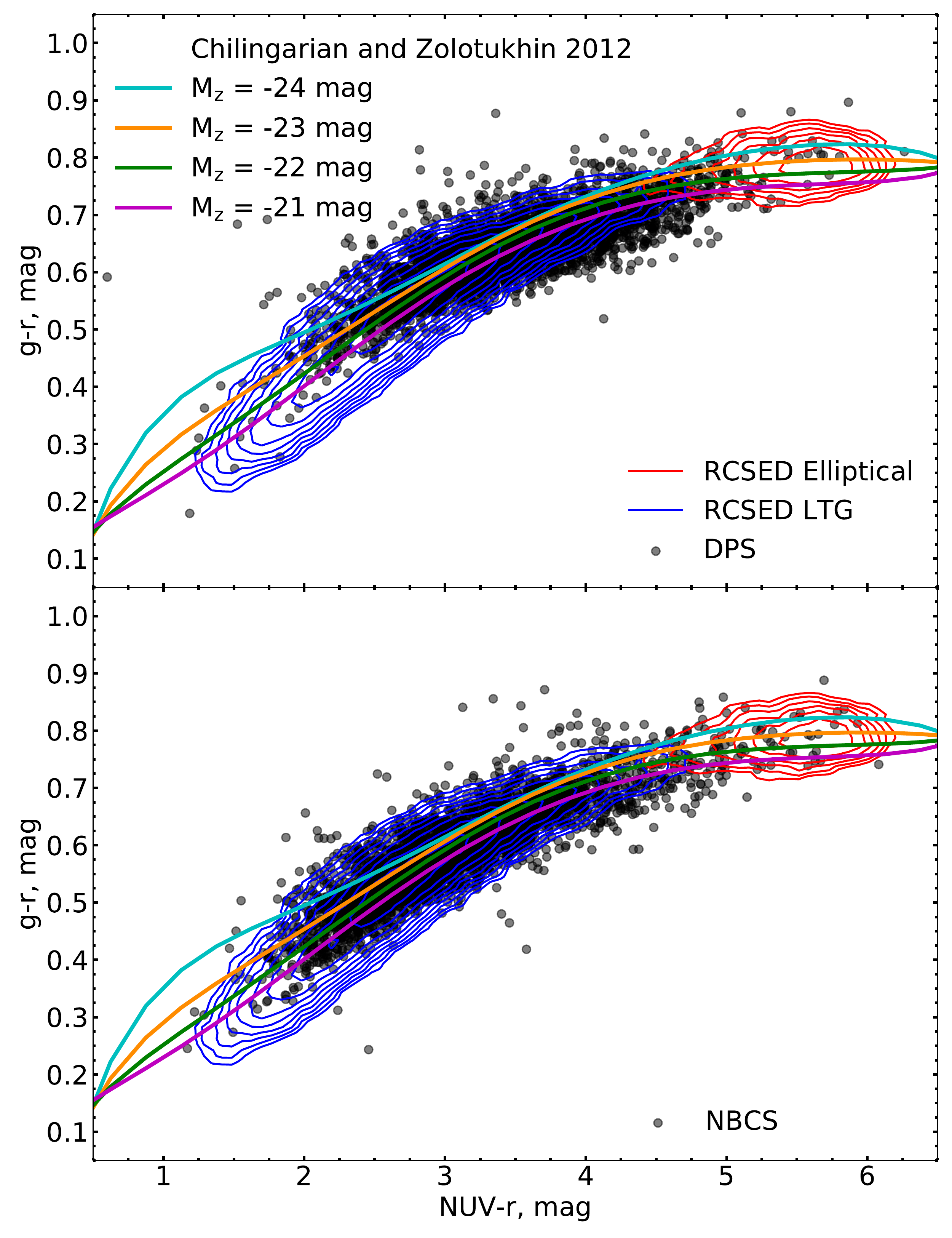}
\caption{Colour–colour relation between $g-r$ and ${\rm NUV}-r$.
We present LTG and elliptical galaxies selected from the RCSED catalogue using \citet{2018MNRAS.476.3661D} as blue and red contours and show the DPS (resp. NBCS) as black dots in the top panel (resp. lower panel). We display the best-fitting polynomial surface equation of constant absolute $z$-band magnitudes with coloured lines \citep{2012MNRAS.419.1727C}.}
\label{fig:color:color}%
\end{figure}
\begin{figure}
\centering 
\includegraphics[width=0.48\textwidth]{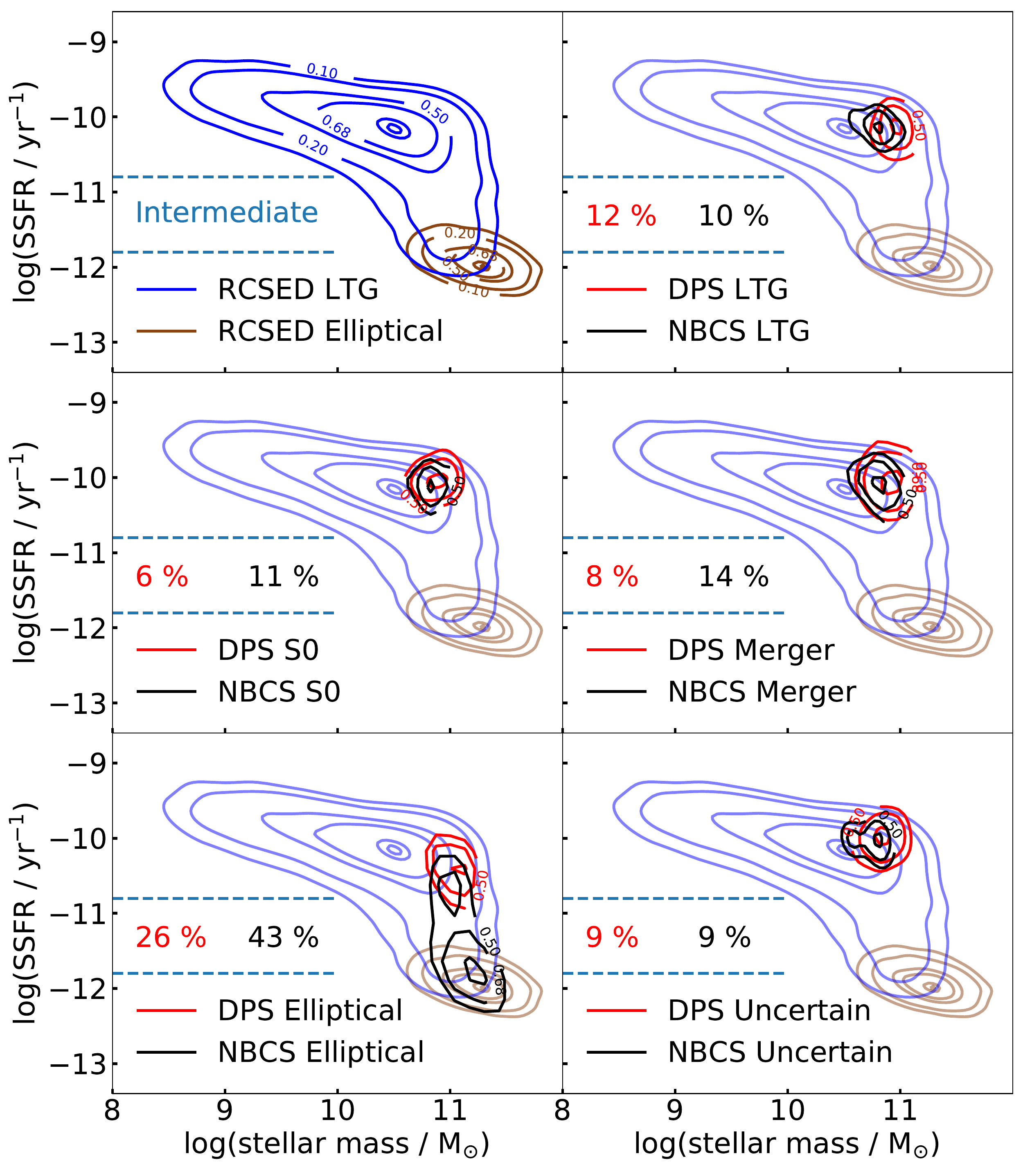}
\caption{Specific star formation rates (SSFR) as a function of stellar mass. The top left panel presents LTG and elliptical galaxies selected from the RCSED catalogue using \citet{2018MNRAS.476.3661D} as blue and brown contours. These contours are also shown as a reminder in the other panels. We show contour lines indicating the $50\%$, $68\%$ and $95\%$ level for different morphological subsamples of the DPS (red) and the NBCS (black). We indicate with blue dashed lines the region of intermediate star formation ($-11.8 < {\rm log(SSFR/M_{\odot})} < -10.8$) \citep{2014SerAJ.189....1S} and display the percentage of the morphological types situated in this zone.}
\label{fig:ssfr:stellar:mass}%
\end{figure}
Star formation (SF) is important to characterise the state of evolution of a galaxy and to estimate its growth rate. In Sect.\,\ref{ssect:bpt} and \ref{ssect:general:characteristics}, we found that the majority of the DPS, CS and NBCS show SF activity. At the same time, we find AGN activities (or galaxies classified as COMP) in the NBCS and the DPS. To further analyse and compare our samples, we use diagnostic colour-colour diagrams, specific star formation rate (SSFR) and stellar mass diagrams and compare the star formation rate (SFR) of the 3" SDSS fiber and of the total galaxy. According to \citet{2007ApJS..173..267S}, the SFR of AGN computed by \citet{2004MNRAS.351.1151B} can be overestimated. We therefore use the NUV-band to compute the SFR and found a good agreement to the SFR measured by \citet{2004MNRAS.351.1151B} for our galaxies classified as AGN. We discuss AGN excitation of galaxies classified as composite galaxies and cross-match our catalogues with radio observations.

In Fig.\,\ref{fig:color:color}, the $g-r$ vs. ${\rm NUV}-r$ colour–colour diagram shows a single sequence with the star-forming blue cloud and the red sequence, characterised by quenched star formation \citep{2001AJ....122.1861S,2002AJ....124..646H,2005MNRAS.363.1257E}, separated by a critical value of ${\rm NUV} - r = 4.75$.
The blue cloud (resp. red sequence) is approximately represented by blue (resp. red) contours for LTGs (resp. elliptical galaxies) from the RCSED \citep[compare with][]{2012MNRAS.419.1727C}. We find the DPS to be situated in the upper region of the blue cloud and only some outliers in the red sequence. We do not find significant effects for the different subsets classified in Sect.\,\ref{sect:classification}. We can thus conclude that the DPS and the NBCS are not quenched and are mainly associated with the main star forming sequence.
The distribution of the DPS is more concentrated in the range centred at ${\rm NUV}-r\sim 3.5$ than the NBCS (${\rm NUV}-r\sim 3.0$), which is centred in the blue cloud. This shift towards redder colours can be explained by considering shorter time scales $\tau$ of an exponential declining star formation history, as shown in \citet{2012MNRAS.419.1727C}. 

However, this is not the case for the DPS as their mean $\tau$ is slightly larger than for the NBCS.
We argue that these redder colours are most probably to the mean extinction estimated to $\rm A_V = 0.25$, as displayed in the last column of Figure \ref{fig:appendix:sfr}. 

We can notice some galaxies in the DPS situated below the two sequences in Fig.\,\ref{fig:color:color}. 
This area is associated with post starburst galaxies which underwent a recent massive starburst that is now quenched \citep{2012MNRAS.419.1727C}. 
Such galaxies can be also situated in the surface, as they can be biased when a strong ${\rm H}{\alpha}\lambda$6565 line contributes to the $r$-band magnitude, shifting the position of the galaxies into the plane. We find that this is not the case for DP galaxies, nor the NBCS. To confirm this, we studied the colour-colour diagram with colours $g-z$ and NUV$-z$, where this off-plane shift disappears.

In Fig.\,\ref{fig:ssfr:stellar:mass}, we show the SSFR, taken from \citet{2004MNRAS.351.1151B}, as a function of the stellar mass, computed by \citet{2003MNRAS.341...33K}. With this diagnostic diagram, \citet{2007ApJS..173..315S} distinguish between the blue and the red sequence in terms of SSFR and stellar mass. \cite{2014SerAJ.189....1S} associated intermediate NUV$-r$ colours known as the "green valley" with intermediate star formation of $-11.8 < {\rm log(SSFR/M_{\odot})} < -10.8$ suggesting that these galaxies might be in transition. In Fig.\,\ref{fig:ssfr:stellar:mass}, we show the distribution of LTGs and elliptical galaxies selected from the RCSED as reference. We display the $50\%$, $68\%$ and $95\%$ contour levels of different morphological types of the DPS and the NBCS. We find, depending on the morphological type, between $6\%$ and $26\%$ (resp. $8\%$ and $43\%$) of the DPS (resp. NBCS) showing intermediate star formation. We only find a significant difference between the DPS and the NBCS for elliptical galaxies: only $26\%$ of the elliptical DP galaxies show intermediate star formation whereas $43\%$ of the elliptical galaxies of the NBCS are situated in this region. We find indeed very similar distributions for all different morphological types of the DPS and thus conclude that SF and the stellar mass are not depending on the morphological type.
\begin{figure*}
\centering 
\includegraphics[width=0.98\textwidth]{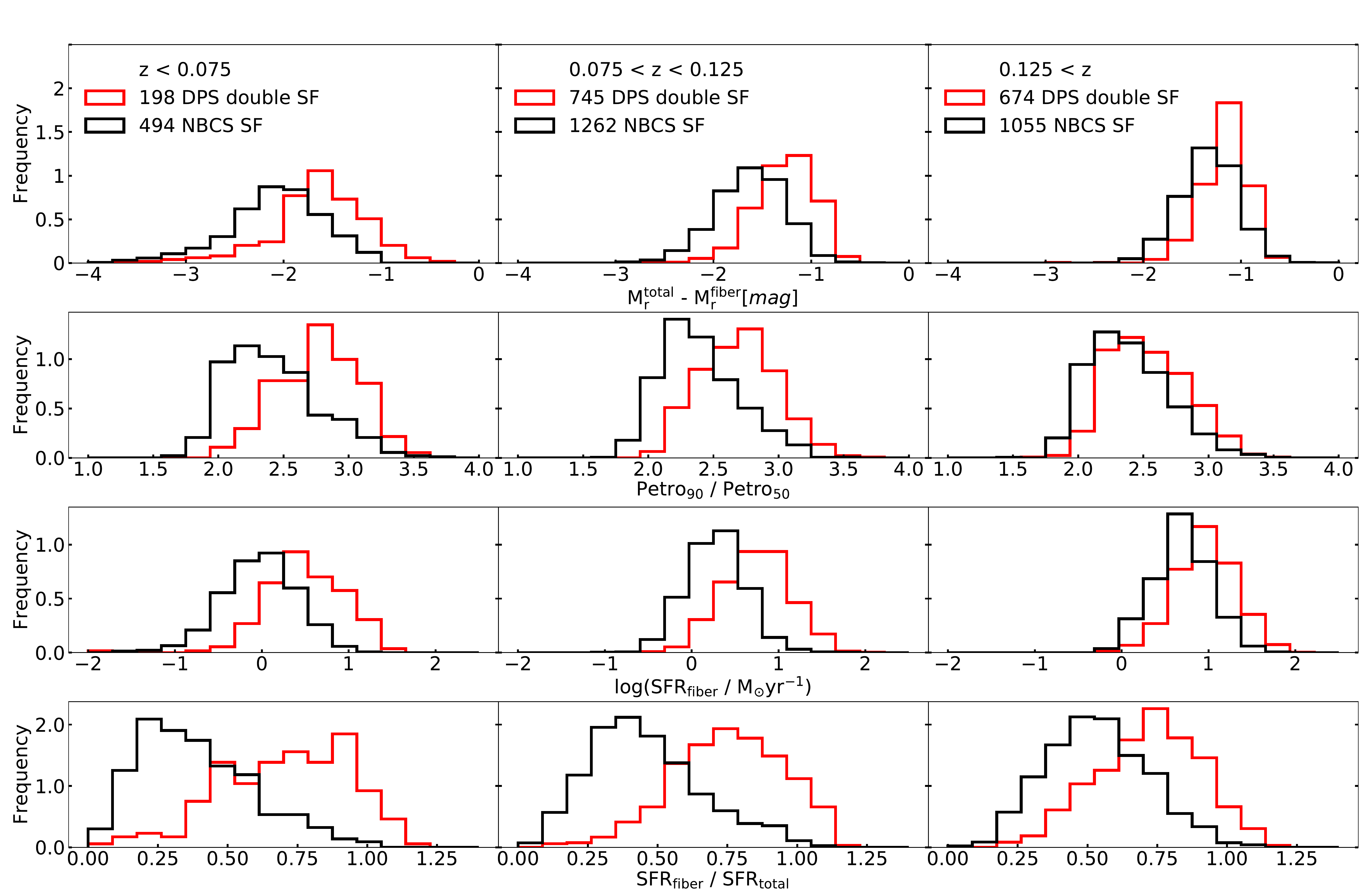}
\caption{Distribution of different observables within the 3" SDSS fiber. We show in three columns three different ranges of redshift. We present galaxies of the DPS which are classified as double SF in red and galaxies classified as SF of the NBCS in black (see Sect.\,\ref{ssect:bpt}) In the top row we show the difference in absolute $r$-band magnitude of the total galaxy and the $3''$ SDSS fiber. The colours are Galactic dust- and k-corrected \citep{2017ApJS..228...14C}. In the second row we present the ratio of the radii comprising $90\%$ and $50\%$ of the Petrosian luminosity \citep{1976ApJ...209L...1P} approximating the central brightness in comparison to the total galaxy. We show the star formation ratio of the 3" SDSS fiber SFR$_{\rm fiber}$ in the third row and the ratio between fiber and total star formation (${\rm SFR_{\rm fiber}} / {\rm SFR_{total}}$ ) in the bottom row \citep{2004MNRAS.351.1151B}.}
\label{fig:fiber:sfr:sf}%
\end{figure*}
\begin{figure}
\centering 
\includegraphics[width=0.48\textwidth]{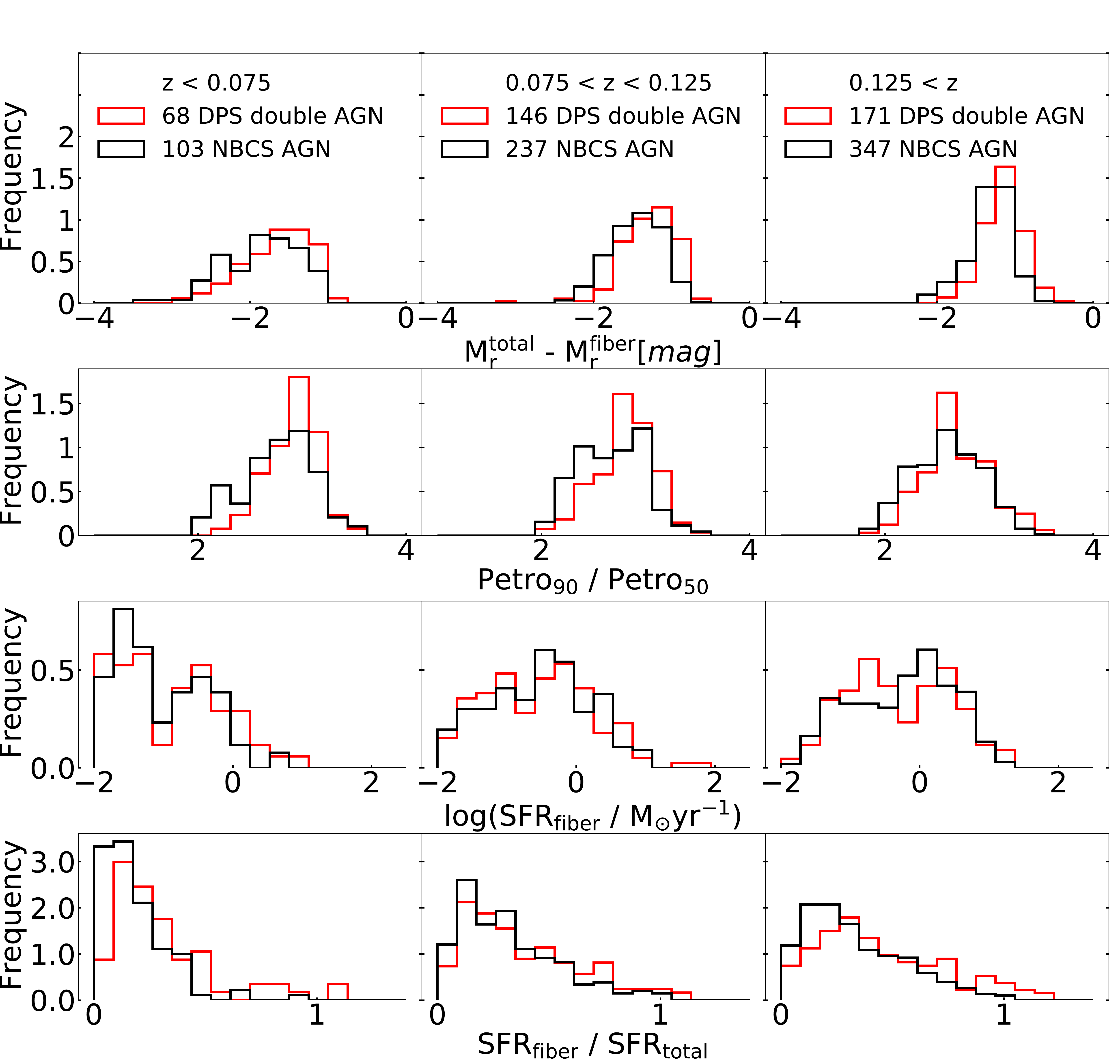}
\caption{Comparison of absolute $r$-band magnitude, Petrosian radii and SFR$_{\rm fiber}$ and SFR$_{\rm total}$ as in Fig.\,\ref{fig:fiber:sfr:sf} but with DP galaxies classified as AGN in both peak components (red) and galaxies of the NBCS classified as AGN (black).}
\label{fig:fiber:sfr:agn}%
\end{figure}
\begin{table}
\caption{Median ratios of fiber and total SFR}\label{table:sfr:ratio}
\vspace{-0.6cm}
\begin{center}
\begin{tabular}{l r c r}
\hline\hline
& & \multicolumn{1}{c}{$\mathcal{R}$} & \\
\hline
Type & ${\rm z < 0.075}$ &  $0.075 < {\rm z} < 0.125$ & ${\rm z > 0.125}$\\
\hline
\multicolumn{4}{c}{DPS} \\
\hline
double SF &  $0.73_{-0.25}^{+0.22}$ & $0.76_{-0.20}^{+0.20}$ & $0.73_{-0.21}^{+0.18}$ \\
double COMP &  $0.56_{-0.26}^{+0.35}$ & $0.67_{-0.28}^{+0.24}$ & $0.64_{-0.22}^{+0.25}$ \\
double AGN &  $0.24_{-0.11}^{+0.23}$ & $0.31_{-0.17}^{+0.36}$ & $0.38_{-0.20}^{+0.38}$ \\
LTG &  $0.56_{-0.30}^{+0.32}$ & $0.58_{-0.29}^{+0.26}$ & $0.54_{-0.24}^{+0.25}$ \\
S0 &  $0.62_{-0.30}^{+0.31}$ & $0.73_{-0.25}^{+0.21}$ & $0.74_{-0.24}^{+0.19}$ \\
Elliptical &  $0.40_{-0.24}^{+0.39}$ & $0.49_{-0.29}^{+0.33}$ & $0.61_{-0.33}^{+0.20}$ \\
Merger &  $0.56_{-0.31}^{+0.33}$ & $0.70_{-0.28}^{+0.28}$ & $0.71_{-0.29}^{+0.26}$ \\
Uncertain &  $0.60_{-0.34}^{+0.34}$ & $0.73_{-0.27}^{+0.22}$ & $0.67_{-0.24}^{+0.20}$ \\
\hline
\multicolumn{4}{c}{NBCS} \\
\hline
SF &  $0.36_{-0.16}^{+0.24}$ & $0.43_{-0.16}^{+0.23}$ & $0.53_{-0.18}^{+0.19}$ \\
COMP &  $0.29_{-0.17}^{+0.34}$ & $0.42_{-0.21}^{+0.34}$ & $0.48_{-0.22}^{+0.26}$ \\
AGN &  $0.14_{-0.09}^{+0.16}$ & $0.28_{-0.16}^{+0.27}$ & $0.28_{-0.16}^{+0.30}$ \\
LTG &  $0.28_{-0.16}^{+0.25}$ & $0.34_{-0.17}^{+0.21}$ & $0.38_{-0.21}^{+0.21}$ \\
S0 &  $0.42_{-0.25}^{+0.24}$ & $0.50_{-0.24}^{+0.25}$ & $0.56_{-0.25}^{+0.22}$ \\
Elliptical &  $0.18_{-0.09}^{+0.14}$ & $0.26_{-0.10}^{+0.20}$ & $0.21_{-0.10}^{+0.27}$ \\
Merger &  $0.31_{-0.15}^{+0.26}$ & $0.45_{-0.25}^{+0.32}$ & $0.49_{-0.24}^{+0.22}$ \\
Uncertain &  $0.30_{-0.16}^{+0.34}$ & $0.45_{-0.18}^{+0.26}$ & $0.48_{-0.19}^{+0.23}$ \\
\hline
\end{tabular}
\end{center}{}
\vspace{0.1cm}
{\justifying \small
{\noindent {\bf Notes:} We show the median and the $68$ percentile of $\mathcal{R} = {\rm SFR_{fiber} / SFR_{total}}$ for different ranges of redshift and galaxy types for the DPS and NBCS.}}
\end{table}

To discuss the SF in a quantitative way, we use SFRs estimated by \citet{2004MNRAS.351.1151B} for the entire galaxies (${\rm SFR_{total}}$) and for the 3" SDSS fiber (${\rm SFR_{fiber}}$). Since the proportion of a galaxy covered by the $3''$ fiber depends on its redshift, we divide the DPS and the NBCS into three groups of redshift ranges: ${\rm z} < 0.075$, $0.075 < {\rm z} < 0.125$ and $0.125 < {\rm z}$. In order to discuss the SF activities in detail, we will especially examine the sub-samples classified as double SF and AGN, using the BPT classification performed in Sect.\,\ref{ssect:bpt}.

Hence, we present the comparison between the DPS and the NBCS with the difference between total and $3''$ absolute $r$-band magnitude, the ratio of the Petrosian radii of $90\%$ and $50\%$, the ${\rm SFR_{fiber}}$ and the ratio $\mathcal{R} = {\rm SFR_{fiber} / SFR_{total}}$. We show these observables for the three redshift groups for galaxies of the DPS classified as double-SF (resp. double-AGN) and SF (resp. AGN) galaxies of the NBCS in Fig.\,\ref{fig:fiber:sfr:sf} (resp. Fig,\,\ref{fig:fiber:sfr:agn}). For double-SF galaxies of the DPS, we find an excess of luminosity in the centre and a higher ${\rm SFR_{fiber}}$ in comparison to SF galaxies {\textbf{of}} the NBCS. This excess is most prominent for galaxies with a redshift ${\rm z < 0.125}$, corresponding to maximal fiber diameter of 7.6 kpc. With respect to ${\rm SFR_{total}}$, we find a clear enhanced-central-SF activity for double-SF galaxies of the DPS in comparison to SF galaxies of the NBCS. We find a median SFR ratio for double-SF galaxies of the DPS with a redshift ${\rm z < 0.075}$ of $\overline{\mathcal{R}}_{\rm DPS} = 0.73_{-0.25}^{+0.22}$. For SF galaxies of the NBCS in the same redshift range we find $\overline{\mathcal{R}}_{\rm NBCS} = 0.36_{-0.16}^{+0.24}$. We find similar values for galaxies with a redshift $0.075 < {\rm z} < 0.125$, whereas this effect is {\textbf slightly weaker} for galaxies with ${\rm z > 0.125}$. 
This is due to the fact that for more distant galaxies the 3" fiber covers the majority of the surface of the galaxies, therefore ${\rm SFR_{fiber}}$ tends to ${\rm SFR_{total}}$. We summarised the median SFR ratios and the 68-percentile for different galaxy types in Table\,\ref{table:sfr:ratio}. We also find an enhanced-central-SF for LTGs, S0 galaxies and galaxies classified as double COMP at lower redshift. 

Double-AGN of the DPS follow the same ${\rm SFR_{fiber}}$ and ${\rm SFR_{total}}$ distributions as the AGNs of the NBCS and no enhancement of central SF activity or central luminosity is observed as shown in Fig.\,\ref{fig:fiber:sfr:agn}. 

The BPT classification in Sect.\,\ref{ssect:bpt} found less SF but more COMP galaxies in the DPS in comparison to the NBCS. Since COMP galaxies are understood to be a mixture of SF and AGN \citep{2006MNRAS.372..961K}, we quantify the AGN excitation using the [OII]$\lambda 3728$ and the [NeV]$\lambda 3426$ emission lines. 

(1) The ratio [OII]$\lambda 3728$/[OIII]$\lambda$5008 can be used to distinguish between a pure AGN and SF. From photo-ionisation models, a value between 0.1 and 0.3 is used for a pure AGN \citep{2015MNRAS.448.3354T}, while above a value of 0.3, the [OII]$\lambda 3728$ flux is thought to be produced in SF sites \citep{2006agna.book.....O}. Using the non-parametric emission-line fit from \citet{2017ApJS..228...14C}, we find a mean ratio of ${\rm [OII]/[OIII]}=0.23\pm 0.14$ for galaxies classified as double AGN. For double SF (resp. COMP) galaxies, we find ${\rm [OII]/[OIII]}=1.57\pm 0.56$ (resp. $1.14\pm 0.54$). This indicates that galaxies which are classified as COMP are dominated by SF. This effect can also be biased by extinction in dusty galaxy centres and must therefore be considered with caution.

(2) We then discuss galaxies with AGN activity based on the detection of the high-ionisation [NeV]$\lambda 3426$ emission line \citep{2010A&A...519A..92G, 2018A&A...620A.193V}. Due to redshift, the [NeV]$\lambda 3426$ line is only detected in SDSS spectra from ${\rm z} > 0.2$. We inspect spectra of the DPS galaxies classified as double-SF, -COMP and -AGN for [NeV]$\lambda 3426$ detection. We find no galaxies classified as double-SF contrary to $62\%$ of the galaxies classified as double-AGN having a detectable [NeV]$\lambda 3426$ line. Only $4\%$ of the $z > 0.2$ galaxies classified as double COMP show this line which supports the evidence that these are SF galaxies.

To discuss further aspects on the connection between AGN excitation and SF in our galaxy samples, we cross-match the DPS and the NBCS with surveys of radio observations. It has been known for a long time that non-thermal radio emission is linearly correlated to the far-infrared flux \citep{1985ApJ...298L...7H,1992ARA&A..30..575C,1996ARA&A..34..749S}, while IR luminosity is a good SF tracer \citep[e.g.][]{2002AJ....124.3135K}. More recently, the radio luminosity at 150\,MHz has been considered as a potential tracer of SF \citep{2017MNRAS.469.3468C,2019A&A...631A.109W}. We use FIRST at 1.4 GHz \citep{1997ApJ...475..479W} and the first data release of the LOFAR Two-metre Sky Survey (LoTSS) \citep{2019A&A...622A...1S} at about 150 MHz. 
Roughly $20\%$ (1154) of the DPS are detected with FIRST, but only $8\%$ (395) of the NBCS. In the observed field of the LoTSS DR1 we find a detection rate of $73\%$ (237) for the DPS and $56\%$ (191) for the NBCS. We find a significant excess of radio sources at 1.4 GHz and a higher detection rate at 150 MHz for the DPS in comparison to the NBCS. We will further explore the radio properties of these galaxies in Maschmann et al. (in prep.). 

\subsection{Kinematic Properties of the double-peak Galaxies}\label{sect:kinematic:properties}

\subsubsection{Velocity Distributions}\label{ssect:velocity:distribution}
\begin{figure}
\centering 
\includegraphics[width=0.48\textwidth]{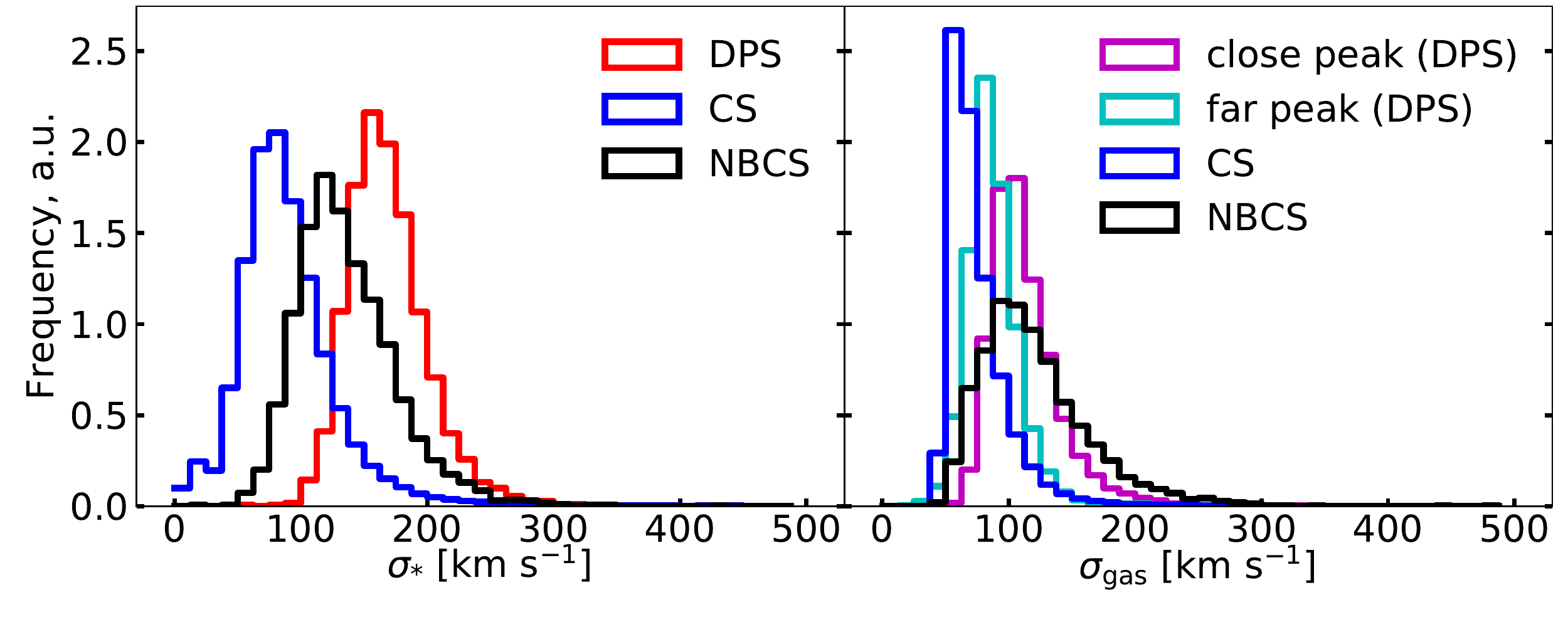}
\caption{Velocity dispersion distributions for the CS (blue), the NBCS (black) and the DPS (red, purple and orange). The left panel displays the stellar velocity dispersion $\sigma_{*}$ computed from \citet{2017ApJS..228...14C}. The second panel shows the gas velocity dispersion $\sigma_{\rm gas}$. We compute the Gaussian emission line dispersion for the CS and the NBCS from \citep{2017ApJS..228...14C}. For the DPS, we display the $\sigma_{1,2}$ of the emission-line components, which is closer to the galaxy stellar velocity in units of $\sigma_{1,2}$ in magenta and the component which is more offset in cyan. For better comparisons, we unify the surface of the histograms of the left (resp. right) panel to 150 (resp. 100).}\label{fig:velocity:distributions}
\end{figure}

To explore the gas and star kinematic properties of our samples, we present basic velocity estimates to assess possible differences between the DPS, the CS and the NBCS. In Fig.\,\ref{fig:velocity:distributions}, we show the stellar velocity dispersion $\sigma_{*}$ and the gas velocity dispersion $\sigma_{\rm gas}$.

The stellar velocity dispersion is computed in \citet{2017ApJS..228...14C}, fitted with a simple stellar population (SSP) and an exponential declining star formation history (EXP-SFH) (see Sect.\,\ref{sect:data}). We compute the $\chi_{\nu}^2$ for both fit functions and select the resulting stellar velocity dispersion from the best fitting function.

To compute the gas velocity dispersion for the CS and the NBCS, we use the velocity dispersion measured for all Balmer and forbidden lines in \citet{2017ApJS..228...14C} and show the one of the strongest line. For DP emission lines, we use the stacked emission line spectra (see Sect.\,\ref{ssect:stack:and:fit}) and compute the $\sigma_{\rm close}$ which corresponds to the DP component closer to the galaxy velocity in units of $\sigma$ and $\sigma_{\rm far}$ to the component which is more offset.

As shown in Fig.\,\ref{fig:velocity:distributions}, we observe a higher $\sigma_{*}$ for the DPS in comparison with the CS and the NBCS. The gas velocity dispersion $\sigma_{\rm gas}$ indicates higher velocities for the NBCS and the DPS in comparison to the CS. This is expected from the Tully-Fisher relation, since these two samples comprise more massive galaxies \citep{1977A&A....54..661T}. 
For the DPS, we find the velocity dispersions, measured for the close peak, to be compatible with the gas velocity dispersions of the NBCS. For the far peak of the DPS, the velocity dispersions are systematically smaller.

\subsubsection{Single Gaussian approximation for DP}\label{ssect:single:gaussian:approximation}
To discuss possible mechanisms behind DP emission lines, we need to compare the gas kinematics of the DPS and the NBCS. In order to do so we use the single Gaussian approximation of the double-peaked emission lines, as computed in Sect.\,\ref{ssect:stack:and:fit}.

\begin{figure}[h]
\centering 
\includegraphics[width=0.48\textwidth]{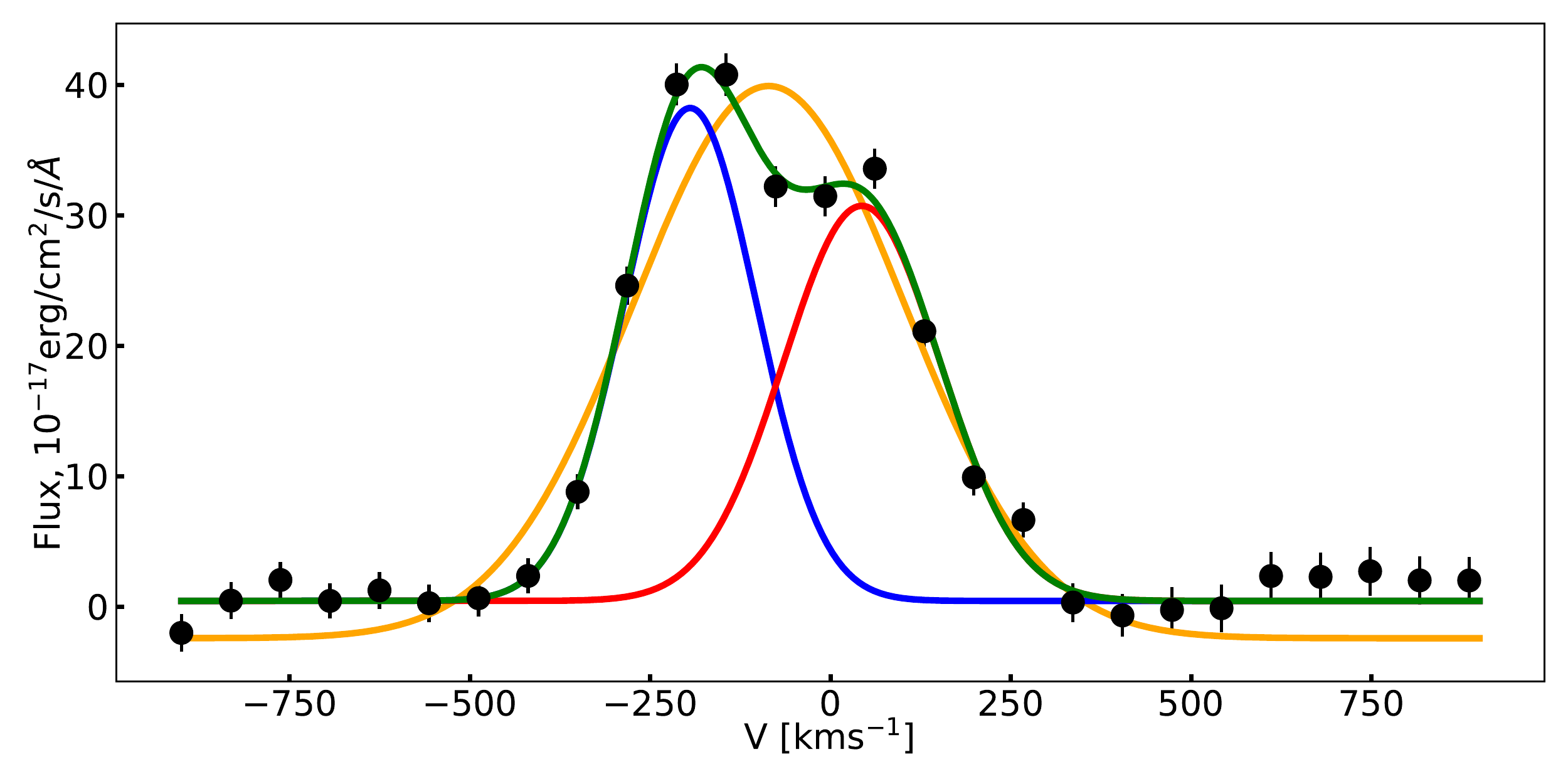}
\caption{Single Gaussian approximation. We compute the two Gaussian functions resulting from the double Gaussian fit procedure to the stacked emission line in \ref{ssect:automated:selection:procedure}. The blueshifted component is represented by a blue and the redshifted by a red line. In orange, we show the best fit of a single Gaussian function.}\label{fig:single:gauss}
\end{figure}

A single Gaussian fit to the complex line shape of DP galaxies provides a measure of the velocity dispersion assuming the DP originates from one system. In Fig.\,\ref{fig:single:gauss} we show one stacked emission line, the double Gaussian fit and the single Gaussian fit performed in Sect.\,\ref{ssect:stack:and:fit}. 

\begin{figure}
\centering 
\includegraphics[width=0.48\textwidth]{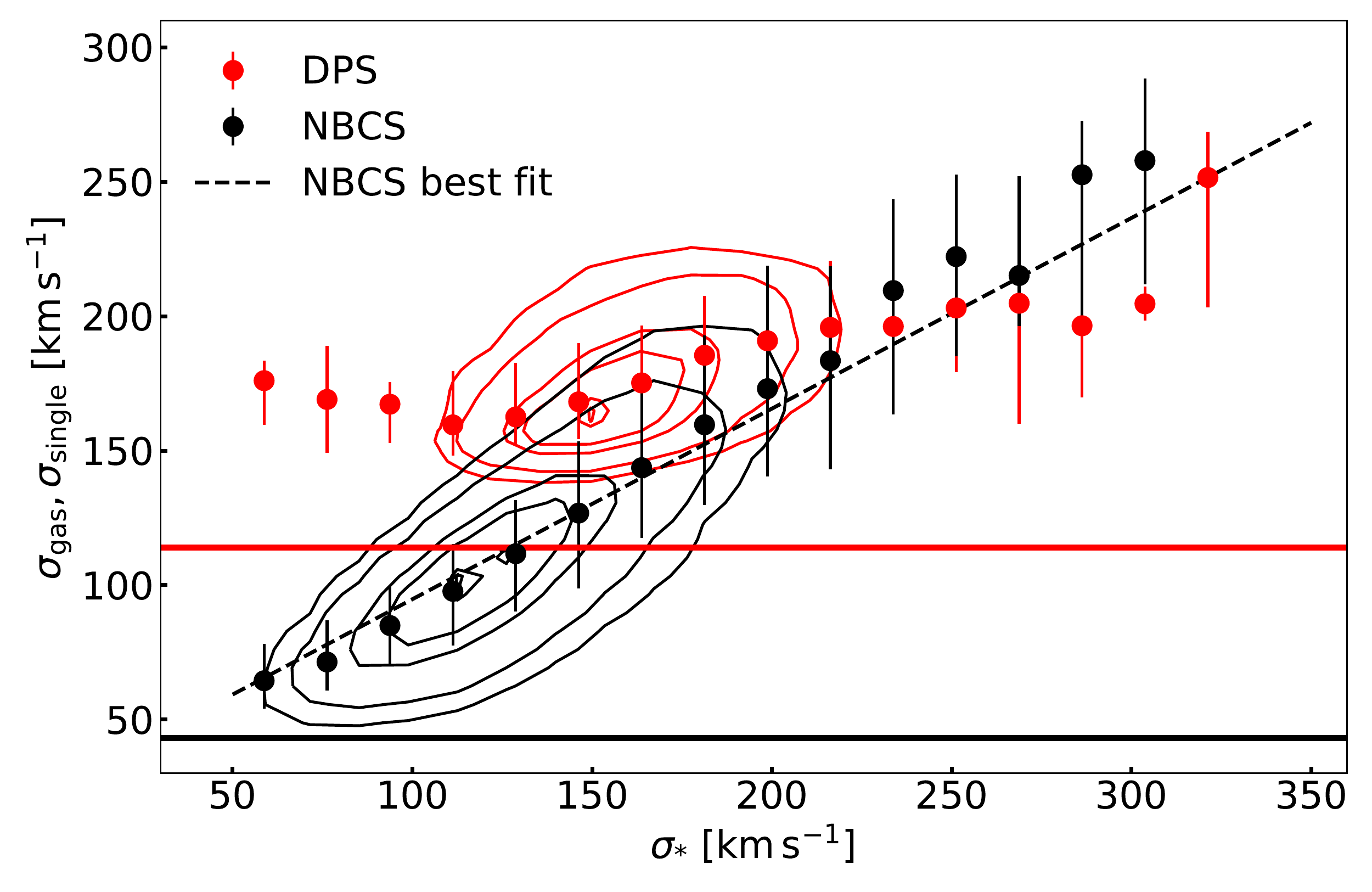}
\caption{Comparison of the stellar velocity distribution $\sigma_{*}$ computed by \citet{2017ApJS..228...14C} with the gas velocity distribution $\sigma_{\rm gas}$ for the NBCS and the velocity distribution of the single Gaussian approximation $\sigma_{\rm single}$ for the DPS. We show in red (resp. black) points the median and the $68\%$ quantile for a binning according to $\sigma_{*}$ of the DPS (resp. NBCS). For the DPS, the limitations in $\sigma_{1,2}$ and $\Delta {\rm v_{DP}}$ of the double Gaussian fit are restricted by the spectral resolution (see Sect.\,\ref{ssect:stack:and:fit}), leading to a restriction of a minimal $\sigma_{\rm single}$, indicated with a red line. We also show the minimal $\sigma_{\rm gas}$ for the NBCS as a black line and the best fit of a linear function as a black dashed line.}\label{fig:compare:star:gas}
\end{figure}

In Fig.\,\ref{fig:compare:star:gas}, we show $\sigma_{*}$ on the x-axis and on the y-axis $\sigma_{\rm gas}$ for the NBCS and $\sigma_{\rm single}$ for the DPS. We display the median and the $68\%$ percentile of $\sigma_{\rm gas}$ and $\sigma_{\rm single}$ with a binning according to $\sigma_{*}$. 
The emission-line parameters $\sigma_{1,2}$ and $\Delta {\rm v_{DP}}$ of the DPS are restricted by a lower limit, defined with the spectral resolution (see Sect.\,\ref{ssect:stack:and:fit}).  We thus find a lower threshold of $114\,\,{\rm km\,s^{-1}}$. Similarly, for the NBCS, we find a minimal $\sigma_{\rm gas}$ of $43\,\,{\rm km\,s^{-1}}$.

By comparing $\sigma_{\rm gas}$ with $\sigma_{*}$ of the NBCS, we find a linear relation described by $ \sigma_{\rm gas} = 0.71\, \sigma_{*} + 23.8 \,{\rm km\,s^{-1}}$ (black dashed line in Fig.\,\ref{fig:compare:star:gas}). This relation 
might be valid for $\sigma_{\rm single}$ and $\sigma_{*}$ of the DPS for higher velocities ($\sigma_{*} > 150\, {\rm km\,s^{-1}}$). In this regime, we can assume a linear relation for the DPS but we still find a shift towards higher velocities in comparison to $\sigma_{\rm gas}$ of the NBCS. This suggests that the gas components of DP galaxies show higher velocities in comparison to $\sigma_{*}$, or that one part of the ionised gas is strongly perturbed or has an external origin. 

\subsubsection{Inclination}\label{ssect:inclination}
\begin{figure*}
  \centering 
 \includegraphics[width=0.98\textwidth]{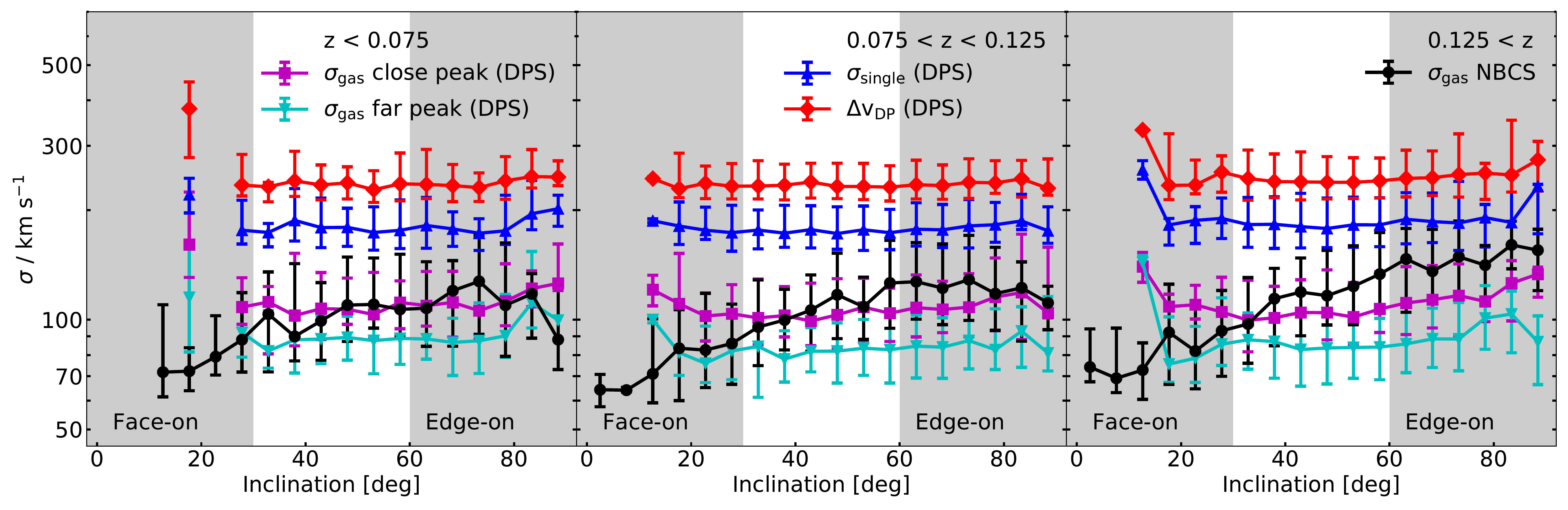}
  \caption{Influence of the inclination on the velocity dispersion. We show different velocity dispersion distributions as a function of the galaxy inclination for different groups of redshift. We only show LTG and S0 galaxies for which we can compute the inclination using \citet{2015MNRAS.446.3943M}. For the DPS, we show $\Delta {\rm V}$ with red dots, $\sigma_{\rm single}$ of the single Gaussian approximation in blue (see sect.\,\ref{ssect:single:gaussian:approximation}) and in magenta (resp. cyan) the $\sigma$ of the emission-line component which is closer (resp. offset) to the stellar velocity in units of $\sigma$. We show the $\sigma_{\rm gas}$ for the NBCS with black dots. All data points represent the medians in the inclination intervals and the error bars the 16th to 84th percentiles.}
  \label{fig:inclination}%
\end{figure*}

A DP emission line shape can of course also originate from a rotating disk. This has been discussed in detail in \citet{2012MNRAS.422.1394E} and shown in simulation in \citet{2019MNRAS.487.3007K}. In order to test this specific scenario, we use a single Gaussian approximation of the DP emission lines (see Sect.\,\ref{ssect:single:gaussian:approximation}) which, in the case of a rotating disc, would be an approximation for the gas velocity dispersion. We assume here that the rotation disc is the main disc of the galaxy, and study the correlation of the velocity dispersion with the inclination angle $i$ defined with the photometric analysis. Hence, we calculate the inclination angle following Eq.\,\ref{eq:inclination} using the minor-to-major axis measured by \citet{2015MNRAS.446.3943M} which provides reasonable results for spiral and S0 galaxies if the disc component exhibits good fit results \citep[see][for more details on fit results]{2015MNRAS.446.3943M}. As a cross-check for the measured inclination, we follow \citet{2010ApJ...709..780Y} and find a dependency between the inclination angle and the Balmer decrement H${\alpha}6565/{\rm H}{\beta}4863$ used to calculate the inner galactic extinction. This trend is comparable to the one observed for the NBCS, but, as mentioned above, the extinction is slightly larger (0.25\,mag) for DPS than for NBCS.

The area covered by the 3" SDSS fiber depends on the redshift. The measured gas velocity dispersion of a rotating disc depends on the area integrated by the spectroscopic observations. We thus define three redshift groups as in Sect.\,\ref{ssect:star:formation:agn} for this analysis: ${\rm z} < 0.075$, $0.075 < {\rm z} < 0.125$ and $0.125 < {\rm z}$. In Fig.\,\ref{fig:inclination}, we present the estimated inclination angle $i$ as a function of various gas velocity measurements for the DPS and the NBCS. The NBCS is represented by the gas velocity dispersion $\sigma_{\rm gas}$ (see Sect.\,\ref{ssect:velocity:distribution}). For the DPS, we display $\sigma_{\rm single}$ from the single Gaussian approximation (Sect.\,\ref{ssect:single:gaussian:approximation}), $\Delta {\rm v_{DP}}$ from the stacking procedure (Sect.\,\ref{ssect:stack:and:fit}) and $\sigma_{\rm close}$ and $\sigma_{\rm far}$ as described in Sect.\,\ref{ssect:velocity:distribution}. For galaxies with a redshift ${z > 0.075}$, we see a clear correlation in the NBCS between $\sigma_{\rm gas}$ and the galaxy inclination $i$. This correlation between $\sigma_{\rm gas}$ and $i$ can be explained by a rotating disc. In contrast, we do not find any dependencies between the displayed velocities of the DPS and $i$ in any redshift range. The absence of such a correlation for the DPS disfavours the rotating (main-) disc scenario.

\subsubsection{Tully-Fisher and Faber-Jackson relations}\label{ssect:TF:FJ}
\begin{figure}
  \centering 
 \includegraphics[width=0.48\textwidth]{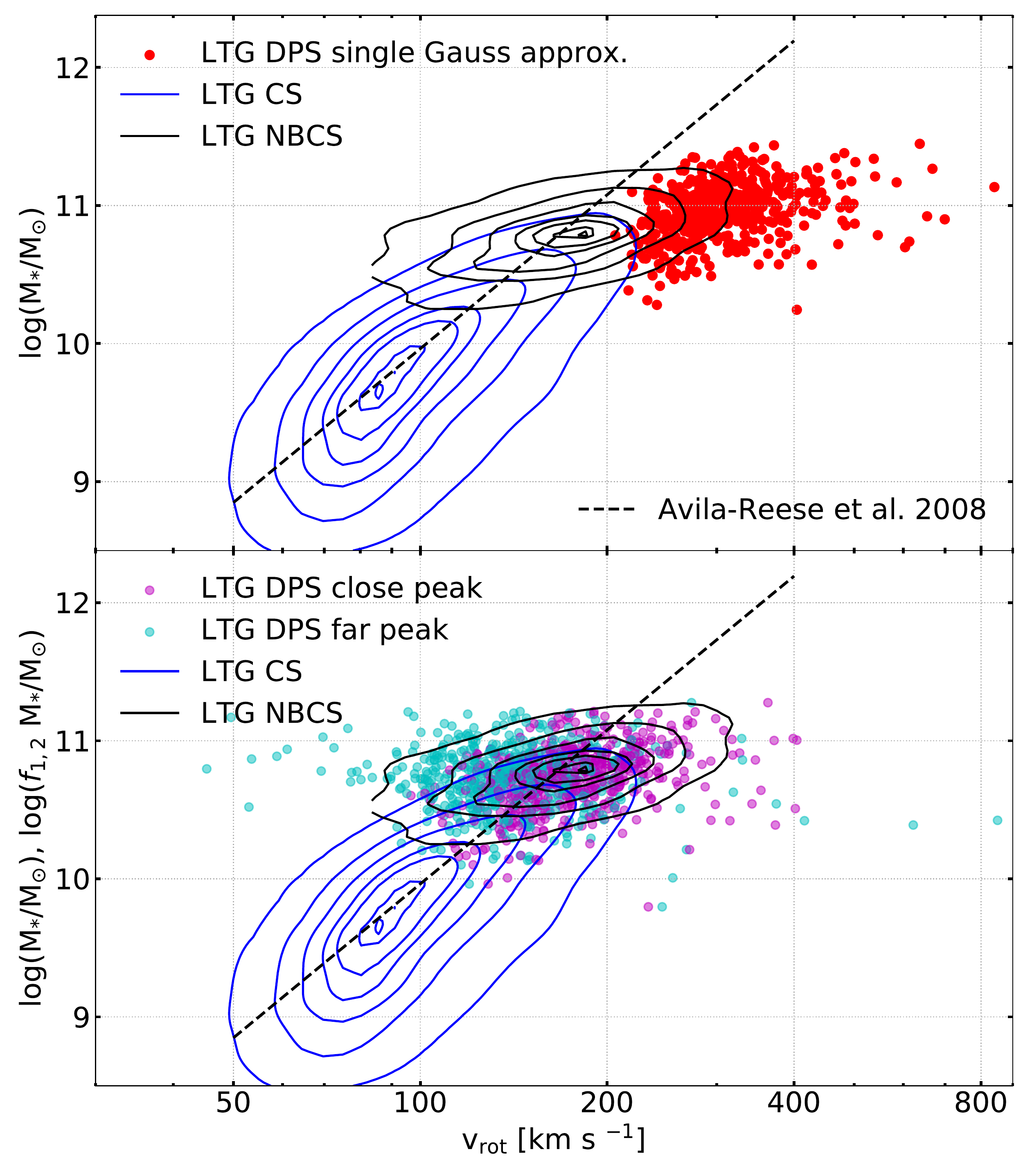}
  \caption{Tully-Fisher relation (TFR) for LTGs. We compute the rotational gas velocity $v_{\rm rot}$ for the NBCS and CS (black and blue contour lines) using the gas velocity dispersion $\sigma_{\rm gas}$ (see Sect.\,\ref{ssect:velocity:distribution}). For the DPS, we use the $\sigma_{\rm single}$ of the single Gaussian approximation in the top panel. In the second panel, we show the TFR for the close and far peak-components of the DPS (see Sect.\,\ref{ssect:velocity:distribution}). The stellar masses are computed by \citet{2003MNRAS.341...33K}. We show the best fit computed by \citet{2008AJ....136.1340A} as a dashed line.}
  \label{fig:tully:fisher:ltg}%
\end{figure}
\begin{figure}
  \centering 
 \includegraphics[width=0.48\textwidth]{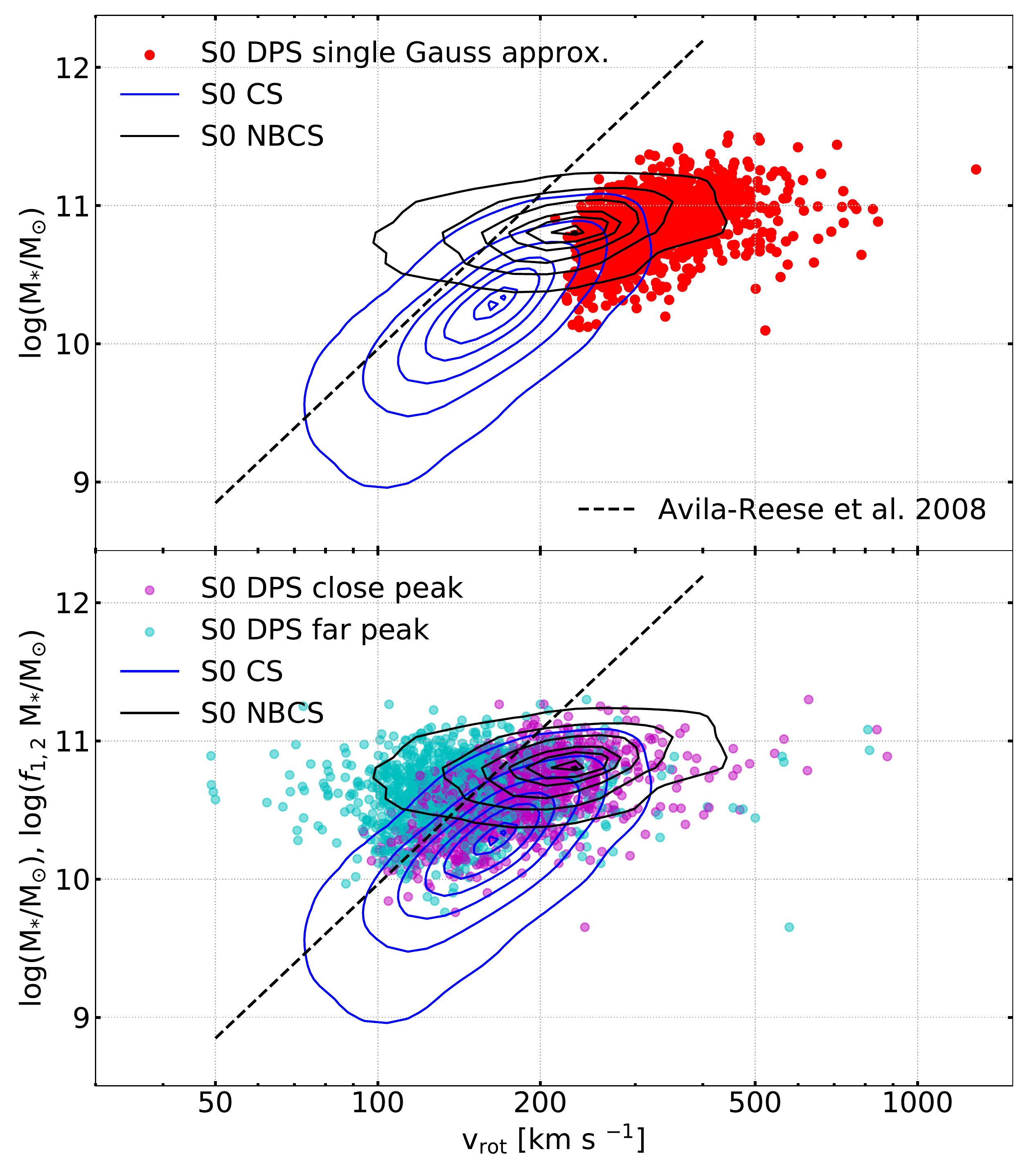}
  \caption{Same as Fig.\,\ref{fig:tully:fisher:ltg}, but for S0 galaxies. The dashed line corresponds to the best fit  of \citet{2008AJ....136.1340A}.}
  \label{fig:tully:fisher:s0}%
\end{figure}
\begin{figure}
  \centering 
 \includegraphics[width=0.48\textwidth]{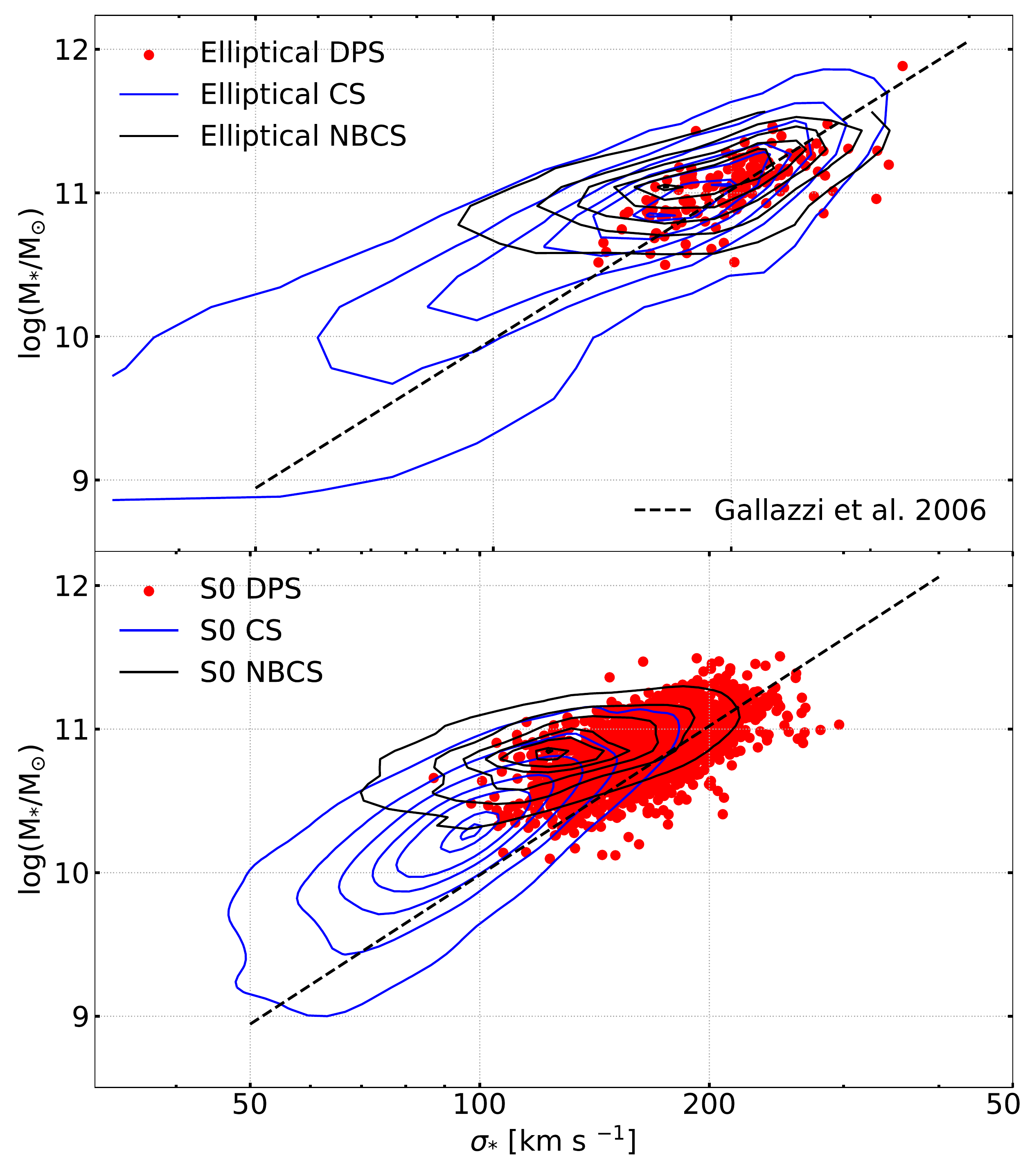}
  \caption{Faber-Jackson relation for elliptical and S0 galaxies using the stellar velocity dispersion taken from \citet{2017ApJS..228...14C} on the $x$-axis and the stellar masses computed by \citet{2003MNRAS.341...33K} on the $y$-axis. We show the best fit computed by \citet{2006MNRAS.370.1106G} in dashed lines.}
  \label{fig:faber:jackson}%
\end{figure}

The velocities measured in galaxies are tightly correlated to their size. This relation was first discovered by \citet{1976ApJ...204..668F} for the stellar velocity dispersion and the absolute magnitude of elliptical galaxies (FJR). They concluded that the luminosity $L$ of a galaxy is consistent with the relation ${L} \propto \sigma_{*}^{4}$. In parallel, a good relation between the full-width at $20\%$ of the maximum of the HI profile, corrected to the inclination of the galaxy, and the galaxy size was found by \citet{1977A&A....54..661T} for spiral galaxies. This Tully-Fisher relation (TFR) is usually understood as the self-regulation of star formation in the disc \citep[e.g.][]{1997ApJ...481..703S}.

The parameters of these relations have been later measured with higher accuracy. Here, we use for the FJR:\\ 
$\log( \sigma_{*} /{\rm  km\,s^{-1})} = -0.90 \pm 0.12 + (0.29 \pm 0.02)\,
\log_{10}(M_{*} / {\rm M}_{\odot})$, measured by \citet{2006MNRAS.370.1106G}. For the TFR, we use:\\ 
$\log_{10}( v_{\rm rot} / {\rm km\,s^{-1}}) = -0.69 + (0.27 \pm 0.01)\, 
\log_{10}(M_{*} /{\rm  M}_{\odot}) $, measured by \citet{2008AJ....136.1340A}, where $v_{\rm rot}$ is the rotation velocity, calculated as  in \citet{2005AJ....130.1037C} and \citet{2018MNRAS.479.2133A}: 
$\mathrm{v_{rot}} = W/ [2 \sin i]$,
where $i$ is the inclination angle (see Eq.\,\ref{eq:inclination}) and W the difference between the 10th and the 90th percentile of the velocity measurements. Since we do not have spatially resolved information on  galaxy kinematics, we approximate from a Gaussian shaped emission lines the 10th and the 90th percentile as ${\rm W = 2.56 \sigma}$ (This is close to the full-widths half maximum of a Gaussian at $2.36\sigma$). We calculate ${\rm v_{rot}}$ for the TFR using $\sigma_{\rm gas}$ for the CS and the NBCS (see Sect.\,\ref{ssect:velocity:distribution}) and $\sigma_{\rm single}$, $\sigma_{\rm close}$ and $\sigma_{\rm far}$ for the DPS (see  Sect.\,\ref{ssect:velocity:distribution} and \ref{ssect:single:gaussian:approximation}). We take the stellar mass ${\rm log(M_{*}/M_{\odot})}$ measured by \citet{2003MNRAS.341...33K} for the TFR and FJR. In the case where we display only one of the DP (the close or the far peak), we approximate the fraction of the stellar mass for each peak as ${\rm log}(f_{1,2}\,{\rm M_{*}/M_{\odot})}$. We assume $f_{1,2}$, the mass fraction of each peak, to be the same fraction as the total emission line flux of the stacked lines. This is a rough approximation but would be our best guess to assume each peak corresponds to one component of a hidden merger.

In Fig.\,\ref{fig:tully:fisher:ltg} (resp. \ref{fig:tully:fisher:s0}), we present the TFR for LTG (resp. S0) of the DPS, the CS and the NBCS. We find good agreement for LTG of the CS and the NBCS with the parameters for the TFR found by \citet{2008AJ....136.1340A}. We find that, for the LTGs of the DPS, the rotation velocity calculated with $\sigma_{\rm single}$ does not follow the TFR. We measure ${\rm v_{rot}}$ to be higher than expected for the measured masses. We find a good agreement using $\sigma_{\rm close}$ whereas rotation velocities calculated with $\sigma_{\rm far}$ are shifted towards lower velocities. 
For S0 galaxies, we observe the same velocity shifts as shown in Fig.\,\ref{fig:tully:fisher:s0}. S0s and LTGs have a similar behaviour: $\sigma_{\rm single}$ is offset with respect to the NBCS distribution considering the two peaks as individual galaxies following this distribution. In both TFRs, we find some outliers with extreme ${\rm v_{rot}}$ values. By individual inspection, we find these velocities to be the result of a small inclination and/or very broad peak component.

In Fig\,\ref{fig:faber:jackson}, we show the FJR (M$_{*}$ vs $\sigma_{*}$) for elliptical and S0 galaxies. As a reference, we use the best fit from \citet{2006MNRAS.370.1106G} as a dashed line. We find good agreement for elliptical galaxies of the DPS, CS and the NBCS. For S0 galaxies, we find in the NBCS and the CS systematically lower $\sigma_{*}$, whereas the DPS shows velocities agreeing with those of elliptical galaxies with the same stellar mass. This is in agreement with the result discussed in Fig.\,\ref{fig:velocity:distributions} that the stellar velocity dispersion is larger for the DP galaxies, supporting again a large system composed of the superposition of two systems.

\subsection{Morphology and galaxy environment}\label{ssect:morph:environment}
We study the environment of the DP galaxies with the galaxy group catalogue provided by \citet{2007ApJ...671..153Y}\footnote{We used the galaxy group catalogue compiled on the SDSS DR 7 available under: https://gax.sjtu.edu.cn/data/Group.html} and \citet{2016A&A...596A..14S}. \citet{2007ApJ...671..153Y} is a halo-based group finder algorithm, determining the group masses and identifying each member of groups using the redshift. About $90\%$ of the DPS and the NBCS are covered by this catalogue. 
\citet{2016A&A...596A..14S} calibrated a group finder algorithm with simulation data providing a higher sensitivity to different kinds of galaxy groups up to a redshift of ${\rm z} < 0.11$ ($47\%$ of the DPS and NBCS). In this redshift range, they cover $97\%$ (resp. $98\%$) of the DPS (resp. NBCS).

We use galaxy groups sizes as discussed in \citet{2009ARA&A..47..159B}: a poor group holds 2 to 4 members, a rich group 5 to 9 and a cluster more than 10. With \citet{2007ApJ...671..153Y}, we find $64\%$ (resp. $66\%$) of the DPS (resp. NBCS) to be isolated, $19\%$ (resp. $17\%$) in poor groups, $3\%$ (resp. $4\%$) in rich groups and $4\%$ (resp. $4\%$) in clusters. To compare this result with \citet{2016A&A...596A..14S}, we looked at the fraction for galaxies with redshift ${\rm z} < 0.11$ and find similar fractions as above. Using \citet{2016A&A...596A..14S}, we find $45\%$ (resp. $45\%$) of the DPS (resp. NBCS) to be isolated, $34\%$ (resp. $33\%$) in poor groups, $11\%$ (resp. $12\%$) in rich groups and $9\%$ (resp. $12\%$) in clusters. We detect fewer isolated galaxies in comparison to \citet{2007ApJ...671..153Y}, which is due to the higher sensitivity of group identification.

In Table\,\ref{table:environment}, we present the fractions of the environmental classification for the morphological types (see Sect.\,\ref{ssect:morph}). Analysing group properties such as number of group members, distance to the closest neighbour or the velocity dispersion of a galaxy group, we only find a difference between the DPS and the NBCS for elliptical galaxies \citep{2016A&A...596A..14S}. In the DPS (resp. NBCS), we find $23\%$ (resp. $41\%$) of the elliptical galaxies to be situated in a dense environment (rich groups or clusters), whereas $72\%$ (resp. $58\%$) are situated in less dense environments (isolated galaxies or poor groups).

S0 galaxies are usually observed more abundant in clusters than in the field \citep{1980ApJ...236..351D}, but this is not the case here. In the DPS and the NBCS, we find about $80\%$ of the S0 galaxies to be situated in less dense environments. We do not find any differences in the fraction of isolated S0 galaxies selected in the DPS and the NBCS, but we clearly see that most of the selected S0 galaxies are isolated. Even though we find an excess of S0 galaxies and a lack of LTGs in the DPS in comparison to the NBCS (see Sect.\,\ref{ssect:morph}), we do not find any connection between the galaxy environment and the morphological type.

\citet{1997ApJ...490..577D} first observed an increase of S0 galaxies over elliptical fraction in groups from ${\rm z} \sim 0.5$ till today, while the spiral fraction before ${\rm z} \sim 0.5$ corresponds to the S0 fraction observed today. This supports the view that spirals evolve into S0 galaxies due to their environment \citep{2000ApJ...542..673F}. This has been supported by subsequent observational works \citep[e.g.][]{2009ApJ...692..298W}. However, this is not the case here as 70$\%$ of the DP-S0 galaxies are isolated according to \citet{2007ApJ...671..153Y}, (or $50\%$ for $z < 0.11$ according to \citet{2016A&A...596A..14S}), and the behaviour of S0 and LTG are not significantly different in the DPS in terms of excitation mechanisms and dynamics. This apparently contradicts the work of \citet{2018ApJ...862..100G}, based on photometric decomposition, who find that the bulges of spirals and S0s are intrinsically different and thus spirals are probably not the progenitors of S0s. 

\subsection{Non-parametric morphological diagnostics}\label{sect:morph:diagnostics}
\begin{figure}[h]
\centering 
\includegraphics[width=0.46\textwidth]{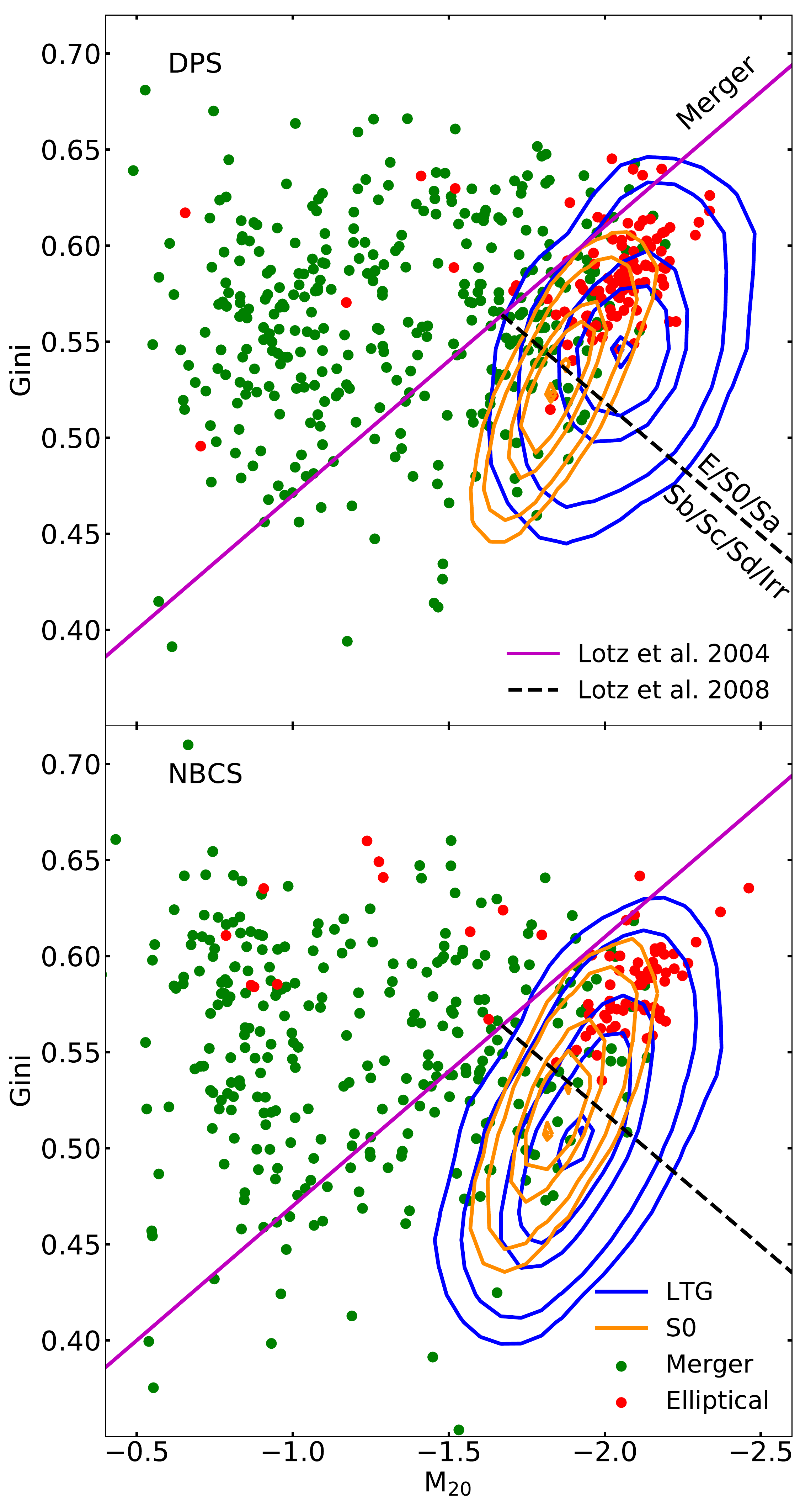}
\caption{Gini vs ${\rm M _{20}}$ coefficients for the DPS (\emph{top panel}) and the NBCS  (\emph{lower panel}). Blue (resp. orange) contours represent galaxies classified as LTG (resp. S0) and red (resp. green) dots denote elliptical (resp. merger) galaxies (See the classification in Sect.\,\ref{ssect:morph}). The purple line separates galaxies classified as merger on the left hand side and galaxies associated with a Hubble type on the right hand side \citep{2004AJ....128..163L}. The lower black dashed line divides elliptical, S0 and Sa galaxies from later types namely Sb, Sc and Irr \citep{2008ApJ...672..177L}.
}
\label{fig:gini:m20}
\end{figure}

To further investigate the properties of the different morphological galaxy types, we compute photometric diagnostics for the DPS and NBCS. We use the python package {\sc statmorph}\footnote{\url{https://statmorph.readthedocs.io}} which calculates non-parametric morphological diagnostics of galaxy images (e.g. CAS-statistics or the Gini coefficient), as well a fit of a 2D S\'ersic profile \citep{2019MNRAS.483.4140R}. Here, we test our galaxy samples using CAS-statistics composed by concentration (C), asymmetry (A) and smoothness (S), the shape-asymmetry (${\rm A_{S}}$), the Gini-coefficient, the second moment of the galaxy’s brightest regions ${\rm M_{20}}$ and the S\'ersic index n. For an exact definition and description of these diagnostics, please see \citet{2019MNRAS.483.4140R} and references therein. We apply {\sc statmorph} to the $r$-band $62^{\prime\prime}\times 62^{\prime\prime}$ Legacy Survey snapshots \citep{2019AJ....157..168D}.

The CAS-statistics were introduced by \citet{2003ApJS..147....1C} to distinguish between different morphological types. As discussed in Sect.\,\ref{ssect:star:formation:agn}, we find for some sub-samples of the DPS a higher concentration, accompanied by a higher central star formation rate, in comparison to the NBCS. However, we do not find any differences in asymmetry or smoothness between the DPS and the NBCS.

To automatically recognise ongoing or past mergers, \citet{2016MNRAS.456.3032P} introduced the shape asymmetry (${\rm A_{S}}$), which specifically detects galaxies with low-surface-brightness tidal features. We observe higher ${\rm A_{S}}$ for galaxies classified as mergers than for other types (as expected), but we find for all sub-samples and redshift ranges similar distributions for the DPS and NBCS. 

\citet{2003ApJ...588..218A} applied the Gini coefficient on galaxy imaging in order to determine the galaxy morphology. The Gini coefficient quantifies the inhomegeneity (or "inequality" when applied to human populations) with which a galaxy’s light is distributed. \citet{2004AJ....128..163L} combined the Gini coefficient with the second moment of the galaxy’s brightest regions ${\rm M_{20}}$ to be able to distinguish between merger and non-merger galaxies. We present this diagnostic in Fig.\,\ref{fig:gini:m20}. Galaxies with an ongoing merger process are located on the left of the Gini-${\rm M_{20}}$ diagram, which is marked by a purple line \citep{2008ApJ...672..177L}. In post coalescence stages, the merger would migrate back to the right where two populations are located: in the upper right part elliptical, S0 and Sa galaxies are located, whereas in the lower right part irregular and spiral galaxies, of type Sb, Sc, Sd or Irr, are located \citet{2008ApJ...672..177L}. These two regions are separated by a black dashed line. Using hydrodynamical simulations, this diagnostic combined with asymmetry is sensitive to merger time-scales of $\sim 200-400$\,Myr for major disc merger and 60\,Myr for minor mergers \citep{2010MNRAS.404..575L}.

\begin{figure}[h]
\centering 
\includegraphics[width=0.46\textwidth]{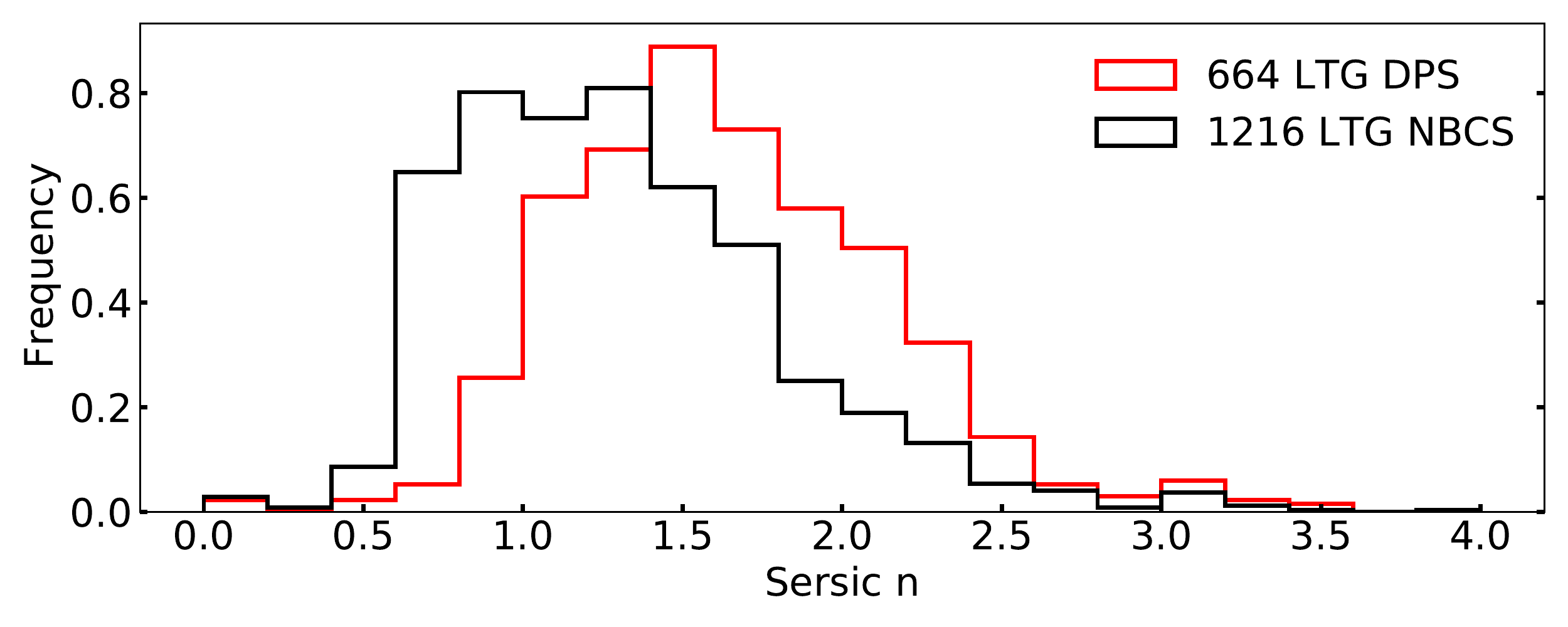}
\caption{Distribution of S\'ersic indices, computed by {\sc statmorph} \citep{2019MNRAS.483.4140R} for the Legacy survey $r$-band images \citep{2019AJ....157..168D}, of LTGs of the DPS (red line) and NBCS (black line). The areas under the histograms are normalised to unity.}
\label{fig:sersic}
\end{figure}

Galaxies of the DPS and the NBCS, classified as merger or elliptical, show good agreement between their classification and their position on the Gini-${\rm M_{20}}$ diagram (Fig.\,\ref{fig:gini:m20}). Galaxies classified as S0 are spread over the region of early and late Hubble types with similar distributions between the DPS and the NBCS.
This diagnostic diagram indicates a difference for LTG of the DPS in comparison to the NBCS: we observe that LTGs of the DPS are mainly ($63\,\%$) situated in the region associated with earlier Hubble type galaxies, whereas LTGs of the NBCS are mainly situated in the region of later Hubble types (only $35\,\%$ in the earlier Hubble type region). One might think that Sa galaxies are over-represented in the DPS.
However, with respect to the full sample, we detect $10\,\%$ of the DPS and $11\,\%$ of the NBCS classified as Sa galaxies. 
We only find $6\,\%$ of the DPS that are classified as Sb-d, whereas $19\,\%$ of the NBCS are of this type. Hence, there is no Sa excess but a deficit of Sb-d galaxies.
In Appendix\,\ref{sect:param:dist}, we find Sb-d of the DPS to be the only morphological sub-sample that shows larger stellar masses ($\rm \sim 0.2$\,dex in $\log(M_*/{\rm M}_\odot)$) in comparison to  LTG galaxies of the NBCS. Additionally, we find the highest difference in extinction of about $\rm 0.3\,mag$ in the $V$ band. We furthermore detect a significant excess in star formation in the centre: for Sb-d of the DPS (resp. NBCS), we find a median ${\rm SFR_{fiber} / SFR_{total}} = 0.59 ^{+0.30}_{-0.31}$ (resp. ${\rm SFR_{fiber} / SFR_{total}} = 0.31 ^{+0.20}_{-0.16}$).

The shift towards earlier Hubble types of LTGs of the DPS, in comparison to the NBCS, is also supported by the measured S\'ersic index, presented in Fig.\,\ref{fig:sersic}. By comparing the S\'ersic index of the DPS with those of the NBCS, we only find a discrepancy for LTGs, all other subsets show similar distributions. This indicates, that LTG of the DPS tend to have larger bulges in comparison to the LTGs of the NBCS.

Interestingly, \citet{2019ApJ...872...76N} tested the position of simulated minor and major mergers in the Gini-${\rm M_{20}}$ diagram (see their Fig. 7) and found that most minor mergers and about half of the major mergers lie below the purple line, and mainly in the ETG/S0/Sa region (above the black dashed line). This corresponds to the trend discussed in \citet{2008ApJ...672..177L}: post-coalescence galaxies are more concentrated in their light distribution, while tidal features vanish.

In Sect. \ref{sect:ld1}, we follow \citet{2019ApJ...872...76N}, who developed a supervised Linear Decomposition Analysis (LDA) in order to compute the predictor coefficients LD1$_{\rm major}$ and LD1$_{\rm minor}$, defined to identify major and minor mergers. To get comparable results, we restrict our analysis to ${\rm z} < 0.075$. We find in cases, such as LTG and S0 galaxies, slightly higher merger rates for the DPS in comparison to the NBCS. This result somehow supports a merging scenario, while it is compatible with the fact DP galaxies might be dominated by  post-coalescence mergers, that would escape such analysis targeting on-going mergers. This result is compatible with the results discussed above based on the Gini-${\rm M_{20}}$ diagram.

\subsection{Summary}
\label{ssect:sum}
We have studied different parameters derived from observations to compare the DPS with the NBCS. The NBCS is defined to have the same redshift and stellar mass distributions as the DPS (see Sect.\,\ref{ssect:bias}).

We find the following similar properties:
\begin{itemize}
\item Distributions of absolute magnitudes, stellar ages, metallicities, specific star formation ratio and [OIII]$\lambda$5008 emission-line luminosities are identical.
\item According to their colour-colour and their specific star formation-stellar mass diagrams, most of DPS and NBCS galaxies follow the star forming main sequence, only a few exhibit a quenched SF and some are located in the red sequence.
\item Double-AGN behave as NBCS-AGN with respect to star-formation: there is no significant central excess.
\item If one considers that each peak corresponds to a component, the DPS and NBCS follow the Tully-Fisher relation.
\item Elliptical galaxies from DPS and NBCS follow the Faber-Jackson relation.
\item The same environment statistics are found for the DPS and the NBCS. Most of S0s are isolated, and the S0s excess observed in the DPS is not due to environment.
\end{itemize}

We detect the following differences:
\begin{itemize}
    \item There is an excess of S0s (resp. a deficit of LTGs) in the DPS with respect to the NBCS.
    \item According to the  Gini-$ \rm M_{20}$ diagram, there appears to also be an excess of Sa galaxies in the DPS compared to the NBCS.
    \item LTG galaxies display higher S\'ersic indices in the DPS  compared to those in the NBCS.
    \item Galaxies classified as LINER are less common in the DPS in comparison to the NBCS. 
    \item DPS galaxies show slightly higher absolute luminosities in the H$_{\alpha}\lambda6565$ emission line. 
    \item Double-SF galaxies exhibit a central starburst stronger than SF galaxies from the NBCS. 
    \item Double-COMP, -LTG and -S0 galaxies also have such an excess but less significant.
    \item Double-COMP are more likely SF galaxies, according to the [OII]$\lambda 3728$ / [OIII]$\lambda 5008$ line ratio, and to the detection rate of the high-ionisation [NeV]$\lambda 3426$ emission line at $z>0.2$.
    \item There is slightly more extinction (0.25\,mag in $V$) in DPS galaxies than in their NBCS counterparts.
    \item All DPS velocity estimators are independent of the inclination, while we do detect a correlation for the NBCS. 
    \item At given stellar masses, stellar velocity dispersions are larger for the DPS than for the NBCS.
    \item The gas velocity dispersion estimator of the close peak component of the DPS is compatible with velocity dispersions measured for the NBCS, whereas the far peak component of the DPS is not.
    \item With a single Gaussian approximation, the LTG and S0 galaxies from the DPS are offset from the NBCS TF relation.
    \item S0 galaxies from the DPS are shifted on the FJ relation from the NBCS. 
\end{itemize}

\section{Discussion}
\label{sect:discussion}
In Sect. \ref{ssect:SFS0}, we argue that S0 galaxies detected as gas-rich DP galaxies belong to a star-forming population identified in previous works. In Sect. \ref{ssect:non-para}, we discuss how our non-parametric analysis revealed that LTG are dominated by Sa galaxies in the DPS. In Sect. \ref{ssect:S0mm}, we propose that the DP galaxies sample is a minor merger sequence which could explain the different results discuss in Sect. \ref{ssect:sum}.
In Sect. \ref{ssect:support}, we discuss here different observations that are compatible with the characteristics of the DPS described as a minor merger sequence. In Sect. \ref{ssect:mima}, we compare the known properties of the DPS galaxies with respect to those of minor and major mergers. To present a complete discussion, we discuss further explanations of the observation of DP in Sect.\,\ref{ssect:alternatives} and find that these only apply to a few galaxies in our sample. 

\subsection{Star-forming S0s}\label{ssect:SFS0}
S0 galaxies are usually described as disc galaxies, which have exhausted their gas content \citep[e.g.][]{2015ARA&A..53...51S} and are proposed as a parallel Hubble sequence \citep{1976ApJ...206..883V,2012ApJS..198....2K}. However, it has been known for several decades that some S0s are not forming stars but host large amounts of HI gas \citep[e.g.][]{1989ApJ...344..685K}. In these early observations of HI-gas-rich S0 galaxies \citep[e.g.][]{1983IAUS..100...99V,1984A&A...133..127K,1985A&A...142....1K,1985A&A...144..202K,1988A&A...191..201V}, inner and outer rings were already detected, and analysed in simulations \citep[e.g.][]{1985ARA&A..23..147A,1996FCPh...17...95B}. The HI gas has often an H$\alpha-[NII]$-counterpart \citep[e.g.][]{1993AJ....106.1405P}. Observations have shown indeed that about $85\%$ of S0 galaxies host optical ionised gas \citep[e.g.][]{1996A&AS..120..463M,2006MNRAS.366.1151S} and $72\%$ of those are isolated \citep[][]{2015AJ....150...24K}. \citet{2009AJ....138..579K} identified a blue sequence of SF S0 galaxies in the low-mass range that might be fading mergers, while massive S0s, with $M>2\times 10^{11}\,M_\odot$, belong to the red sequence, but up to 2$\%$ of these massive S0 are SF. In the blue S0 sequence, they identify central blue colour gradients that are interpreted as reminiscent of mergers \citep[e.g.][]{2004ApJ...611L..89K,2006AJ....131.2004K}. \citet{2016ApJ...831...63X} find about $8\%$ of an S0 sample exhibits central star formation.  They also found that the majority of these SF S0 galaxies have a stellar mass below $10^{10.6} M_\odot$. \citet{2020MNRAS.tmp.1540T} did not find such a mass separation. In their study of these two S0 populations composed of SF and quiescent galaxies, they found that SF S0 galaxies avoid high-galaxy density and that their SFR and spectral characteristics are entirely similar to those seen in LTGs.

Numerous observations show further evidence of various tidal disturbances in the outer parts of S0: outer rings \citep[e.g.][]{2014A&A...562A.121C,2018A&A...620L...7S}, polar rings \citep[e.g.][]{1987ApJ...314..457V,1990AJ....100.1489W}, circum-nuclear polar rings \citep[e.g.][]{2004AJ....127.2641S,2009A&A...504..389C,2016AJ....152...73S}, counter-rotations \citep[e.g.][]{2013ApJ...769..105K,2014MNRAS.438.2798K,2014MNRAS.439..334I,2015AJ....150...24K,2016MNRAS.461.2068K,2018A&A...616A..22P} and cylindrical rotation \citep[e.g.][]{2019MNRAS.488.1012M}. In \citet{2014MNRAS.438.2798K,2015AJ....150...24K}, \citet{2019AJ....158....5P} and \citet{2019ApJS..244....6S}, the authors discuss the origin of this counter-rotating gas and argue that S0s might accrete gas probably from filaments or from minor mergers. In parallel, numerous works have shown that S0s can be produced by galaxy mergers including major mergers \citep{1998ApJ...502L.133B,2015A&A...579L...2Q,2017A&A...604A.105T,2018A&A...617A.113E} as well as minor mergers \citep{2005A&A...437...69B,2011MNRAS.415.1783B}.

Here, we clearly detect intermediate-mass S0 galaxies in the mass range $3\times 10^{10} - 3\times 10^{11}\,M_\odot$ with optical ionised gas, including a large fraction of star forming galaxies. Beside their optical morphology, their kinematic behaviour is similar to those of LTGs. We find similar distributions for LTGs and S0 galaxies in the TFR in Sect.\,\ref{ssect:TF:FJ}. We argue that the close peak might represent the central galaxy, while the far peak might correspond to a smaller companion or to the result of a recent gas accretion.

\subsection{Non-parametric merger identification}\label{ssect:non-para}

In the Sect.\,\ref{sect:morph:diagnostics}, we compute several non-parametric diagnostics such as CAS-statistics, the shape asymmetry ${\rm A_{s}}$, the Gini and ${\rm M_{20}}$ coefficients and the S\'ersic index n to investigate different approaches of merger identification.

Morphological asymmetry is more prevalent in galaxies within close pairs \citep{2016MNRAS.461.2589P}.
Moreover, SF is enhanced in close pairs with small differences in line-of-sight velocities  
\citep{2011MNRAS.412..591P,2013MNRAS.433L..59P}.
We studied galaxy asymmetries in the DPS, relying on the parameters of  \citet{2016MNRAS.461.2589P}. We do not find any asymmetry difference between the DPS and the NBCS nor any connection between SF and asymmetry. In a scenario where the central SF enhancement was due to a merging process, the absence of a significant asymmetry in the DPS would favour later stages of merging processes or mergers hidden inside the fibre.
\citet{2016MNRAS.456.3032P} introduced the shape asymmetry to identify later stages of merger in comparison to the asymmetry. We do not find any difference between the DPS and the NBCS using this diagnostics. 

We conducted a merger selection based on the Gini-${\rm M_{20}}$ disgnostics, proposed by \citet{2004AJ....128..163L, 2008ApJ...672..177L} to separate mergers from non-mergers and to distinguish between late and early Hubble types. We find good agreements with the predictions for S0, merger and elliptical galaxies of the DPS and the NBCS. LTGs of the DPS reveal a systematic effect: they are situated in the region of earlier Hubble types in comparison to LTGs of the NBCS (see Fig.\,\ref{fig:gini:m20}). This is in agreement with the fact that we find a tendency towards higher S\'ersic index $n$ for LTGs of the DPS in comparison to the NBCS, supporting the idea of a sequence of galaxies thickening towards S0. Last, this is in agreement with the trend discussed in Appendix \ref{sect:ld1}
with the small positive bias for minor and major mergers observed in the predictor coefficient distributions, as defined by \citet{2019ApJ...872...76N}. These authors also show that numerous simulated (major and minor) mergers lie in the non-merger regions of the Gini-M$_{20}$ diagram.

Comparing all non-parametric methods, designed to detect merger rates, we do not find a direct relation between on-going mergers and DP structure at $z<0.075$. Even though, these diagnostics are developed to detect late and post-coalescence stages of mergers \citep{2016MNRAS.456.3032P,2008ApJ...672..177L,2010MNRAS.404..575L,2010MNRAS.404..590L,2019ApJ...872...76N}, we might see an even later stage, where traces such as fading tidal features are already quite weak and double nucleus beyond spatial resolution. \citet{2019ApJ...872...76N} defined post-coalescence when the two galactic nuclei are separated by less than 1\,kpc, which correspond to the SDSS resolution at $z \le 0.03$. Indeed as the DP feature is based on the central 3" spectra, it is probable that the discussed diagnostics are not sensitive.  

\subsection{S0s as a part of the minor merger sequence}\label{ssect:S0mm}
\citet{1996ApJ...460..121W} found that in 1:10 minor mergers, the discs are not destroyed but evolve to an earlier Hubble type, while a core of $45\%$ of the satellite initial mass can reach the central kpc in 1\,Gyr. This scenario might account for our DPS: high stellar velocity dispersions, two gas components and a similar kinematic behaviour for the different morphological types. This is supported by numerous hydrodynamical simulations. \citet{2005A&A...437...69B} have shown that intermediate mergers with 1:4 to 1:10 mass ratio can produce S0 galaxies. A similar situation is observed in GALMER simulations \citep{2010A&A...518A..61C}. Hence, while it is now well-known that S0s can be formed by minor and major mergers depending on the orientation and kinematics, \citet{2007A&A...476.1179B} have studied the evolution of galaxies due to repeated minor mergers, as expected in the hierarchical growth of galaxies. Multiple sequential mergers feed the main progenitor, and gradually change its morphology from a spiral to an elliptical-like system. In addition, this is compatible with the work of  \citet{2015ApJ...813...23V} who proposed a two-channel evolution scheme, to account for the evolution of mass-size relation: they first grow inside-out with gas accretion and gas-rich minor mergers, until they quench, and continue to grow by dry minor mergers. \citet{2013ApJ...778L...2C} proposed a similar modelling to account for the evolution of slow and fast rotators, as further discussed in \citet{2016ARA&A..54..597C}.
This scenario is fully compatible with the properties of Sa and S0 galaxies and such a sequential gas-rich minor merger sequence accounts for the main properties of the DPS galaxies, identified here. 

The increased central star formation and enhanced central extinction are also well accounted for by minor mergers and galaxy–galaxy interactions \citep[e.g.][]{2008MNRAS.385.1903L,2011MNRAS.416.2182E}. Using hydrodynamical simulations, \citet{2015A&A...575A..16M} produced rejenuvated S0 galaxies and a central SF enhancement with minor mergers with gas-rich galaxies.
More generally, this work is compatible with the results of \citet{2008MNRAS.385.1903L}, who find that the majority of galaxies with high SSFR have a companion or exhibit tidal features.

\subsection{Support from other observations}\label{ssect:support}
Studies of S0s performed with planetary nebulae provide characteristics comparable with our findings on the TF relation. With a spectroscopic/kinematic analysis of planetary nebulae in six S0 galaxies, \citet{2013MNRAS.432.1010C} found that S0s are supported by random motions in addition to their rotating discs. S0 galaxies lie about 1 magnitude below the TFR for spiral galaxies while their spheroids lie 1 magnitude above the FJR for elliptical galaxies. This supports previous findings on these S0 characteristics \citep{2006MNRAS.373.1125B,2013MNRAS.433.2667R}, that we observe in the DPS for S0s but also for LTGs. \citet{2016MNRAS.455..214D} also observed a break for high velocities in the TF relation that they interpret as an additional baryonic mass present in the central part. 

Similarly to \citet{2009AJ....138..579K}, \cite{Bait+17} and \citet{2018MNRAS.481.5580F} identified two separate populations of S0 galaxies. Beside old, massive and metal rich galaxies with a bulge older than the disc (probably due to an inside-out quenching), they discuss less massive and more metal poor population having bulges with more recent star formation than their disc. They might have undergone a bulge rejuvenation (or disc fading), or compaction. They argued that environment is not playing a major role, and proposed a faded spiral scenario that forms low mass S0s while other processes such as mergers form the more massive S0s. As we only detect massive S0s, this is compatible with our findings. We have a population of massive star-forming galaxies, that are evolving through different stages of mergers. S0 galaxies represent 36$\%$ of the whole DP population, they are not faded. Beside their morphology, we cannot disentangle them from LTGs.

\subsection{Minor versus major mergers}
\label{ssect:mima}
As discussed in \citet{2005A&A...437...69B}, intermediate mass mergers with ratio 1:4 to 1:10 produce Sa to S0 galaxies, while equal mass mergers produce mainly ellipticals \citep[e.g.][]{1991ApJ...370L..65B}, and 1:3-1:4 can produce S0-like systems \citep[e.g.][]{1998ApJ...502L.133B,2018A&A...617A.113E}. Last, the scenario of single minor mergers, with mass ratio larger than 1:10, produces spiral galaxies, while sequential multiple minor mergers can also lead to Sa/S0 galaxies \citep{2007A&A...476.1179B}. 

On the one hand, the fact that the two peaks have typical intensity ratio in the range $1-3$ might favour the idea of two galaxy nuclei with progenitors with 1:1 to 1:4 mass ratios. However, the absence of differences in the rate of  morphological mergers between DPS and NBCS is not expected if the majority were such major mergers, as well as the absence of any differences in the distribution of shape-asymmetry and asymmetry. Major mergers can produce large starbursts \citep[e.g.][]{1991ApJ...370L..65B}, even though this is not always the case \citep[e.g.][]{2007A&A...468...61D}. However, \citet{2020MNRAS.493.3716H} found that strong enhancements of SFR are dominated by major mergers. They estimated that a starburst of 50\,$M_\odot\,$yr$
^{-1}$ has  four times higher chances to occur in a major merger than in a minor merger. We have here $1.3\,\%$ of the sample with a SFR larger than 50\,$M_\odot\,$yr$^{-1}$.

On the other hand, as discussed above multiple sequential gas-rich minor mergers and/or gas accretion might also produce the DPS galaxies \citep[e.g.][]{1996ApJ...460..121W}. They could also account for the excess of S0 galaxies and the prevalence of Sa galaxies in LTG, as well as relatively morphologically regular DPS galaxies (Mazzilli-Ciraulo et al., in prep.). 
The intensity ratio of the two peaks (between 1 and 3) can also be biased by the excitation of the AGN triggered by a merger (Maschmann et al., in prep.). 

As discussed in Appendix\,\ref{sect:param:dist}, the 0.2 dex higher typical stellar masses of DPS galaxies, their 0.25 mag typical excess extinction in $V$ and their central enhancement of star formation for Sb-d galaxies, all in comparison to the NBCS, suggest a past merger scenario for them. A visual inspection of the selected galaxies reveals that they are not dominated by edge-on or strongly inclined galaxies. However, Sb-d galaxies constitute $6\,\%$ of the DPS ($19\,\%$ of the NBCS).

As discussed by \citet{2020MNRAS.493.3716H} and references therein, minor mergers are expected much more numerous than major ones, e.g. \citet{2014MNRAS.440.2944K} estimated that 40$\%$ of SF observed in local spirals is directly triggered by minor mergers. In addition, major mergers are known to trigger stronger SF than their minor counterparts. The distributions of SFR$_{\rm fiber}$  estimated in the SF-DPS galaxies peak between 2 and 10\,$\rm M_\odot\,$yr$^{-1}$ and is about a factor of 2 larger than for the NBCS. In numerical simulations, \citet{2020MNRAS.493.3716H} found an SFR enhancement of a factor two in $\rm 10.0 \le log (M_* /M_\odot ) \le 11.4 $ SF post-merger galaxies.

Elliptical galaxies of the DPS are situated in the star forming main sequence (See Sect.\,\ref{ssect:star:formation:agn}), whereas their counterparts of the NBCS are quenched. We quantify this in appendix\,\ref{sect:param:dist}: elliptical DP galaxies show a 4 times higher SFR in comparison to single peaked elliptical galaxies. We furthermore find that elliptical galaxies with a DP have on average a 3 Gyr younger stellar population in comparison with those showing a single peak. This is consistent with the detection of molecular gas in star forming early type galaxies by \citet{2007MNRAS.377.1795C}, who conclude that the molecular gas might has been accreted from the environment and shows properties rather independent from the old, pre-existing stellar component. 
Such a recent gas accretion might explain the emission line shape and the ongoing star formation that we detect.

DP galaxies have numerous characteristics of post-mergers with enhanced central star formation and extinction. The nature of the DP feature is still elusive, and could correspond to gas clumps as well as to relics of two galaxy nuclei. We argue that DPS galaxies, most of which exhibit an ordinary morphology, are more likely to be linked to gas-rich minor mergers or gas accretion than to major mergers, which would impact more strongly their morphologies. Indeed, we do detect the same number of morphological mergers in the DPS and NBCS. The sequential multiple minor merger scenario accounts for the S0 morphological excess clearly detected in the DPS, while we cannot exclude that a few of these DP galaxies originate from major mergers.

\subsection{Alternatives}\label{ssect:alternatives}
Supermassive black holes hosted in galaxy centres are expected to have a significant feedback during the AGN phase \citep{2012ARA&A..50..455F}. This effect is measured to be stronger in massive galaxies  \citep{2015ARA&A..53...51S,2018ApJ...866L...4Z}, than in smaller galaxies dominated by star formation feedback. Galaxy collisions are usually thought to trigger AGN feedback \citep{2005Natur.433..604D}.
Gas outflow are found to be associated to emission-line asymmetries \citep{1981ApJ...247..403H, 1985MNRAS.213....1W}. This is studied on large data samples with fitting procedures using single- and double-Gaussian functions in e.g. \citet{2005ApJ...627..721G} and \citet{2016ApJ...817..108W}. Using observations based on spectroscopic integral field units, \citet{2016ApJ...819..148K} observed AGN driven outflow components with velocities between $~300$ and $600\, {\rm km\,s^{-1}}$ with velocity dispersion up to $~800\, {\rm km\,s^{-1}}$. In the study on major mergers, \citet{2013ApJ...768...75R} have shown that galaxies with a QSO have the highest projected outflow velocities of at least $~1450\, {\rm km\,s^{-1}}$. In the system F08572+3915:NW, they found velocities up to $~3350\,{\rm km\,s^{-1}}$. Galaxies without an AGN still reach projected velocities up to $~1000\,{\rm km\,s^{-1}}$. They concluded that QSOs play a key role in accelerating gas outflows. Beside these extreme velocities, an outflow with low projected velocities of several hundred ${\rm km\,s^{-1}}$ has been observed in NGC 5929 \citep{2014ApJ...780L..24R}.

In order to discuss an AGN outflow creating emission-lines with a DP structure, we select galaxies with a broader off-centred emission-line component. Therefore, these galaxies have one peak component associated with the stellar velocity, computed by \citet{2017ApJS..228...14C} and a second peak component more offset in units of $\sigma_{1,2}$ (as discussed in Sect.\,\ref{ssect:velocity:distribution}). Furthermore, we demand the second peak component to show a velocity dispersion larger than $200\, {\rm km\,s^{-1}}$ which is similar to lower velocity dispersion of the outflow component found by \citet{2016ApJ...819..148K}. We thus select 68 galaxies showing a broad off-centred component as described in the discussed literature above. We show an example in the upper panel of Fig.\,\ref{fig:Spec}. 

Using the non-parametric emission line for the BPT classification (see Sect.\,\ref{ssect:bpt}), we find only $1\%$ of these outflow candidates to be classified as SF galaxies. We find $50\%$ to be classified as AGN, $24\%$ as COMP and $13\%$ as LINER. Even though we find a connection to AGN activity for outflow candidates, these galaxies make only about $1\%$ of the DPS. The lack of sensitivity to AGN outflow candidates can be explained by the amplitude criteria $1/3 < {\rm A_1 / A_2} < 3$ applied in the selection procedure (see Sect.\,\ref{ssect:automated:selection:procedure}). Outflow components show mostly very broad component with a lower amplitude described as wings \citep[e.g][]{1981ApJ...247..403H, 1985MNRAS.213....1W, 2005ApJ...627..721G}. These types of asymmetries are systematically filtered out in this work. 

We also discussed an outflow scenario to explain off-centred weak [OIII]$\lambda$5008 in \citet{2019A&A...627L...3M}. In summary, we find evidence for AGN-driven gas outflow in some galaxies, but in comparison to the large sample of DP galaxies we conclude that this scenario provides well-suited arguments only for a small fraction of them.

\section{Conclusions}\label{sect:conclusion}
We identified double-peak emission-line galaxies with an automated procedure. They are quite rare as they constitute $0.8\%$ of the RCSED/SDSS catalogue. 
We compared the double-peak galaxies with their counter-parts from the no-bias control sample and found several significant differences. There is an excess of S0 galaxies, that cannot be accounted for by the environment.
The S0-DP galaxies correspond to the star-forming S0 galaxies identified in isolated environments as discussed in \citet{2020MNRAS.tmp.1540T}.
In parallel, the Gini-$M_{20}$ diagram reveals that LTG are mainly Sa galaxies in the DPS with respect to the LTG of the NBCS. Similarly, the DP LTG exhibit a larger S\'ersic index than their NBCS counterparts.

On one hand, we found that LTG and S0 galaxies behave similarly on the TF relation but are off-centred towards large velocity dispersions.  On the other hand, if two gas components are considered, we show that the close component behaves as expected while the far peak is offset and might correspond to a smaller component or a large gas clump. Other results further support this scenario: (1) the absence of any dependency on the galaxy inclination, (2) larger stellar velocity dispersions, (3) a systematic central enhancement of star formation and (4) a central enhancement of the extinction.

We argue that this double-peak sample constitutes a sequence of multiple sequential minor mergers, which could explain the similar behaviour observed for the different morphological types. It is a sequence in the sense that the impact of sequential minor mergers is to increase the size of the bulge, leading to  larger fractions of S0 galaxies, while the majority of disc galaxies are Sa. It is difficult to disentangle gas-rich minor mergers from gas accretion, but both will have similar effects. The absence of excess of proper morphological mergers supports the view that the impact is small, typical of minor mergers. The specificity of these DP galaxies is that the spectroscopic signature is inside the 3" SDSS-fiber, hence, it is somehow a post-coalescence stage not detected in morphological studies. Last, this sample also constitutes a time sequence of mergers as their spread in redshift gathers galaxies observed with resolution between 1 kpc and 12 kpc.

\begin{acknowledgements}
This work is based on the master thesis of DM.
We acknowledge Fran\c coise Combes, Ana\"elle Halle, Barbara Mazzilli-Ciraulo and Anton Afanasiev for fruitful discussions.
We thank the anonymous referee for suggestions on further investigating in non-parametric morphological diagnostics and enlarging the discussion on the merger rate, which supported our analysis and made it more solid.
We thank Ivan Zolotukhin for his support in the early stage of this work.
DM thanks Gabriele Gramelsberger, Dawid Kasprowicz, Lisa Sch\"uttler, Daniel Wenz, Lukas B\"ohres and Frederic Kerksieck for supporting this work and for constructive discussions on software structure.
We thank David Patton for providing the asymmetry parameter for SDSS galaxies and Helena Dom\'inguez-S\'anchez and Kanak Saha for helping with the morphological classification. We also thank Marina Trevisan for nice feedback on the discussion.

IC is supported by Telescope Data Center at Smithsonian Astrophysical Observatory. 
IC and IK acknowledge the Russian Science Foundation grant 19-12-00281 and the Program of development of M.V. Lomonosov Moscow State University for the Leading Scientific School ``Physics of stars, relativistic objects and galaxies''.

Rene Brun and Fons Rademakers, 
ROOT - An Object Oriented Data Analysis Framework, 
Proceedings AIHENP'96 Workshop, Lausanne, Sep. 1996, Nucl. Inst. \& Meth. in Phys. Res. A 389 (1997) 81-86. See also http://root.cern.ch/.

Funding for the Sloan Digital Sky Survey IV has been provided by the Alfred P. Sloan Foundation, the U.S. Department of Energy Office of Science, and the Participating Institutions. SDSS-IV acknowledges
support and resources from the Center for High-Performance Computing at
the University of Utah. The SDSS web site is www.sdss.org.

SDSS-IV is managed by the Astrophysical Research Consortium for the 
Participating Institutions of the SDSS Collaboration including the 
Brazilian Participation Group, the Carnegie Institution for Science, 
Carnegie Mellon University, the Chilean Participation Group, the French Participation Group, Harvard-Smithsonian Center for Astrophysics, 
Instituto de Astrof\'isica de Canarias, The Johns Hopkins University, Kavli Institute for the Physics and Mathematics of the Universe (IPMU) / 
University of Tokyo, the Korean Participation Group, Lawrence Berkeley National Laboratory, 
Leibniz Institut f\"ur Astrophysik Potsdam (AIP),  
Max-Planck-Institut f\"ur Astronomie (MPIA Heidelberg), 
Max-Planck-Institut f\"ur Astrophysik (MPA Garching), 
Max-Planck-Institut f\"ur Extraterrestrische Physik (MPE), 
National Astronomical Observatories of China, New Mexico State University, 
New York University, University of Notre Dame, 
Observat\'ario Nacional / MCTI, The Ohio State University, 
Pennsylvania State University, Shanghai Astronomical Observatory, 
United Kingdom Participation Group,
Universidad Nacional Aut\'onoma de M\'exico, University of Arizona, 
University of Colorado Boulder, University of Oxford, University of Portsmouth, 
University of Utah, University of Virginia, University of Washington, University of Wisconsin, 
Vanderbilt University, and Yale University.

The Legacy Surveys (http://legacysurvey.org/) consist of three individual and complementary projects: the Dark Energy Camera Legacy Survey (DECaLS; NOAO Proposal ID \# 2014B-0404; PIs: David Schlegel and Arjun Dey), the Beijing-Arizona Sky Survey (BASS; NOAO Proposal ID \# 2015A-0801; PIs: Zhou Xu and Xiaohui Fan), and the Mayall $z$-band Legacy Survey (MzLS; NOAO Proposal ID \# 2016A-0453; PI: Arjun Dey). DECaLS, BASS and MzLS together include data obtained, respectively, at the Blanco telescope, Cerro Tololo Inter-American Observatory, National Optical Astronomy Observatory (NOAO); the Bok telescope, Steward Observatory, University of Arizona; and the Mayall telescope, Kitt Peak National Observatory, NOAO. The Legacy Surveys project is honoured to be permitted to conduct astronomical research on Iolkam Du`ag (Kitt Peak), a mountain with particular significance to the Tohono O'odham Nation.
\end{acknowledgements}
\bibliographystyle{aa}
\bibliography{Mybiblio}

\begin{thebibliography}{186}
\expandafter\ifx\csname natexlab\endcsname\relax\def\natexlab#1{#1}\fi

\bibitem[{{Abazajian} {et~al.}(2009){Abazajian}, {Adelman-McCarthy},
  {Ag{\"u}eros}, {Allam}, {Allende Prieto}, {An}, {Anderson}, {Anderson},
  {Annis}, {Bahcall}, {Bailer-Jones}, {Barentine}, {Bassett}, {Becker},
  {Beers}, {Bell}, {Belokurov}, {Berlind}, {Berman}, {Bernardi}, {Bickerton},
  {Bizyaev}, {Blakeslee}, {Blanton}, {Bochanski}, {Boroski}, {Brewington},
  {Brinchmann}, {Brinkmann}, {Brunner}, {Budav{\'a}ri}, {Carey}, {Carliles},
  {Carr}, {Castander}, {Cinabro}, {Connolly}, {Csabai}, {Cunha}, {Czarapata},
  {Davenport}, {de Haas}, {Dilday}, {Doi}, {Eisenstein}, {Evans}, {Evans},
  {Fan}, {Friedman}, {Frieman}, {Fukugita}, {G{\"a}nsicke}, {Gates},
  {Gillespie}, {Gilmore}, {Gonzalez}, {Gonzalez}, {Grebel}, {Gunn},
  {Gy{\"o}ry}, {Hall}, {Harding}, {Harris}, {Harvanek}, {Hawley}, {Hayes},
  {Heckman}, {Hendry}, {Hennessy}, {Hindsley}, {Hoblitt}, {Hogan}, {Hogg},
  {Holtzman}, {Hyde}, {Ichikawa}, {Ichikawa}, {Im}, {Ivezi{\'c}}, {Jester},
  {Jiang}, {Johnson}, {Jorgensen}, {Juri{\'c}}, {Kent}, {Kessler}, {Kleinman},
  {Knapp}, {Konishi}, {Kron}, {Krzesinski}, {Kuropatkin}, {Lampeitl},
  {Lebedeva}, {Lee}, {Lee}, {French Leger}, {L{\'e}pine}, {Li}, {Lima}, {Lin},
  {Long}, {Loomis}, {Loveday}, {Lupton}, {Magnier}, {Malanushenko},
  {Malanushenko}, {Mand elbaum}, {Margon}, {Marriner}, {Mart{\'\i}nez-Delgado},
  {Matsubara}, {McGehee}, {McKay}, {Meiksin}, {Morrison}, {Mullally}, {Munn},
  {Murphy}, {Nash}, {Nebot}, {Neilsen}, {Newberg}, {Newman}, {Nichol},
  {Nicinski}, {Nieto-Santisteban}, {Nitta}, {Okamura}, {Oravetz}, {Ostriker},
  {Owen}, {Padmanabhan}, {Pan}, {Park}, {Pauls}, {Peoples}, {Percival}, {Pier},
  {Pope}, {Pourbaix}, {Price}, {Purger}, {Quinn}, {Raddick}, {Re Fiorentin},
  {Richards}, {Richmond}, {Riess}, {Rix}, {Rockosi}, {Sako}, {Schlegel},
  {Schneider}, {Scholz}, {Schreiber}, {Schwope}, {Seljak}, {Sesar}, {Sheldon},
  {Shimasaku}, {Sibley}, {Simmons}, {Sivarani}, {Allyn Smith}, {Smith},
  {Smol{\v{c}}i{\'c}}, {Snedden}, {Stebbins}, {Steinmetz}, {Stoughton},
  {Strauss}, {SubbaRao}, {Suto}, {Szalay}, {Szapudi}, {Szkody}, {Tanaka},
  {Tegmark}, {Teodoro}, {Thakar}, {Tremonti}, {Tucker}, {Uomoto}, {Vanden
  Berk}, {Vandenberg}, {Vidrih}, {Vogeley}, {Voges}, {Vogt}, {Wadadekar},
  {Watters}, {Weinberg}, {West}, {White}, {Wilhite}, {Wonders}, {Yanny},
  {Yocum}, {York}, {Zehavi}, {Zibetti}, \& {Zucker}}]{2009ApJS..182..543A}
{Abazajian}, K.~N., {Adelman-McCarthy}, J.~K., {Ag{\"u}eros}, M.~A., {et~al.}
  2009, \apjs, 182, 543

\bibitem[{{Abraham} {et~al.}(2003){Abraham}, {van den Bergh}, \&
  {Nair}}]{2003ApJ...588..218A}
{Abraham}, R.~G., {van den Bergh}, S., \& {Nair}, P. 2003, \apj, 588, 218

\bibitem[{{Acker} {et~al.}(1989){Acker}, {K{\"o}ppen}, {Samland}, \&
  {Stenholm}}]{1989Msngr..58...44A}
{Acker}, A., {K{\"o}ppen}, J., {Samland}, M., \& {Stenholm}, B. 1989, The
  Messenger, 58, 44

\bibitem[{{Aquino-Ort{\'\i}z} {et~al.}(2018){Aquino-Ort{\'\i}z}, {Valenzuela},
  {S{\'a}nchez}, {Hern{\'a}ndez-Toledo}, {{\'A}vila-Reese}, {van de Ven},
  {Rodr{\'\i}guez-Puebla}, {Zhu}, {Mancillas}, {Cano-D{\'\i}az}, \&
  {Garc{\'\i}a-Benito}}]{2018MNRAS.479.2133A}
{Aquino-Ort{\'\i}z}, E., {Valenzuela}, O., {S{\'a}nchez}, S.~F., {et~al.} 2018,
  \mnras, 479, 2133

\bibitem[{{Athanassoula} \& {Bosma}(1985)}]{1985ARA&A..23..147A}
{Athanassoula}, E. \& {Bosma}, A. 1985, \araa, 23, 147

\bibitem[{{Avila-Reese} {et~al.}(2008){Avila-Reese}, {Zavala}, {Firmani}, \&
  {Hern{\'a}ndez-Toledo}}]{2008AJ....136.1340A}
{Avila-Reese}, V., {Zavala}, J., {Firmani}, C., \& {Hern{\'a}ndez-Toledo},
  H.~M. 2008, \aj, 136, 1340

\bibitem[{{Bait} {et~al.}(2017){Bait}, {Barway}, \& {Wadadekar}}]{Bait+17}
{Bait}, O., {Barway}, S., \& {Wadadekar}, Y. 2017, \mnras, 471, 2687

\bibitem[{{Baldwin} {et~al.}(1981){Baldwin}, {Phillips}, \&
  {Terlevich}}]{1981PASP...93....5B}
{Baldwin}, J.~A., {Phillips}, M.~M., \& {Terlevich}, R. 1981, \pasp, 93, 5

\bibitem[{{Balogh} {et~al.}(1998){Balogh}, {Schade}, {Morris}, {Yee},
  {Carlberg}, \& {Ellingson}}]{1998ApJ...504L..75B}
{Balogh}, M.~L., {Schade}, D., {Morris}, S.~L., {et~al.} 1998, \apjl, 504, L75

\bibitem[{{Barnes} \& {Hernquist}(1991)}]{1991ApJ...370L..65B}
{Barnes}, J.~E. \& {Hernquist}, L.~E. 1991, \apjl, 370, L65

\bibitem[{{Bedregal} {et~al.}(2006){Bedregal}, {Arag{\'o}n-Salamanca}, \&
  {Merrifield}}]{2006MNRAS.373.1125B}
{Bedregal}, A.~G., {Arag{\'o}n-Salamanca}, A., \& {Merrifield}, M.~R. 2006,
  \mnras, 373, 1125

\bibitem[{{Begelman} {et~al.}(1980){Begelman}, {Blandford}, \&
  {Rees}}]{1980Natur.287..307B}
{Begelman}, M.~C., {Blandford}, R.~D., \& {Rees}, M.~J. 1980, \nat, 287, 307

\bibitem[{{Bekki}(1998)}]{1998ApJ...502L.133B}
{Bekki}, K. 1998, \apjl, 502, L133

\bibitem[{{Bekki} \& {Couch}(2011)}]{2011MNRAS.415.1783B}
{Bekki}, K. \& {Couch}, W.~J. 2011, \mnras, 415, 1783

\bibitem[{{Blanton} \& {Moustakas}(2009)}]{2009ARA&A..47..159B}
{Blanton}, M.~R. \& {Moustakas}, J. 2009, \araa, 47, 159

\bibitem[{{Bond} {et~al.}(1996){Bond}, {Kofman}, \&
  {Pogosyan}}]{1996Natur.380..603B}
{Bond}, J.~R., {Kofman}, L., \& {Pogosyan}, D. 1996, \nat, 380, 603

\bibitem[{{Bothun} \& {Dressler}(1986)}]{1986ApJ...301...57B}
{Bothun}, G.~D. \& {Dressler}, A. 1986, \apj, 301, 57

\bibitem[{{Bournaud} {et~al.}(2005){Bournaud}, {Jog}, \&
  {Combes}}]{2005A&A...437...69B}
{Bournaud}, F., {Jog}, C.~J., \& {Combes}, F. 2005, \aap, 437, 69

\bibitem[{{Bournaud} {et~al.}(2007){Bournaud}, {Jog}, \&
  {Combes}}]{2007A&A...476.1179B}
{Bournaud}, F., {Jog}, C.~J., \& {Combes}, F. 2007, \aap, 476, 1179

\bibitem[{{Brinchmann} {et~al.}(2004){Brinchmann}, {Charlot}, {White},
  {Tremonti}, {Kauffmann}, {Heckman}, \& {Brinkmann}}]{2004MNRAS.351.1151B}
{Brinchmann}, J., {Charlot}, S., {White}, S.~D.~M., {et~al.} 2004, \mnras, 351,
  1151

\bibitem[{{Buta} \& {Combes}(1996)}]{1996FCPh...17...95B}
{Buta}, R. \& {Combes}, F. 1996, \fcp, 17, 95

\bibitem[{{Calistro Rivera} {et~al.}(2017){Calistro Rivera}, {Williams},
  {Hardcastle}, {Duncan}, {R{\"o}ttgering}, {Best}, {Br{\"u}ggen}, {Chy{\.z}y},
  {Conselice}, {de Gasperin}, {Engels}, {G{\"u}rkan}, {Intema}, {Jarvis},
  {Mahony}, {Miley}, {Morabito}, {Prandoni}, {Sabater}, {Smith}, {Tasse}, {van
  der Werf}, \& {White}}]{2017MNRAS.469.3468C}
{Calistro Rivera}, G., {Williams}, W.~L., {Hardcastle}, M.~J., {et~al.} 2017,
  \mnras, 469, 3468

\bibitem[{{Cappellari}(2013)}]{2013ApJ...778L...2C}
{Cappellari}, M. 2013, \apjl, 778, L2

\bibitem[{{Cappellari}(2016)}]{2016ARA&A..54..597C}
{Cappellari}, M. 2016, \araa, 54, 597

\bibitem[{{Catinella} {et~al.}(2005){Catinella}, {Haynes}, \&
  {Giovanelli}}]{2005AJ....130.1037C}
{Catinella}, B., {Haynes}, M.~P., \& {Giovanelli}, R. 2005, \aj, 130, 1037

\bibitem[{{Catinella} {et~al.}(2012){Catinella}, {Kauffmann}, {Schiminovich},
  {Lemonias}, {Scannapieco}, {Wang}, {Fabello}, {Hummels}, {Moran}, {Wu},
  {Cooper}, {Giovanelli}, {Haynes}, {Heckman}, \&
  {Saintonge}}]{2012MNRAS.420.1959C}
{Catinella}, B., {Kauffmann}, G., {Schiminovich}, D., {et~al.} 2012, \mnras,
  420, 1959

\bibitem[{{Chilingarian} {et~al.}(2010{\natexlab{a}}){Chilingarian}, {Di
  Matteo}, {Combes}, {Melchior}, \& {Semelin}}]{2010A&A...518A..61C}
{Chilingarian}, I.~V., {Di Matteo}, P., {Combes}, F., {Melchior}, A.~L., \&
  {Semelin}, B. 2010{\natexlab{a}}, \aap, 518, A61

\bibitem[{{Chilingarian} {et~al.}(2018){Chilingarian}, {Katkov}, {Zolotukhin},
  {Grishin}, {Beletsky}, {Boutsia}, \& {Osip}}]{2018ApJ...863....1C}
{Chilingarian}, I.~V., {Katkov}, I.~Y., {Zolotukhin}, I.~Y., {et~al.} 2018,
  \apj, 863, 1

\bibitem[{{Chilingarian} {et~al.}(2010{\natexlab{b}}){Chilingarian},
  {Melchior}, \& {Zolotukhin}}]{2010MNRAS.405.1409C}
{Chilingarian}, I.~V., {Melchior}, A.-L., \& {Zolotukhin}, I.~Y.
  2010{\natexlab{b}}, \mnras, 405, 1409

\bibitem[{{Chilingarian} {et~al.}(2009){Chilingarian}, {Novikova}, {Cayatte},
  {Combes}, {Di Matteo}, \& {Zasov}}]{2009A&A...504..389C}
{Chilingarian}, I.~V., {Novikova}, A.~P., {Cayatte}, V., {et~al.} 2009, \aap,
  504, 389

\bibitem[{{Chilingarian} \& {Zolotukhin}(2012)}]{2012MNRAS.419.1727C}
{Chilingarian}, I.~V. \& {Zolotukhin}, I.~Y. 2012, \mnras, 419, 1727

\bibitem[{{Chilingarian} {et~al.}(2017){Chilingarian}, {Zolotukhin}, {Katkov},
  {Melchior}, {Rubtsov}, \& {Grishin}}]{2017ApJS..228...14C}
{Chilingarian}, I.~V., {Zolotukhin}, I.~Y., {Katkov}, I.~Y., {et~al.} 2017,
  \apjs, 228, 14

\bibitem[{{Combes} {et~al.}(2007){Combes}, {Young}, \&
  {Bureau}}]{2007MNRAS.377.1795C}
{Combes}, F., {Young}, L.~M., \& {Bureau}, M. 2007, \mnras, 377, 1795

\bibitem[{{Comerford} {et~al.}(2009){Comerford}, {Gerke}, {Newman}, {Davis},
  {Yan}, {Cooper}, {Faber}, {Koo}, {Coil}, {Rosario}, \&
  {Dutton}}]{2009ApJ...698..956C}
{Comerford}, J.~M., {Gerke}, B.~F., {Newman}, J.~A., {et~al.} 2009, \apj, 698,
  956

\bibitem[{{Comerford} {et~al.}(2012){Comerford}, {Gerke}, {Stern}, {Cooper},
  {Weiner}, {Newman}, {Madsen}, \& {Barrows}}]{2012ApJ...753...42C}
{Comerford}, J.~M., {Gerke}, B.~F., {Stern}, D., {et~al.} 2012, \apj, 753, 42

\bibitem[{{Comerford} {et~al.}(2018){Comerford}, {Nevin}, {Stemo},
  {M{\"u}ller-S{\'a}nchez}, {Barrows}, {Cooper}, \&
  {Newman}}]{2018ApJ...867...66C}
{Comerford}, J.~M., {Nevin}, R., {Stemo}, A., {et~al.} 2018, \apj, 867, 66

\bibitem[{{Comerford} {et~al.}(2015){Comerford}, {Pooley}, {Barrows}, {Greene},
  {Zakamska}, {Madejski}, \& {Cooper}}]{2015ApJ...806..219C}
{Comerford}, J.~M., {Pooley}, D., {Barrows}, R.~S., {et~al.} 2015, \apj, 806,
  219

\bibitem[{{Comerford} {et~al.}(2013){Comerford}, {Schluns}, {Greene}, \&
  {Cool}}]{2013ApJ...777...64C}
{Comerford}, J.~M., {Schluns}, K., {Greene}, J.~E., \& {Cool}, R.~J. 2013,
  \apj, 777, 64

\bibitem[{{Comer{\'o}n} {et~al.}(2014){Comer{\'o}n}, {Salo}, {Laurikainen},
  {Knapen}, {Buta}, {Herrera-Endoqui}, {Laine}, {Holwerda}, {Sheth}, {Regan},
  {Hinz}, {Mu{\~n}oz-Mateos}, {Gil de Paz}, {Men{\'e}ndez-Delmestre},
  {Seibert}, {Mizusawa}, {Kim}, {Erroz-Ferrer}, {Gadotti}, {Athanassoula},
  {Bosma}, \& {Ho}}]{2014A&A...562A.121C}
{Comer{\'o}n}, S., {Salo}, H., {Laurikainen}, E., {et~al.} 2014, \aap, 562,
  A121

\bibitem[{{Condon}(1992)}]{1992ARA&A..30..575C}
{Condon}, J.~J. 1992, \araa, 30, 575

\bibitem[{{Conselice}(2003)}]{2003ApJS..147....1C}
{Conselice}, C.~J. 2003, \apjs, 147, 1

\bibitem[{{Cortesi} {et~al.}(2013){Cortesi}, {Merrifield}, {Coccato},
  {Arnaboldi}, {Gerhard}, {Bamford}, {Napolitano}, {Romanowsky}, {Douglas},
  {Kuijken}, {Capaccioli}, {Freeman}, {Saha}, \&
  {Chies-Santos}}]{2013MNRAS.432.1010C}
{Cortesi}, A., {Merrifield}, M.~R., {Coccato}, L., {et~al.} 2013, \mnras, 432,
  1010

\bibitem[{{Cui} {et~al.}(2001){Cui}, {Xia}, {Deng}, {Mao}, \&
  {Zou}}]{2001AJ....122...63C}
{Cui}, J., {Xia}, X.~Y., {Deng}, Z.~G., {Mao}, S., \& {Zou}, Z.~L. 2001, \aj,
  122, 63

\bibitem[{{Davis} {et~al.}(2016){Davis}, {Greene}, {Ma}, {Pand ya},
  {Blakeslee}, {McConnell}, \& {Thomas}}]{2016MNRAS.455..214D}
{Davis}, T.~A., {Greene}, J., {Ma}, C.-P., {et~al.} 2016, \mnras, 455, 214

\bibitem[{{De Propris} {et~al.}(2005){De Propris}, {Liske}, {Driver}, {Allen},
  \& {Cross}}]{2005AJ....130.1516D}
{De Propris}, R., {Liske}, J., {Driver}, S.~P., {Allen}, P.~D., \& {Cross}, N.
  J.~G. 2005, \aj, 130, 1516

\bibitem[{{Deane} {et~al.}(2014){Deane}, {Paragi}, {Jarvis}, {Coriat},
  {Bernardi}, {Fender}, {Frey}, {Heywood}, {Kl{\"o}ckner}, {Grainge}, \&
  {Rumsey}}]{2014Natur.511...57D}
{Deane}, R.~P., {Paragi}, Z., {Jarvis}, M.~J., {et~al.} 2014, \nat, 511, 57

\bibitem[{{Dey} {et~al.}(2019){Dey}, {Schlegel}, {Lang}, {Blum}, {Burleigh},
  {Fan}, {Findlay}, {Finkbeiner}, {Herrera}, {Juneau}, {Landriau}, {Levi},
  {McGreer}, {Meisner}, {Myers}, {Moustakas}, {Nugent}, {Patej}, {Schlafly},
  {Walker}, {Valdes}, {Weaver}, {Y{\`e}che}, {Zou}, {Zhou}, {Abareshi},
  {Abbott}, {Abolfathi}, {Aguilera}, {Alam}, {Allen}, {Alvarez}, {Annis},
  {Ansarinejad}, {Aubert}, {Beechert}, {Bell}, {BenZvi}, {Beutler}, {Bielby},
  {Bolton}, {Brice{\~n}o}, {Buckley-Geer}, {Butler}, {Calamida}, {Carlberg},
  {Carter}, {Casas}, {Castander}, {Choi}, {Comparat}, {Cukanovaite}, {Delubac},
  {DeVries}, {Dey}, {Dhungana}, {Dickinson}, {Ding}, {Donaldson}, {Duan},
  {Duckworth}, {Eftekharzadeh}, {Eisenstein}, {Etourneau}, {Fagrelius},
  {Farihi}, {Fitzpatrick}, {Font-Ribera}, {Fulmer}, {G{\"a}nsicke},
  {Gaztanaga}, {George}, {Gerdes}, {Gontcho}, {Gorgoni}, {Green}, {Guy},
  {Harmer}, {Hernandez}, {Honscheid}, {Huang}, {James}, {Jannuzi}, {Jiang},
  {Joyce}, {Karcher}, {Karkar}, {Kehoe}, {Kneib}, {Kueter-Young}, {Lan},
  {Lauer}, {Le Guillou}, {Le Van Suu}, {Lee}, {Lesser}, {Perreault Levasseur},
  {Li}, {Mann}, {Marshall}, {Mart{\'{\i}}nez-V{\'a}zquez}, {Martini}, {du Mas
  des Bourboux}, {McManus}, {Meier}, {M{\'e}nard}, {Metcalfe},
  {Mu{\~n}oz-Guti{\'e}rrez}, {Najita}, {Napier}, {Narayan}, {Newman}, {Nie},
  {Nord}, {Norman}, {Olsen}, {Paat}, {Palanque-Delabrouille}, {Peng},
  {Poppett}, {Poremba}, {Prakash}, {Rabinowitz}, {Raichoor}, {Rezaie},
  {Robertson}, {Roe}, {Ross}, {Ross}, {Rudnick}, {Safonova}, {Saha},
  {S{\'a}nchez}, {Savary}, {Schweiker}, {Scott}, {Seo}, {Shan}, {Silva},
  {Slepian}, {Soto}, {Sprayberry}, {Staten}, {Stillman}, {Stupak}, {Summers},
  {Sien Tie}, {Tirado}, {Vargas-Maga{\~n}a}, {Vivas}, {Wechsler}, {Williams},
  {Yang}, {Yang}, {Yapici}, {Zaritsky}, {Zenteno}, {Zhang}, {Zhang}, {Zhou}, \&
  {Zhou}}]{2019AJ....157..168D}
{Dey}, A., {Schlegel}, D.~J., {Lang}, D., {et~al.} 2019, \aj, 157, 168

\bibitem[{{Di Matteo} {et~al.}(2007){Di Matteo}, {Combes}, {Melchior}, \&
  {Semelin}}]{2007A&A...468...61D}
{Di Matteo}, P., {Combes}, F., {Melchior}, A.~L., \& {Semelin}, B. 2007, \aap,
  468, 61

\bibitem[{{Di Matteo} {et~al.}(2005){Di Matteo}, {Springel}, \&
  {Hernquist}}]{2005Natur.433..604D}
{Di Matteo}, T., {Springel}, V., \& {Hernquist}, L. 2005, \nat, 433, 604

\bibitem[{{Dom{\'\i}nguez} {et~al.}(2013){Dom{\'\i}nguez}, {Siana}, {Henry},
  {Scarlata}, {Bedregal}, {Malkan}, {Atek}, {Ross}, {Colbert}, {Teplitz},
  {Rafelski}, {McCarthy}, {Bunker}, {Hathi}, {Dressler}, {Martin}, \&
  {Masters}}]{2013ApJ...763..145D}
{Dom{\'\i}nguez}, A., {Siana}, B., {Henry}, A.~L., {et~al.} 2013, \apj, 763,
  145

\bibitem[{{Dom{\'{\i}}nguez S{\'a}nchez} {et~al.}(2018){Dom{\'{\i}}nguez
  S{\'a}nchez}, {Huertas-Company}, {Bernardi}, {Tuccillo}, \&
  {Fischer}}]{2018MNRAS.476.3661D}
{Dom{\'{\i}}nguez S{\'a}nchez}, H., {Huertas-Company}, M., {Bernardi}, M.,
  {Tuccillo}, D., \& {Fischer}, J.~L. 2018, \mnras, 476, 3661

\bibitem[{{Dressler}(1980)}]{1980ApJ...236..351D}
{Dressler}, A. 1980, \apj, 236, 351

\bibitem[{{Dressler} {et~al.}(1997){Dressler}, {Oemler}, {Couch}, {Smail},
  {Ellis}, {Barger}, {Butcher}, {Poggianti}, \&
  {Sharples}}]{1997ApJ...490..577D}
{Dressler}, A., {Oemler}, Augustus, J., {Couch}, W.~J., {et~al.} 1997, \apj,
  490, 577

\bibitem[{{Drory} {et~al.}(2015){Drory}, {MacDonald}, {Bershady}, {Bundy},
  {Gunn}, {Law}, {Smith}, {Stoll}, {Tremonti}, {Wake}, {Yan}, {Weijmans},
  {Byler}, {Cherinka}, {Cope}, {Eigenbrot}, {Harding}, {Holder}, {Huehnerhoff},
  {Jaehnig}, {Jansen}, {Klaene}, {Paat}, {Percival}, \&
  {Sayres}}]{2015AJ....149...77D}
{Drory}, N., {MacDonald}, N., {Bershady}, M.~A., {et~al.} 2015, \aj, 149, 77

\bibitem[{{Eliche-Moral} {et~al.}(2018){Eliche-Moral},
  {Rodr{\'\i}guez-P{\'e}rez}, {Borlaff}, {Querejeta}, \&
  {Tapia}}]{2018A&A...617A.113E}
{Eliche-Moral}, M.~C., {Rodr{\'\i}guez-P{\'e}rez}, C., {Borlaff}, A.,
  {Querejeta}, M., \& {Tapia}, T. 2018, \aap, 617, A113

\bibitem[{{Elitzur} {et~al.}(2012){Elitzur}, {Asensio Ramos}, \&
  {Ceccarelli}}]{2012MNRAS.422.1394E}
{Elitzur}, M., {Asensio Ramos}, A., \& {Ceccarelli}, C. 2012, \mnras, 422, 1394

\bibitem[{{Ellis} {et~al.}(2005){Ellis}, {Driver}, {Allen}, {Liske},
  {Bland-Hawthorn}, \& {De Propris}}]{2005MNRAS.363.1257E}
{Ellis}, S.~C., {Driver}, S.~P., {Allen}, P.~D., {et~al.} 2005, \mnras, 363,
  1257

\bibitem[{{Ellison} {et~al.}(2011{\natexlab{a}}){Ellison}, {Nair}, {Patton},
  {Scudder}, {Mendel}, \& {Simard}}]{2011MNRAS.416.2182E}
{Ellison}, S.~L., {Nair}, P., {Patton}, D.~R., {et~al.} 2011{\natexlab{a}},
  \mnras, 416, 2182

\bibitem[{{Ellison} {et~al.}(2011{\natexlab{b}}){Ellison}, {Patton}, {Mendel},
  \& {Scudder}}]{2011MNRAS.418.2043E}
{Ellison}, S.~L., {Patton}, D.~R., {Mendel}, J.~T., \& {Scudder}, J.~M.
  2011{\natexlab{b}}, \mnras, 418, 2043

\bibitem[{{Ellison} {et~al.}(2008){Ellison}, {Patton}, {Simard}, \&
  {McConnachie}}]{2008AJ....135.1877E}
{Ellison}, S.~L., {Patton}, D.~R., {Simard}, L., \& {McConnachie}, A.~W. 2008,
  \aj, 135, 1877

\bibitem[{{Ellison} {et~al.}(2010){Ellison}, {Patton}, {Simard}, {McConnachie},
  {Baldry}, \& {Mendel}}]{2010MNRAS.407.1514E}
{Ellison}, S.~L., {Patton}, D.~R., {Simard}, L., {et~al.} 2010, \mnras, 407,
  1514

\bibitem[{{Faber} \& {Jackson}(1976)}]{1976ApJ...204..668F}
{Faber}, S.~M. \& {Jackson}, R.~E. 1976, \apj, 204, 668

\bibitem[{{Fabian}(2012)}]{2012ARA&A..50..455F}
{Fabian}, A.~C. 2012, \araa, 50, 455

\bibitem[{{Fasano} {et~al.}(2000){Fasano}, {Poggianti}, {Couch}, {Bettoni},
  {Kj{\ae}rgaard}, \& {Moles}}]{2000ApJ...542..673F}
{Fasano}, G., {Poggianti}, B.~M., {Couch}, W.~J., {et~al.} 2000, \apj, 542, 673

\bibitem[{Fisher(1954)}]{Fisher:724001}
Fisher, R.~A. 1954, {Statistical methods for research workers; 20th ed.}
  (Edinburgh: Oliver and Boyd)

\bibitem[{{Fraser-McKelvie} {et~al.}(2018){Fraser-McKelvie},
  {Arag{\'o}n-Salamanca}, {Merrifield}, {Tabor}, {Bernardi}, {Drory}, {Parikh},
  \& {Argudo-Fern{\'a}ndez}}]{2018MNRAS.481.5580F}
{Fraser-McKelvie}, A., {Arag{\'o}n-Salamanca}, A., {Merrifield}, M., {et~al.}
  2018, \mnras, 481, 5580

\bibitem[{{Fu} {et~al.}(2015){Fu}, {Myers}, {Djorgovski}, {Yan}, {Wrobel}, \&
  {Stockton}}]{2015ApJ...799...72F}
{Fu}, H., {Myers}, A.~D., {Djorgovski}, S.~G., {et~al.} 2015, \apj, 799, 72

\bibitem[{{Gallazzi} {et~al.}(2006){Gallazzi}, {Charlot}, {Brinchmann}, \&
  {White}}]{2006MNRAS.370.1106G}
{Gallazzi}, A., {Charlot}, S., {Brinchmann}, J., \& {White}, S. D.~M. 2006,
  \mnras, 370, 1106

\bibitem[{{Gao} {et~al.}(2018){Gao}, {Ho}, {Barth}, \&
  {Li}}]{2018ApJ...862..100G}
{Gao}, H., {Ho}, L.~C., {Barth}, A.~J., \& {Li}, Z.-Y. 2018, \apj, 862, 100

\bibitem[{{Ge} {et~al.}(2012){Ge}, {Hu}, {Wang}, {Bai}, \&
  {Zhang}}]{2012ApJS..201...31G}
{Ge}, J.-Q., {Hu}, C., {Wang}, J.-M., {Bai}, J.-M., \& {Zhang}, S. 2012, \apjs,
  201, 31

\bibitem[{{Genzel} {et~al.}(2001){Genzel}, {Tacconi}, {Rigopoulou}, {Lutz}, \&
  {Tecza}}]{2001ApJ...563..527G}
{Genzel}, R., {Tacconi}, L.~J., {Rigopoulou}, D., {Lutz}, D., \& {Tecza}, M.
  2001, \apj, 563, 527

\bibitem[{{Gilli} {et~al.}(2010){Gilli}, {Vignali}, {Mignoli}, {Iwasawa},
  {Comastri}, \& {Zamorani}}]{2010A&A...519A..92G}
{Gilli}, R., {Vignali}, C., {Mignoli}, M., {et~al.} 2010, \aap, 519, A92

\bibitem[{{Goulding} {et~al.}(2019){Goulding}, {Pardo}, {Greene}, {Mingarelli},
  {Nyland}, \& {Strauss}}]{2019ApJ...879L..21G}
{Goulding}, A.~D., {Pardo}, K., {Greene}, J.~E., {et~al.} 2019, \apjl, 879, L21

\bibitem[{{Green} {et~al.}(2010){Green}, {Myers}, {Barkhouse}, {Mulchaey},
  {Bennert}, {Cox}, \& {Aldcroft}}]{2010ApJ...710.1578G}
{Green}, P.~J., {Myers}, A.~D., {Barkhouse}, W.~A., {et~al.} 2010, \apj, 710,
  1578

\bibitem[{{Greene} \& {Ho}(2005)}]{2005ApJ...627..721G}
{Greene}, J.~E. \& {Ho}, L.~C. 2005, \apj, 627, 721

\bibitem[{{Hani} {et~al.}(2020){Hani}, {Gosain}, {Ellison}, {Patton}, \&
  {Torrey}}]{2020MNRAS.493.3716H}
{Hani}, M.~H., {Gosain}, H., {Ellison}, S.~L., {Patton}, D.~R., \& {Torrey}, P.
  2020, \mnras, 493, 3716

\bibitem[{{Heckman} {et~al.}(1981){Heckman}, {Miley}, {van Breugel}, \&
  {Butcher}}]{1981ApJ...247..403H}
{Heckman}, T.~M., {Miley}, G.~K., {van Breugel}, W.~J.~M., \& {Butcher}, H.~R.
  1981, \apj, 247, 403

\bibitem[{{Helou} {et~al.}(1985){Helou}, {Soifer}, \&
  {Rowan-Robinson}}]{1985ApJ...298L...7H}
{Helou}, G., {Soifer}, B.~T., \& {Rowan-Robinson}, M. 1985, \apjl, 298, L7

\bibitem[{{Hogg} {et~al.}(2002){Hogg}, {Blanton}, {Strateva}, {Bahcall},
  {Brinkmann}, {Csabai}, {Doi}, {Fukugita}, {Hennessy}, {Ivezi{\'c}}, {Knapp},
  {Lamb}, {Lupton}, {Munn}, {Nichol}, {Schlegel}, {Schneider}, \&
  {York}}]{2002AJ....124..646H}
{Hogg}, D.~W., {Blanton}, M., {Strateva}, I., {et~al.} 2002, \aj, 124, 646

\bibitem[{{Ilyina} {et~al.}(2014){Ilyina}, {Sil'chenko}, \&
  {Afanasiev}}]{2014MNRAS.439..334I}
{Ilyina}, M.~A., {Sil'chenko}, O.~K., \& {Afanasiev}, V.~L. 2014, \mnras, 439,
  334

\bibitem[{{Kannappan}(2004)}]{2004ApJ...611L..89K}
{Kannappan}, S.~J. 2004, \apjl, 611, L89

\bibitem[{{Kannappan} {et~al.}(2009){Kannappan}, {Guie}, \&
  {Baker}}]{2009AJ....138..579K}
{Kannappan}, S.~J., {Guie}, J.~M., \& {Baker}, A.~J. 2009, \aj, 138, 579

\bibitem[{{Karouzos} {et~al.}(2016){Karouzos}, {Woo}, \&
  {Bae}}]{2016ApJ...819..148K}
{Karouzos}, M., {Woo}, J.-H., \& {Bae}, H.-J. 2016, \apj, 819, 148

\bibitem[{{Katkov} {et~al.}(2015){Katkov}, {Kniazev}, \&
  {Sil'chenko}}]{2015AJ....150...24K}
{Katkov}, I.~Y., {Kniazev}, A.~Y., \& {Sil'chenko}, O.~K. 2015, \aj, 150, 24

\bibitem[{{Katkov} {et~al.}(2013){Katkov}, {Sil'chenko}, \&
  {Afanasiev}}]{2013ApJ...769..105K}
{Katkov}, I.~Y., {Sil'chenko}, O.~K., \& {Afanasiev}, V.~L. 2013, \apj, 769,
  105

\bibitem[{{Katkov} {et~al.}(2014){Katkov}, {Sil'chenko}, \&
  {Afanasiev}}]{2014MNRAS.438.2798K}
{Katkov}, I.~Y., {Sil'chenko}, O.~K., \& {Afanasiev}, V.~L. 2014, \mnras, 438,
  2798

\bibitem[{{Katkov} {et~al.}(2016){Katkov}, {Sil'chenko}, {Chilingarian},
  {Uklein}, \& {Egorov}}]{2016MNRAS.461.2068K}
{Katkov}, I.~Y., {Sil'chenko}, O.~K., {Chilingarian}, I.~V., {Uklein}, R.~I.,
  \& {Egorov}, O.~V. 2016, \mnras, 461, 2068

\bibitem[{{Kauffmann} {et~al.}(2003){Kauffmann}, {Heckman}, {White}, {Charlot},
  {Tremonti}, {Brinchmann}, {Bruzual}, {Peng}, {Seibert}, {Bernardi},
  {Blanton}, {Brinkmann}, {Castander}, {Cs{\'a}bai}, {Fukugita}, {Ivezic},
  {Munn}, {Nichol}, {Padmanabhan}, {Thakar}, {Weinberg}, \&
  {York}}]{2003MNRAS.341...33K}
{Kauffmann}, G., {Heckman}, T.~M., {White}, S. D.~M., {et~al.} 2003, \mnras,
  341, 33

\bibitem[{{Kaviraj}(2014)}]{2014MNRAS.440.2944K}
{Kaviraj}, S. 2014, \mnras, 440, 2944

\bibitem[{{Kennicutt}(1989)}]{1989ApJ...344..685K}
{Kennicutt}, Robert~C., J. 1989, \apj, 344, 685

\bibitem[{{Kewley} {et~al.}(2001){Kewley}, {Dopita}, {Sutherland}, {Heisler},
  \& {Trevena}}]{2001ApJ...556..121K}
{Kewley}, L.~J., {Dopita}, M.~A., {Sutherland}, R.~S., {Heisler}, C.~A., \&
  {Trevena}, J. 2001, \apj, 556, 121

\bibitem[{{Kewley} {et~al.}(2006{\natexlab{a}}){Kewley}, {Geller}, \&
  {Barton}}]{2006AJ....131.2004K}
{Kewley}, L.~J., {Geller}, M.~J., \& {Barton}, E.~J. 2006{\natexlab{a}}, \aj,
  131, 2004

\bibitem[{{Kewley} {et~al.}(2002){Kewley}, {Geller}, {Jansen}, \&
  {Dopita}}]{2002AJ....124.3135K}
{Kewley}, L.~J., {Geller}, M.~J., {Jansen}, R.~A., \& {Dopita}, M.~A. 2002,
  \aj, 124, 3135

\bibitem[{{Kewley} {et~al.}(2006{\natexlab{b}}){Kewley}, {Groves}, {Kauffmann},
  \& {Heckman}}]{2006MNRAS.372..961K}
{Kewley}, L.~J., {Groves}, B., {Kauffmann}, G., \& {Heckman}, T.
  2006{\natexlab{b}}, \mnras, 372, 961

\bibitem[{{Knapp} {et~al.}(1984){Knapp}, {van Driel}, {Schwarz}, {van Woerden},
  \& {Gallagher}}]{1984A&A...133..127K}
{Knapp}, G.~R., {van Driel}, W., {Schwarz}, U.~J., {van Woerden}, H., \&
  {Gallagher}, J.~S., I. 1984, \aap, 133, 127

\bibitem[{{Knapp} {et~al.}(1985){Knapp}, {van Driel}, \& {van
  Woerden}}]{1985A&A...142....1K}
{Knapp}, G.~R., {van Driel}, W., \& {van Woerden}, H. 1985, \aap, 142, 1

\bibitem[{{Kohandel} {et~al.}(2019){Kohandel}, {Pallottini}, {Ferrara},
  {Zanella}, {Behrens}, {Carniani}, {Gallerani}, \&
  {Vallini}}]{2019MNRAS.487.3007K}
{Kohandel}, M., {Pallottini}, A., {Ferrara}, A., {et~al.} 2019, \mnras, 487,
  3007

\bibitem[{{Kormendy} \& {Bender}(2012)}]{2012ApJS..198....2K}
{Kormendy}, J. \& {Bender}, R. 2012, \apjs, 198, 2

\bibitem[{{Koss} {et~al.}(2012){Koss}, {Mushotzky}, {Treister}, {Veilleux},
  {Vasudevan}, \& {Trippe}}]{2012ApJ...746L..22K}
{Koss}, M., {Mushotzky}, R., {Treister}, E., {et~al.} 2012, \apjl, 746, L22

\bibitem[{{Koss} {et~al.}(2018){Koss}, {Blecha}, {Bernhard}, {Hung}, {Lu},
  {Trakhtenbrot}, {Treister}, {Weigel}, {Sartori}, {Mushotzky}, {Schawinski},
  {Ricci}, {Veilleux}, \& {Sanders}}]{2018Natur.563..214K}
{Koss}, M.~J., {Blecha}, L., {Bernhard}, P., {et~al.} 2018, \nat, 563, 214

\bibitem[{{Koss} {et~al.}(2016){Koss}, {Glidden}, {Balokovi{\'c}}, {Stern},
  {Lamperti}, {Assef}, {Bauer}, {Ballantyne}, {Boggs}, {Craig}, {Farrah},
  {F{\"u}rst}, {Gand hi}, {Gehrels}, {Hailey}, {Harrison}, {Markwardt},
  {Masini}, {Ricci}, {Treister}, {Walton}, \& {Zhang}}]{2016ApJ...824L...4K}
{Koss}, M.~J., {Glidden}, A., {Balokovi{\'c}}, M., {et~al.} 2016, \apjl, 824,
  L4

\bibitem[{{Krumm} {et~al.}(1985){Krumm}, {van Driel}, \& {van
  Woerden}}]{1985A&A...144..202K}
{Krumm}, N., {van Driel}, W., \& {van Woerden}, H. 1985, \aap, 144, 202

\bibitem[{{Laigle} {et~al.}(2018){Laigle}, {Pichon}, {Arnouts}, {McCracken},
  {Dubois}, {Devriendt}, {Slyz}, {Le Borgne}, {Benoit-L{\'e}vy}, {Hwang},
  {Ilbert}, {Kraljic}, {Malavasi}, {Park}, \& {Vibert}}]{2018MNRAS.474.5437L}
{Laigle}, C., {Pichon}, C., {Arnouts}, S., {et~al.} 2018, \mnras, 474, 5437

\bibitem[{{Lee} {et~al.}(2018){Lee}, {Krolewski}, {White}, {Schlegel},
  {Nugent}, {Hennawi}, {M{\"u}ller}, {Pan}, {Prochaska}, {Font-Ribera},
  {Suzuki}, {Glazebrook}, {Kacprzak}, {Kartaltepe}, {Koekemoer}, {Le
  F{\`e}vre}, {Lemaux}, {Maier}, {Nanayakkara}, {Rich}, {Sanders}, {Salvato},
  {Tasca}, \& {Tran}}]{2018ApJS..237...31L}
{Lee}, K.-G., {Krolewski}, A., {White}, M., {et~al.} 2018, \apjs, 237, 31

\bibitem[{{Li} {et~al.}(2008){Li}, {Kauffmann}, {Heckman}, {Jing}, \&
  {White}}]{2008MNRAS.385.1903L}
{Li}, C., {Kauffmann}, G., {Heckman}, T.~M., {Jing}, Y.~P., \& {White}, S.
  D.~M. 2008, \mnras, 385, 1903

\bibitem[{{Liu} {et~al.}(2013){Liu}, {Civano}, {Shen}, {Green}, {Greene}, \&
  {Strauss}}]{2013ApJ...762..110L}
{Liu}, X., {Civano}, F., {Shen}, Y., {et~al.} 2013, \apj, 762, 110

\bibitem[{{Liu} {et~al.}(2010){Liu}, {Shen}, {Strauss}, \&
  {Greene}}]{2010ApJ...708..427L}
{Liu}, X., {Shen}, Y., {Strauss}, M.~A., \& {Greene}, J.~E. 2010, \apj, 708,
  427

\bibitem[{{Liu} {et~al.}(2011){Liu}, {Shen}, {Strauss}, \&
  {Hao}}]{2011ApJ...737..101L}
{Liu}, X., {Shen}, Y., {Strauss}, M.~A., \& {Hao}, L. 2011, \apj, 737, 101

\bibitem[{{Lotz} {et~al.}(2008){Lotz}, {Davis}, {Faber}, {Guhathakurta},
  {Gwyn}, {Huang}, {Koo}, {Le Floc'h}, {Lin}, {Newman}, {Noeske}, {Papovich},
  {Willmer}, {Coil}, {Conselice}, {Cooper}, {Hopkins}, {Metevier}, {Primack},
  {Rieke}, \& {Weiner}}]{2008ApJ...672..177L}
{Lotz}, J.~M., {Davis}, M., {Faber}, S.~M., {et~al.} 2008, \apj, 672, 177

\bibitem[{{Lotz} {et~al.}(2010{\natexlab{a}}){Lotz}, {Jonsson}, {Cox}, \&
  {Primack}}]{2010MNRAS.404..590L}
{Lotz}, J.~M., {Jonsson}, P., {Cox}, T.~J., \& {Primack}, J.~R.
  2010{\natexlab{a}}, \mnras, 404, 590

\bibitem[{{Lotz} {et~al.}(2010{\natexlab{b}}){Lotz}, {Jonsson}, {Cox}, \&
  {Primack}}]{2010MNRAS.404..575L}
{Lotz}, J.~M., {Jonsson}, P., {Cox}, T.~J., \& {Primack}, J.~R.
  2010{\natexlab{b}}, \mnras, 404, 575

\bibitem[{{Lotz} {et~al.}(2004){Lotz}, {Primack}, \&
  {Madau}}]{2004AJ....128..163L}
{Lotz}, J.~M., {Primack}, J., \& {Madau}, P. 2004, \aj, 128, 163

\bibitem[{{Macchetto} {et~al.}(1996){Macchetto}, {Pastoriza}, {Caon}, {Sparks},
  {Giavalisco}, {Bender}, \& {Capaccioli}}]{1996A&AS..120..463M}
{Macchetto}, F., {Pastoriza}, M., {Caon}, N., {et~al.} 1996, \aaps, 120, 463

\bibitem[{{Mapelli} {et~al.}(2015){Mapelli}, {Rampazzo}, \&
  {Marino}}]{2015A&A...575A..16M}
{Mapelli}, M., {Rampazzo}, R., \& {Marino}, A. 2015, \aap, 575, A16

\bibitem[{{Maschmann} \& {Melchior}(2019)}]{2019A&A...627L...3M}
{Maschmann}, D. \& {Melchior}, A.-L. 2019, \aap, 627, L3

\bibitem[{{Meert} {et~al.}(2015){Meert}, {Vikram}, \&
  {Bernardi}}]{2015MNRAS.446.3943M}
{Meert}, A., {Vikram}, V., \& {Bernardi}, M. 2015, \mnras, 446, 3943

\bibitem[{Mendenhall \& Sincich(2011)}]{mendenhall2011second}
Mendenhall, W. \& Sincich, T. 2011, A Second Course in Statistics: Regression
  Analysis (Pearson Education)

\bibitem[{{Mo} {et~al.}(1998){Mo}, {Mao}, \& {White}}]{1998MNRAS.295..319M}
{Mo}, H.~J., {Mao}, S., \& {White}, S. D.~M. 1998, \mnras, 295, 319

\bibitem[{{Molaeinezhad} {et~al.}(2019){Molaeinezhad}, {Zhu},
  {Falc{\'o}n-Barroso}, {van de Ven}, {M{\'e}ndez-Abreu}, {Balcells},
  {Aguerri}, {Vazdekis}, {Khosroshahi}, \& {Peletier}}]{2019MNRAS.488.1012M}
{Molaeinezhad}, A., {Zhu}, L., {Falc{\'o}n-Barroso}, J., {et~al.} 2019, \mnras,
  488, 1012

\bibitem[{{M{\"u}ller-S{\'a}nchez} {et~al.}(2015){M{\"u}ller-S{\'a}nchez},
  {Comerford}, {Nevin}, {Barrows}, {Cooper}, \& {Greene}}]{2015ApJ...813..103M}
{M{\"u}ller-S{\'a}nchez}, F., {Comerford}, J.~M., {Nevin}, R., {et~al.} 2015,
  \apj, 813, 103

\bibitem[{{Nevin} {et~al.}(2019){Nevin}, {Blecha}, {Comerford}, \&
  {Greene}}]{2019ApJ...872...76N}
{Nevin}, R., {Blecha}, L., {Comerford}, J., \& {Greene}, J. 2019, \apj, 872, 76

\bibitem[{{Nevin} {et~al.}(2016){Nevin}, {Comerford}, {M{\"u}ller-S{\'a}nchez},
  {Barrows}, \& {Cooper}}]{2016ApJ...832...67N}
{Nevin}, R., {Comerford}, J., {M{\"u}ller-S{\'a}nchez}, F., {Barrows}, R., \&
  {Cooper}, M. 2016, \apj, 832, 67

\bibitem[{{Nevin} {et~al.}(2018){Nevin}, {Comerford}, {M{\"u}ller-S{\'a}nchez},
  {Barrows}, \& {Cooper}}]{2018MNRAS.473.2160N}
{Nevin}, R., {Comerford}, J.~M., {M{\"u}ller-S{\'a}nchez}, F., {Barrows}, R.,
  \& {Cooper}, M.~C. 2018, \mnras, 473, 2160

\bibitem[{{Osterbrock} \& {Ferland}(2006)}]{2006agna.book.....O}
{Osterbrock}, D.~E. \& {Ferland}, G.~J. 2006, {Astrophysics of gaseous nebulae
  and active galactic nuclei}

\bibitem[{{Patton} {et~al.}(2011){Patton}, {Ellison}, {Simard}, {McConnachie},
  \& {Mendel}}]{2011MNRAS.412..591P}
{Patton}, D.~R., {Ellison}, S.~L., {Simard}, L., {McConnachie}, A.~W., \&
  {Mendel}, J.~T. 2011, \mnras, 412, 591

\bibitem[{{Patton} {et~al.}(2016){Patton}, {Qamar}, {Ellison}, {Bluck},
  {Simard}, {Mendel}, {Moreno}, \& {Torrey}}]{2016MNRAS.461.2589P}
{Patton}, D.~R., {Qamar}, F.~D., {Ellison}, S.~L., {et~al.} 2016, \mnras, 461,
  2589

\bibitem[{{Patton} {et~al.}(2013){Patton}, {Torrey}, {Ellison}, {Mendel}, \&
  {Scudder}}]{2013MNRAS.433L..59P}
{Patton}, D.~R., {Torrey}, P., {Ellison}, S.~L., {Mendel}, J.~T., \& {Scudder},
  J.~M. 2013, \mnras, 433, L59

\bibitem[{{Pawlik} {et~al.}(2016){Pawlik}, {Wild}, {Walcher}, {Johansson},
  {Villforth}, {Rowlands}, {Mendez-Abreu}, \& {Hewlett}}]{2016MNRAS.456.3032P}
{Pawlik}, M.~M., {Wild}, V., {Walcher}, C.~J., {et~al.} 2016, \mnras, 456, 3032

\bibitem[{{Petrosian}(1976)}]{1976ApJ...209L...1P}
{Petrosian}, V. 1976, \apjl, 210, L53

\bibitem[{{Pfeifle} {et~al.}(2019){Pfeifle}, {Satyapal}, {Manzano-King},
  {Cann}, {Sexton}, {Rothberg}, {Canalizo}, {Ricci}, {Blecha}, {Ellison},
  {Gliozzi}, {Secrest}, {Constantin}, \& {Harvey}}]{2019ApJ...883..167P}
{Pfeifle}, R.~W., {Satyapal}, S., {Manzano-King}, C., {et~al.} 2019, \apj, 883,
  167

\bibitem[{{Pimbblet} {et~al.}(2002){Pimbblet}, {Smail}, {Kodama}, {Couch},
  {Edge}, {Zabludoff}, \& {O'Hely}}]{2002MNRAS.331..333P}
{Pimbblet}, K.~A., {Smail}, I., {Kodama}, T., {et~al.} 2002, \mnras, 331, 333

\bibitem[{{Pizzella} {et~al.}(2018){Pizzella}, {Morelli}, {Coccato}, {Corsini},
  {Dalla Bont{\`a}}, {Fabricius}, \& {Saglia}}]{2018A&A...616A..22P}
{Pizzella}, A., {Morelli}, L., {Coccato}, L., {et~al.} 2018, \aap, 616, A22

\bibitem[{{Pogge} \& {Eskridge}(1993)}]{1993AJ....106.1405P}
{Pogge}, R.~W. \& {Eskridge}, P.~B. 1993, \aj, 106, 1405

\bibitem[{{Proshina} {et~al.}(2019){Proshina}, {Kniazev}, \&
  {Sil'chenko}}]{2019AJ....158....5P}
{Proshina}, I.~S., {Kniazev}, A.~Y., \& {Sil'chenko}, O.~K. 2019, \aj, 158, 5

\bibitem[{{Querejeta} {et~al.}(2015){Querejeta}, {Eliche-Moral}, {Tapia},
  {Borlaff}, {van de Ven}, {Lyubenova}, {Martig}, {Falc{\'o}n-Barroso}, \&
  {M{\'e}ndez-Abreu}}]{2015A&A...579L...2Q}
{Querejeta}, M., {Eliche-Moral}, M.~C., {Tapia}, T., {et~al.} 2015, \aap, 579,
  L2

\bibitem[{{Rawle} {et~al.}(2013){Rawle}, {Lucey}, {Smith}, \&
  {Head}}]{2013MNRAS.433.2667R}
{Rawle}, T.~D., {Lucey}, J.~R., {Smith}, R.~J., \& {Head}, J.~T.~C.~G. 2013,
  \mnras, 433, 2667

\bibitem[{{Riffel} {et~al.}(2014){Riffel}, {Storchi-Bergmann}, \&
  {Riffel}}]{2014ApJ...780L..24R}
{Riffel}, R.~A., {Storchi-Bergmann}, T., \& {Riffel}, R. 2014, \apjl, 780, L24

\bibitem[{{Rodriguez-Gomez} {et~al.}(2019){Rodriguez-Gomez}, {Snyder}, {Lotz},
  {Nelson}, {Pillepich}, {Springel}, {Genel}, {Weinberger}, {Tacchella},
  {Pakmor}, {Torrey}, {Marinacci}, {Vogelsberger}, {Hernquist}, \&
  {Thilker}}]{2019MNRAS.483.4140R}
{Rodriguez-Gomez}, V., {Snyder}, G.~F., {Lotz}, J.~M., {et~al.} 2019, \mnras,
  483, 4140

\bibitem[{{Rupke} \& {Veilleux}(2013)}]{2013ApJ...768...75R}
{Rupke}, D. S.~N. \& {Veilleux}, S. 2013, \apj, 768, 75

\bibitem[{{Salim}(2014)}]{2014SerAJ.189....1S}
{Salim}, S. 2014, Serbian Astronomical Journal, 189, 1

\bibitem[{{Salim} {et~al.}(2016){Salim}, {Lee}, {Janowiecki}, {da Cunha},
  {Dickinson}, {Boquien}, {Burgarella}, {Salzer}, \&
  {Charlot}}]{2016ApJS..227....2S}
{Salim}, S., {Lee}, J.~C., {Janowiecki}, S., {et~al.} 2016, \apjs, 227, 2

\bibitem[{{Salim} {et~al.}(2007){Salim}, {Rich}, {Charlot}, {Brinchmann},
  {Johnson}, {Schiminovich}, {Seibert}, {Mallery}, {Heckman}, {Forster},
  {Friedman}, {Martin}, {Morrissey}, {Neff}, {Small}, {Wyder}, {Bianchi},
  {Donas}, {Lee}, {Madore}, {Milliard}, {Szalay}, {Welsh}, \&
  {Yi}}]{2007ApJS..173..267S}
{Salim}, S., {Rich}, R.~M., {Charlot}, S., {et~al.} 2007, \apjs, 173, 267

\bibitem[{{Sanders} \& {Mirabel}(1996)}]{1996ARA&A..34..749S}
{Sanders}, D.~B. \& {Mirabel}, I.~F. 1996, \araa, 34, 749

\bibitem[{{Sarron} {et~al.}(2019){Sarron}, {Adami}, {Durret}, \&
  {Laigle}}]{2019A&A...632A..49S}
{Sarron}, F., {Adami}, C., {Durret}, F., \& {Laigle}, C. 2019, \aap, 632, A49

\bibitem[{{Sarzi} {et~al.}(2006){Sarzi}, {Falc{\'o}n-Barroso}, {Davies},
  {Bacon}, {Bureau}, {Cappellari}, {de Zeeuw}, {Emsellem}, {Fathi},
  {Krajnovi{\'c}}, {Kuntschner}, {McDermid}, \&
  {Peletier}}]{2006MNRAS.366.1151S}
{Sarzi}, M., {Falc{\'o}n-Barroso}, J., {Davies}, R.~L., {et~al.} 2006, \mnras,
  366, 1151

\bibitem[{{Saulder} {et~al.}(2016){Saulder}, {van Kampen}, {Chilingarian},
  {Mieske}, \& {Zeilinger}}]{2016A&A...596A..14S}
{Saulder}, C., {van Kampen}, E., {Chilingarian}, I.~V., {Mieske}, S., \&
  {Zeilinger}, W.~W. 2016, \aap, 596, A14

\bibitem[{{Schawinski} {et~al.}(2007){Schawinski}, {Thomas}, {Sarzi},
  {Maraston}, {Kaviraj}, {Joo}, {Yi}, \& {Silk}}]{2007MNRAS.382.1415S}
{Schawinski}, K., {Thomas}, D., {Sarzi}, M., {et~al.} 2007, \mnras, 382, 1415

\bibitem[{{Schiminovich} {et~al.}(2007){Schiminovich}, {Wyder}, {Martin},
  {Johnson}, {Salim}, {Seibert}, {Treyer}, {Budav{\'a}ri}, {Hoopes},
  {Zamojski}, {Barlow}, {Forster}, {Friedman}, {Morrissey}, {Neff}, {Small},
  {Bianchi}, {Donas}, {Heckman}, {Lee}, {Madore}, {Milliard}, {Rich}, {Szalay},
  {Welsh}, \& {Yi}}]{2007ApJS..173..315S}
{Schiminovich}, D., {Wyder}, T.~K., {Martin}, D.~C., {et~al.} 2007, \apjs, 173,
  315

\bibitem[{{Schirmer} {et~al.}(2013){Schirmer}, {Diaz}, {Holhjem}, {Levenson},
  \& {Winge}}]{2013ApJ...763...60S}
{Schirmer}, M., {Diaz}, R., {Holhjem}, K., {Levenson}, N.~A., \& {Winge}, C.
  2013, \apj, 763, 60

\bibitem[{{Schneider} {et~al.}(2010){Schneider}, {Richards}, {Hall}, {Strauss},
  {Anderson}, {Boroson}, {Ross}, {Shen}, {Brandt}, {Fan}, {Inada}, {Jester},
  {Knapp}, {Krawczyk}, {Thakar}, {Vanden Berk}, {Voges}, {Yanny}, {York},
  {Bahcall}, {Bizyaev}, {Blanton}, {Brewington}, {Brinkmann}, {Eisenstein},
  {Frieman}, {Fukugita}, {Gray}, {Gunn}, {Hibon}, {Ivezi{\'c}}, {Kent}, {Kron},
  {Lee}, {Lupton}, {Malanushenko}, {Malanushenko}, {Oravetz}, {Pan}, {Pier},
  {Price}, {Saxe}, {Schlegel}, {Simmons}, {Snedden}, {SubbaRao}, {Szalay}, \&
  {Weinberg}}]{2010AJ....139.2360S}
{Schneider}, D.~P., {Richards}, G.~T., {Hall}, P.~B., {et~al.} 2010, \aj, 139,
  2360

\bibitem[{{Shimwell} {et~al.}(2019){Shimwell}, {Tasse}, {Hardcastle}, {Mechev},
  {Williams}, {Best}, {R{\"o}ttgering}, {Callingham}, {Dijkema}, {de Gasperin},
  {Hoang}, {Hugo}, {Mirmont}, {Oonk}, {Prandoni}, {Rafferty}, {Sabater},
  {Smirnov}, {van Weeren}, {White}, {Atemkeng}, {Bester}, {Bonnassieux},
  {Br{\"u}ggen}, {Brunetti}, {Chy{\.z}y}, {Cochrane}, {Conway}, {Croston},
  {Danezi}, {Duncan}, {Haverkorn}, {Heald}, {Iacobelli}, {Intema}, {Jackson},
  {Jamrozy}, {Jarvis}, {Lakhoo}, {Mevius}, {Miley}, {Morabito}, {Morganti},
  {Nisbet}, {Orr{\'u}}, {Perkins}, {Pizzo}, {Schrijvers}, {Smith}, {Vermeulen},
  {Wise}, {Alegre}, {Bacon}, {van Bemmel}, {Beswick}, {Bonafede}, {Botteon},
  {Bourke}, {Brienza}, {Calistro Rivera}, {Cassano}, {Clarke}, {Conselice},
  {Dettmar}, {Drabent}, {Dumba}, {Emig}, {En{\ss}lin}, {Ferrari}, {Garrett},
  {G{\'e}nova-Santos}, {Goyal}, {G{\"u}rkan}, {Hale}, {Harwood}, {Heesen},
  {Hoeft}, {Horellou}, {Jackson}, {Kokotanekov}, {Kondapally},
  {Kunert-Bajraszewska}, {Mahatma}, {Mahony}, {Mandal}, {McKean}, {Merloni},
  {Mingo}, {Miskolczi}, {Mooney}, {Nikiel-Wroczy{\'n}ski}, {O'Sullivan},
  {Quinn}, {Reich}, {Roskowi{\'n}ski}, {Rowlinson}, {Savini}, {Saxena},
  {Schwarz}, {Shulevski}, {Sridhar}, {Stacey}, {Urquhart}, {van der Wiel},
  {Varenius}, {Webster}, \& {Wilber}}]{2019A&A...622A...1S}
{Shimwell}, T.~W., {Tasse}, C., {Hardcastle}, M.~J., {et~al.} 2019, \aap, 622,
  A1

\bibitem[{{Sil'chenko} {et~al.}(2018){Sil'chenko}, {Kostiuk}, {Burenkov}, \&
  {Parul}}]{2018A&A...620L...7S}
{Sil'chenko}, O., {Kostiuk}, I., {Burenkov}, A., \& {Parul}, H. 2018, \aap,
  620, L7

\bibitem[{{Sil'chenko}(2016)}]{2016AJ....152...73S}
{Sil'chenko}, O.~K. 2016, \aj, 152, 73

\bibitem[{{Sil'chenko} \& {Afanasiev}(2004)}]{2004AJ....127.2641S}
{Sil'chenko}, O.~K. \& {Afanasiev}, V.~L. 2004, \aj, 127, 2641

\bibitem[{{Silk}(1997)}]{1997ApJ...481..703S}
{Silk}, J. 1997, \apj, 481, 703

\bibitem[{{Sil{\textquoteright}chenko}
  {et~al.}(2019){Sil{\textquoteright}chenko}, {Moiseev}, \&
  {Egorov}}]{2019ApJS..244....6S}
{Sil{\textquoteright}chenko}, O.~K., {Moiseev}, A.~V., \& {Egorov}, O.~V. 2019,
  \apjs, 244, 6

\bibitem[{{Smith} {et~al.}(2010){Smith}, {Shields}, {Bonning}, {McMullen},
  {Rosario}, \& {Salviander}}]{2010ApJ...716..866S}
{Smith}, K.~L., {Shields}, G.~A., {Bonning}, E.~W., {et~al.} 2010, \apj, 716,
  866

\bibitem[{{Somerville} \& {Dav{\'e}}(2015)}]{2015ARA&A..53...51S}
{Somerville}, R.~S. \& {Dav{\'e}}, R. 2015, \araa, 53, 51

\bibitem[{{Strateva} {et~al.}(2001){Strateva}, {Ivezi{\'c}}, {Knapp},
  {Narayanan}, {Strauss}, {Gunn}, {Lupton}, {Schlegel}, {Bahcall}, {Brinkmann},
  {Brunner}, {Budav{\'a}ri}, {Csabai}, {Castander}, {Doi}, {Fukugita},
  {Gy{\H{o}}ry}, {Hamabe}, {Hennessy}, {Ichikawa}, {Kunszt}, {Lamb}, {McKay},
  {Okamura}, {Racusin}, {Sekiguchi}, {Schneider}, {Shimasaku}, \&
  {York}}]{2001AJ....122.1861S}
{Strateva}, I., {Ivezi{\'c}}, {\v{Z}}., {Knapp}, G.~R., {et~al.} 2001, \aj,
  122, 1861

\bibitem[{{Tammour} {et~al.}(2015){Tammour}, {Gallagher}, \&
  {Richards}}]{2015MNRAS.448.3354T}
{Tammour}, A., {Gallagher}, S.~C., \& {Richards}, G. 2015, \mnras, 448, 3354

\bibitem[{{Tapia} {et~al.}(2017){Tapia}, {Eliche-Moral}, {Aceves},
  {Rodr{\'\i}guez-P{\'e}rez}, {Borlaff}, \& {Querejeta}}]{2017A&A...604A.105T}
{Tapia}, T., {Eliche-Moral}, M.~C., {Aceves}, H., {et~al.} 2017, \aap, 604,
  A105

\bibitem[{{Tous} {et~al.}(2020){Tous}, {Solanes}, \&
  {Perea}}]{2020MNRAS.tmp.1540T}
{Tous}, J.~L., {Solanes}, J.~M., \& {Perea}, J.~D. 2020, \mnras
  [\eprint[arXiv]{2005.09016}]

\bibitem[{{Tremonti} {et~al.}(2004){Tremonti}, {Heckman}, {Kauffmann},
  {Brinchmann}, {Charlot}, {White}, {Seibert}, {Peng}, {Schlegel}, {Uomoto},
  {Fukugita}, \& {Brinkmann}}]{2004ApJ...613..898T}
{Tremonti}, C.~A., {Heckman}, T.~M., {Kauffmann}, G., {et~al.} 2004, \apj, 613,
  898

\bibitem[{{Tully} \& {Fisher}(1977)}]{1977A&A....54..661T}
{Tully}, R.~B. \& {Fisher}, J.~R. 1977, \aap, 54, 661

\bibitem[{{van den Bergh}(1976)}]{1976ApJ...206..883V}
{van den Bergh}, S. 1976, \apj, 206, 883

\bibitem[{{van Dokkum} {et~al.}(2015){van Dokkum}, {Nelson}, {Franx}, {Oesch},
  {Momcheva}, {Brammer}, {F{\"o}rster Schreiber}, {Skelton}, {Whitaker}, {van
  der Wel}, {Bezanson}, {Fumagalli}, {Illingworth}, {Kriek}, {Leja}, \&
  {Wuyts}}]{2015ApJ...813...23V}
{van Dokkum}, P.~G., {Nelson}, E.~J., {Franx}, M., {et~al.} 2015, \apj, 813, 23

\bibitem[{{van Driel} {et~al.}(1988){van Driel}, {van Woerden}, {Gallagher}, \&
  {Schwarz}}]{1988A&A...191..201V}
{van Driel}, W., {van Woerden}, H., {Gallagher}, J.~S., I., \& {Schwarz}, U.~J.
  1988, \aap, 191, 201

\bibitem[{{van Gorkom} {et~al.}(1987){van Gorkom}, {Schechter}, \&
  {Kristian}}]{1987ApJ...314..457V}
{van Gorkom}, J.~H., {Schechter}, P.~L., \& {Kristian}, J. 1987, \apj, 314, 457

\bibitem[{{van Woerden} {et~al.}(1983){van Woerden}, {van Driel}, \&
  {Schwarz}}]{1983IAUS..100...99V}
{van Woerden}, H., {van Driel}, W., \& {Schwarz}, U.~J. 1983, in IAU Symposium,
  Vol. 100, Internal Kinematics and Dynamics of Galaxies, ed.
  E.~{Athanassoula}, 99--104

\bibitem[{{Vergani} {et~al.}(2018){Vergani}, {Garilli}, {Polletta},
  {Franzetti}, {Scodeggio}, {Zamorani}, {Haines}, {Bolzonella}, {Guzzo},
  {Granett}, {de la Torre}, {Abbas}, {Adami}, {Bottini}, {Cappi}, {Cucciati},
  {Davidzon}, {De Lucia}, {Fritz}, {Gargiulo}, {Hawken}, {Iovino}, {Krywult},
  {Le Brun}, {Le F{\`e}vre}, {Maccagni}, {Ma{\l}ek}, {Marulli}, {Pollo},
  {Tasca}, {Tojeiro}, {Zanichelli}, {Arnouts}, {Bel}, {Branchini}, {Coupon},
  {Ilbert}, {Moutard}, \& {Moscardini}}]{2018A&A...620A.193V}
{Vergani}, D., {Garilli}, B., {Polletta}, M., {et~al.} 2018, \aap, 620, A193

\bibitem[{{Walker} {et~al.}(1996){Walker}, {Mihos}, \&
  {Hernquist}}]{1996ApJ...460..121W}
{Walker}, I.~R., {Mihos}, J.~C., \& {Hernquist}, L. 1996, \apj, 460, 121

\bibitem[{{Wang} {et~al.}(2009){Wang}, {Chen}, {Hu}, {Mao}, {Zhang}, \&
  {Bian}}]{2009ApJ...705L..76W}
{Wang}, J.-M., {Chen}, Y.-M., {Hu}, C., {et~al.} 2009, \apjl, 705, L76

\bibitem[{{Wang} {et~al.}(2019{\natexlab{a}}){Wang}, {Gao}, {Duncan},
  {Williams}, {Rowan-Robinson}, {Sabater}, {Shimwell}, {Bonato},
  {Calistro-Rivera}, {Chy{\.z}y}, {Farrah}, {G{\"u}rkan}, {Hardcastle},
  {McCheyne}, {Prandoni}, {Read}, {R{\"o}ttgering}, \&
  {Smith}}]{2019A&A...631A.109W}
{Wang}, L., {Gao}, F., {Duncan}, K.~J., {et~al.} 2019{\natexlab{a}}, \aap, 631,
  A109

\bibitem[{{Wang} {et~al.}(2019{\natexlab{b}}){Wang}, {Luo}, {Song}, {Shen},
  {Feng}, {Wang}, {Wang}, {Li}, {Du}, {Hou}, {Guo}, {Kong}, \&
  {Zhang}}]{2019MNRAS.482.1889W}
{Wang}, M.~X., {Luo}, A.~L., {Song}, Y.~H., {et~al.} 2019{\natexlab{b}},
  \mnras, 482, 1889

\bibitem[{{White} {et~al.}(1997){White}, {Becker}, {Helfand}, \&
  {Gregg}}]{1997ApJ...475..479W}
{White}, R.~L., {Becker}, R.~H., {Helfand}, D.~J., \& {Gregg}, M.~D. 1997,
  \apj, 475, 479

\bibitem[{{Whitmore} {et~al.}(1993){Whitmore}, {Gilmore}, \&
  {Jones}}]{1993ApJ...407..489W}
{Whitmore}, B.~C., {Gilmore}, D.~M., \& {Jones}, C. 1993, \apj, 407, 489

\bibitem[{{Whitmore} {et~al.}(1990){Whitmore}, {Lucas}, {McElroy},
  {Steiman-Cameron}, {Sackett}, \& {Olling}}]{1990AJ....100.1489W}
{Whitmore}, B.~C., {Lucas}, R.~A., {McElroy}, D.~B., {et~al.} 1990, \aj, 100,
  1489

\bibitem[{{Whittle}(1985)}]{1985MNRAS.213....1W}
{Whittle}, M. 1985, \mnras, 213, 1

\bibitem[{{Willett} {et~al.}(2013){Willett}, {Lintott}, {Bamford}, {Masters},
  {Simmons}, {Casteels}, {Edmondson}, {Fortson}, {Kaviraj}, {Keel}, {Melvin},
  {Nichol}, {Raddick}, {Schawinski}, {Simpson}, {Skibba}, {Smith}, \&
  {Thomas}}]{2013MNRAS.435.2835W}
{Willett}, K.~W., {Lintott}, C.~J., {Bamford}, S.~P., {et~al.} 2013, \mnras,
  435, 2835

\bibitem[{{Wilman} {et~al.}(2009){Wilman}, {Oemler}, {Mulchaey}, {McGee},
  {Balogh}, \& {Bower}}]{2009ApJ...692..298W}
{Wilman}, D.~J., {Oemler}, A., J., {Mulchaey}, J.~S., {et~al.} 2009, \apj, 692,
  298

\bibitem[{{Woo} {et~al.}(2016){Woo}, {Bae}, {Son}, \&
  {Karouzos}}]{2016ApJ...817..108W}
{Woo}, J.-H., {Bae}, H.-J., {Son}, D., \& {Karouzos}, M. 2016, \apj, 817, 108

\bibitem[{{Woo} {et~al.}(2004){Woo}, {Urry}, {Lira}, {van der Marel}, \&
  {Maza}}]{2004ApJ...617..903W}
{Woo}, J.-H., {Urry}, C.~M., {Lira}, P., {van der Marel}, R.~P., \& {Maza}, J.
  2004, \apj, 617, 903

\bibitem[{{Xiao} {et~al.}(2016){Xiao}, {Gu}, {Chen}, \&
  {Zhou}}]{2016ApJ...831...63X}
{Xiao}, M.-Y., {Gu}, Q.-S., {Chen}, Y.-M., \& {Zhou}, L. 2016, \apj, 831, 63

\bibitem[{{Yang} {et~al.}(2007){Yang}, {Mo}, {van den Bosch}, {Pasquali}, {Li},
  \& {Barden}}]{2007ApJ...671..153Y}
{Yang}, X., {Mo}, H.~J., {van den Bosch}, F.~C., {et~al.} 2007, \apj, 671, 153

\bibitem[{{Yip} {et~al.}(2010){Yip}, {Szalay}, {Wyse}, {Dobos}, {Budav{\'a}ri},
  \& {Csabai}}]{2010ApJ...709..780Y}
{Yip}, C.-W., {Szalay}, A.~S., {Wyse}, R. F.~G., {et~al.} 2010, \apj, 709, 780

\bibitem[{{Zhang} {et~al.}(2018){Zhang}, {Zaritsky}, {Werk}, \&
  {Behroozi}}]{2018ApJ...866L...4Z}
{Zhang}, H., {Zaritsky}, D., {Werk}, J., \& {Behroozi}, P. 2018, \apjl, 866, L4

\end{thebibliography}
\appendix
\section{Statistical significance of contingency tables}\label{ssect:fishers:exact:test}
\begin{table}
\caption{Contingency table for galaxy types}\label{table:fishers:test}
\vspace{-0.6cm}
\begin{center}
\begin{tabular}{l r r r}
\hline\hline
Type & DPS & NBCS & P-value \\
\hline
SF & 2534(44.7\%)& 2811(54.8\%)& 1.66e-25 \\
COMP & 2153(38.0\%)& 1226(23.9\%)& 1.13e-56 \\
AGN & 630(11.1\%)& 687(13.4\%)& 3.27e-04 \\
LINER & 174(3.1\%)& 308(6.0\%)& 1.56e-13 \\
\hline
LTG & 914(16.1\%)& 1539(30.0\%)& 2.90e-66 \\
ETG & 167(2.9\%)& 134(2.6\%)& 2.93e-01 \\
S0 & 2027(35.8\%)& 1009(19.7\%)& 2.11e-78 \\
Merger & 589(10.4\%)& 487(9.5\%)& 1.23e-01 \\
\hline
\end{tabular}
\end{center}{}
\vspace{0.1cm}
{\justifying \small
{\noindent {\bf Notes:} We present the fraction of BPT and morphological classification for the DPS and the NBCS and the corresponding p-value.}}
\end{table}
In Sect.\,\ref{ssect:bpt} and \ref{ssect:morph}, we identify some differences between the DPS and the NBCS. Due to the number of different categories, it is difficult to quantitatively conclude how significant these differences are. Therefore, we use Fisher's exact test \citep{Fisher:724001} to check some selected categories and evaluate a p-value to determine how significant the differences between the DPS and the NBCS are. The p-value represents the probability that the measured fraction of one galaxy type is originating from the same parent sample

From the BPT classification, we use the SF, COMP, AGN and LINER categories. For the DPS, we use the non-parametric fit (see Sect.\,\ref{ssect:bpt}). For the morphological classification, we examine LTG, elliptical, S0 and galaxy mergers.

For Fisher's exact test, we take the number of the above mentioned categories ${\rm c}_{\rm DPS}$ and ${\rm c}_{\rm NBCS}$ of each sample and consider the objects which are not in this category as the counterpart $\overline{\rm c}_{\rm DPS, NBCS}$. We calculate the p-value following \citep{Fisher:724001} as:
\begin{equation}
    {\rm p}_{\rm c} = 
    \frac{
    \binom{{\rm c}_{\rm DPS} + {\rm c}_{\rm NBCS}}{{\rm c}_{\rm DPS}}
    \binom{\overline{\rm c}_{\rm DPS} + \overline{\rm c}_{\rm NBCS}}{\overline{\rm c}_{\rm DPS}}}
    {\binom{{\rm n}_{\rm DPS} + {\rm n}_{\rm NBCS}}{{\rm c}_{\rm DPS} + \overline{\rm c}_{\rm DPS}}}  
\end{equation}
where ${\rm n}_{\rm DPS}$ and ${\rm n}_{\rm NBCS}$ are the total numbers of objects in the samples and $\binom{a}{b}$ denotes the binomial coefficient.

In Table\,\ref{table:fishers:test}, we present the different classification with their percentage of occurrence and the corresponding p-value. For a sub-sample with a p-value smaller than 0.05, we consider the two parent samples as significantly different. From this table, we can conclude that the lack of SF or LINER and the excess of COMP galaxies in the DPS is significant. Also the fraction of LTG and S0 galaxies are significantly different for the two samples. We also find that the two samples show very similar fractions of galaxy merger and elliptical galaxies.

\section{Non-parametric merger indicators}\label{sect:ld1}
\begin{figure}[h]
\centering 
\includegraphics[width=0.46\textwidth]{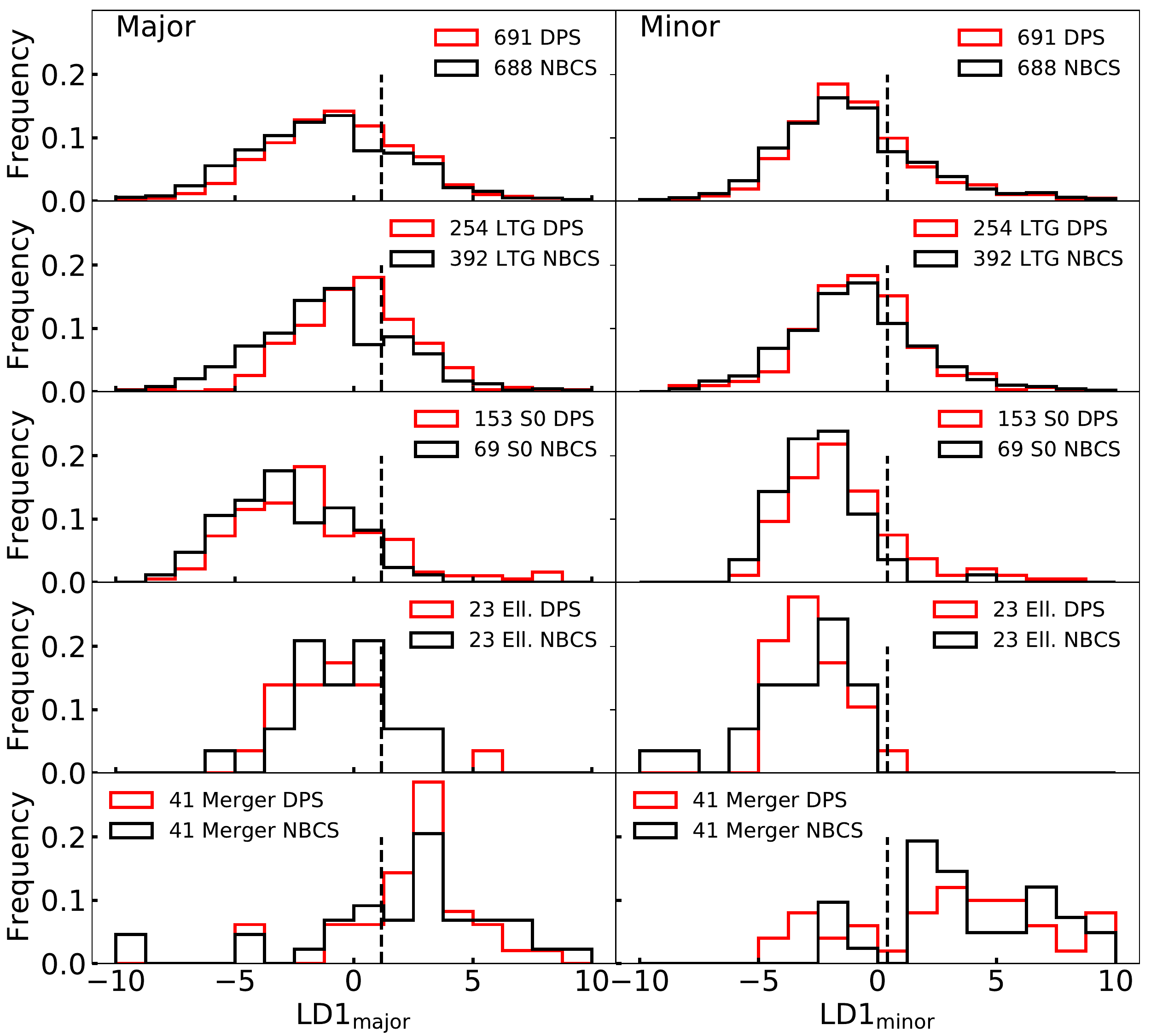}
\caption{Distribution of predictor coefficients LD1$_{\rm major}$ on the left panels and LD1$_{\rm minor}$ on the right panels \citep{2019ApJ...872...76N} for galaxies with ${\rm z} < 0.075$. We mark the decision boundary for LD1$_{\rm major}$ (resp. LD1$_{\rm minor}$) with black dashed lines at $>1.16$ (resp. $>0.42$). We show the DPS in red lines and NBCS in black lines. We show all galaxies in the top panels and present LTG, S0, elliptical and merger galaxies in the lower panels respectively.}
\label{fig:ld1}
\end{figure}
\begin{table}
\caption{Merger rate estimation using LD1$_{\rm major}$ and LD1$_{\rm minor}$}\label{table:ld1}
\vspace{-0.6cm}
\begin{center}
\begin{tabular}{l r r r r}
\hline\hline
Type & \multicolumn{2}{c}{DPS} & \multicolumn{2}{c}{NBCS} \\
 & major & minor & major & minor\\
\hline
total & 26.9\%  & 18.5\% & 25.1\%  & 21.1\% \\
LTG & 31.5\%  & 18.9\% & 24.5\%  & 20.4\% \\
S0 & 15.7\%  & 11.8\% & 4.3\%  & 1.4\% \\
Elliptical & 21.7\%  & 4.3\% & 21.7\%  & 0.0\% \\
Merger & 78.0\%  & 70.7\% & 75.6\%  & 87.8\% \\
\hline
\end{tabular}
\end{center}{}
\vspace{0.1cm}
{\justifying \small
{\noindent {\bf Notes:} We present the computed major and minor merger rate using the predictor coefficients LD1$_{\rm major}$ and LD1$_{\rm minor}$ for galaxies with $ z < 0.075$ of the DPS and the NBCS \citep{2019ApJ...872...76N}. The distributions are shown in Fig.\,\ref{fig:ld1}. We present also the merger rates for different morphological types (See Sect.\,\ref{ssect:morph}).}}
\end{table}

It is challenging to identify late merger stages directly from imaging observations, since merging signatures are depend strongly on the mass ratio and the initial conditions. As discussed in Sect.\,\ref{sect:morph:diagnostics}, the combination of Gini-${\rm M_{20}}$ diagnostics is a powerful tool \citep{2008ApJ...672..177L}, but which is strongly modulated by time \citep{2010MNRAS.404..575L,2010MNRAS.404..590L}. A promising method to achieve a higher sensitivity to detect a larger range of minor and major merger stages was proposed by \citet{2019ApJ...872...76N}, using a supervised Linear Decomposition Analysis (LDA). As input, they used non-parametric imaging predictors such as the CAS-statistics, the Gini and ${\rm M_{20}}$ coefficients and the shape asymmetry (${\rm A_{S}}$) computed by {\sc statmorph} \citep{2019MNRAS.483.4140R}. Using hydodynamical simulations of galaxy merger with different mass ratios, they reached a high accuracy of $85\,\%$ (resp. $81\,\%$) and a precision of $97\,\%$ (resp. $94\,\%$) for major (resp. minor) merger scenarios.

\citet{2019ApJ...872...76N}  focused on galaxies such as those observed with the SDSS MaNGA survey \citep{2015AJ....149...77D}, whose  average redshift is $\langle z \rangle \sim 0.03$. To keep a statistically significant sample, we restrict the merger identification analysis on subsets of the DPS and the NBCS with $ z < 0.075$. We use the predictor inputs computed by {\sc statmorph} as described in Sect.\,\ref{sect:morph:diagnostics} and compute the predictor coefficients LD1$_{\rm major}$ and LD1$_{\rm minor}$ for major and minor mergers as described in  \citet{2019ApJ...872...76N}: 

\begin{align}
\rm 
LD1_{major} = &\,\,  -0.81 +0.69 \,{\rm Gini} + 3.84 \,C + 5.78 \,A + 13.14 \,A_{S} \nonumber \\
& - 3.68 \,{\rm Gini} \times A_{S} - 6.5 \,C \times A_{S} - 6.12 \,A \times A_{S} \nonumber 
\end{align}
\begin{align}
\rm 
LD1_{minor} = &\,\, -0.87 +8.64 \,{\rm Gini} + 14.22 \,C + 5.21\, A + 2.53 \,A_{S} \nonumber \\
& - 20.33 \,{\rm Gini} \times C - 4.32 \,A \times A_{S} \nonumber 
\end{align}
We standardise the predictor inputs: we subtract the mean for each predictor and divide it by the standard deviation. To compute the means and the standard deviations, we use all the galaxies we want to test i.e. the DPS and the NBCS. To separate mergers and non-mergers, the decision boundary for LD1$_{\rm major}$ (resp. LD1$_{\rm minor}$) is $>1.16$ (resp. $>0.42$).

Since we only consider galaxies with a redshift of $ z < 0.075$, we classify only 23 elliptical and 41 merger galaxies of the DPS and the NBCS, which is due to the redshift distribution of these sub-samples (see Fig.\,\ref{fig:redshift_dist}). This makes it difficult to conclude on the measured merger rate of these two sub-samples.

In Fig.\,\ref{fig:ld1}, we show the predictor coefficients LD1$_{\rm major}$ and LD1$_{\rm minor}$ for different morphological types of the DPS and the NBCS and display the decision boundary for each predictor as a black dashed line. We present the merger rates for DPS and the NBCS and different morphological classifications in Table\,\ref{table:ld1}. In comparison to the merger rate found in Sect.\,\ref{ssect:morph}, we indeed find a slightly higher major merger rate of $27\,\%$ for the DPS and $25\,\%$ for the NBCS and a slightly higher minor merger rate of $19\,\%$ for the DPS and $21\,\%$ for the NBCS. This diagnostic is also in good agreement with the merger selection based on \citet{2018MNRAS.476.3661D}(See Sect.\,\ref{ssect:morph}): visual merger are classified as major merger with a rate of $78\,\%$ (DPS) and $76\,\%$ (NBCS) using the LDA-method. 

LTGs, which are detected at lower redshift in the DPS and the NBCS, build up the largest sub-sample with 254 of the DPS and 392 of the NBCS. We find indeed a higher major-merger rate of $32\,\%$ for the DPS in comparison to the NBCS with $25\,\%$, whereas we find a slightly lower minor-merger rate for the DPS of $19\,\%$ in comparison to $21\,\%$ for the NBCS. For S0 galaxies, we find a higher major ($16\,\%$) and minor ($12\,\%$) merger rate for the DPS in comparison to the NBCS ($4\,\%$ and $1\,\%$ respectively). 

In our merger scenario for double peak emission line galaxies (see Sect.\,\ref{sect:conclusion}), we do not find the expected merger rates using the LDA-method. This may be explained by DPS galaxies being past the coalescence phase, while the method developed by  \citet{2019ApJ...872...76N} is not sensitive to faded and later stages of mergers. 

\section{Characteristic parameter distributions}\label{sect:param:dist}
\begin{figure*}[h]
\centering 
\includegraphics[width=0.98\textwidth]{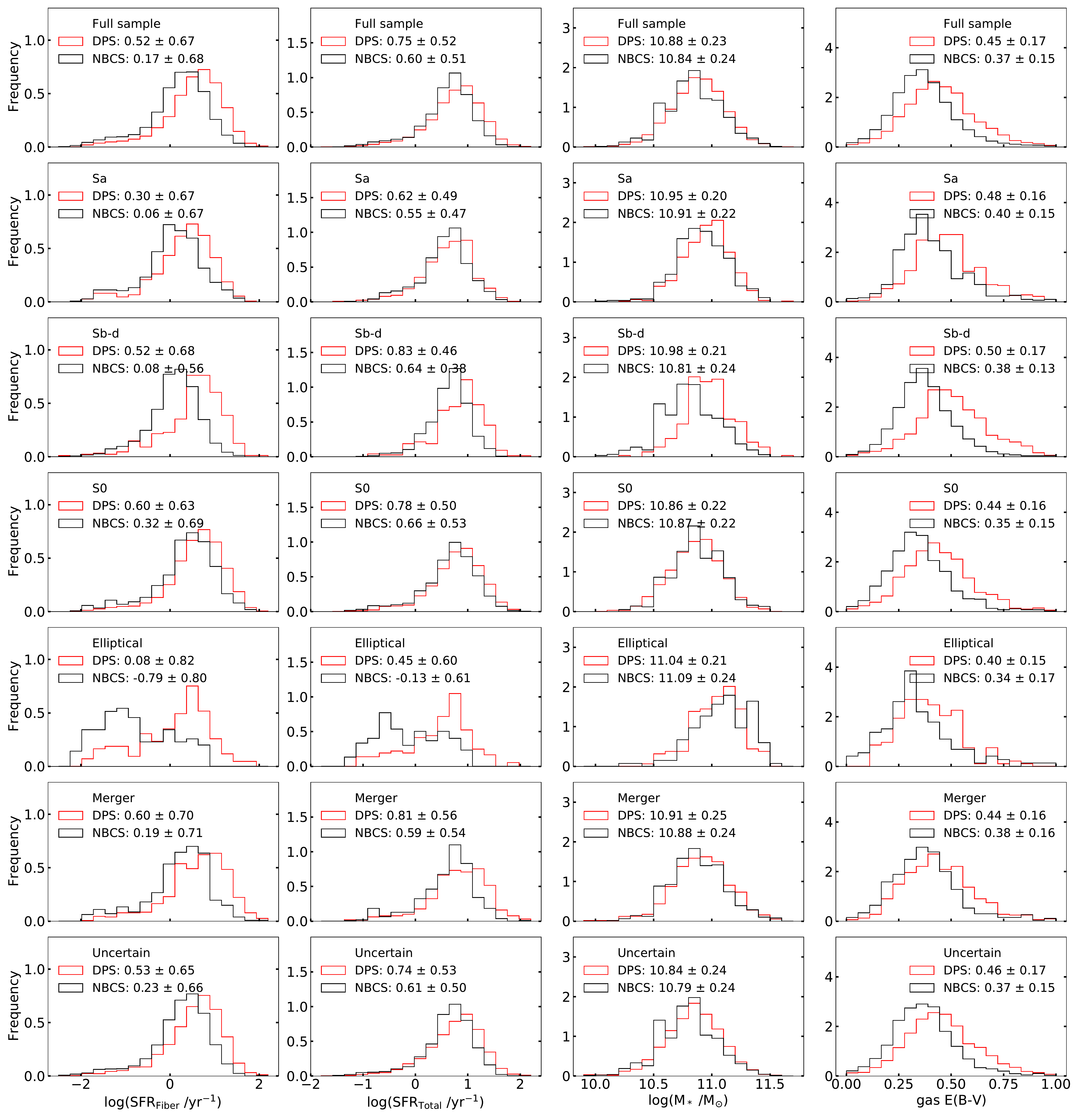}
\caption{Each column presents distributions of different parameters of the DPS (red) and the NBCS (black). We show different morphological classifications in each row: in the top row the full samples, followed by Sa, Sb-d, S0s, elliptical, merger and galaxies without classification (see Sect.\,\ref{ssect:morph}). From left to right: the first column shows the star formation rate inside the $3^{\prime\prime}$ SDSS fiber (SFR$_{\rm Fiber}$), the second column the total star formation rate (SFR$_{\rm Total}$) \citep{2004MNRAS.351.1151B} and the third column the stellar mass $\rm M_{*}$ \citep{2003MNRAS.341...33K}. In the last column, we present the extinction computed with the Balmer decrements H$\alpha$/H$\beta$ \citep{2013ApJ...763..145D}. We use the non-parametric emission line fit to estimate emission line fluxes taken from \citet{2017ApJS..228...14C}. We display the mean and the standard deviation for each distribution.}
\label{fig:appendix:sfr}%
\end{figure*}

As discussed in Sect.\,\ref{ssect:star:formation:agn}, we detect higher SFRs inside the SDSS fiber for the DPS in comparison to the NBCS. We find comparable total SFRs and about the same stellar mass distribution. As discussed in Sect.\,\ref{ssect:general:characteristics}, we detect more extinction for the DPS in comparison to the NBCS with an excess of about 0.25\,mag in the $V$ band. We furthermore find comparable stellar ages for the DPS and the NBCS, comparable time scales and about the same stellar and gas metallicities.

With respect to the morphological classifications, we detect significant differences between the DPS and the NBCS in two subsamples, namely Sb-d and elliptical galaxies. 
Sb-d galaxies of the DPS show higher mean stellar masses ($\rm \sim 0.2$\,dex in $\log(M_* / M_{\odot})$) and extinction ($\sim0.3$\,mag in $V$) than Sb-d galaxies of the NBCS (Fig.\,\ref{fig:appendix:sfr}). Furthermore, we detect a significant excess of SF in the centre of DP Sb-d galaxies. We detect ${\rm SFR_{fiber} / SFR_{total}} = 0.59 ^{+0.30}_{-0.31}$ for Sb-d of the DPS and ${\rm SFR_{fiber} / SFR_{total}} = 0.31 ^{+0.20}_{-0.16}$ for the NBCS.
In the case of elliptical galaxies, we detect a 4 times higher SFR in the DPS in comparison to their NBCS counterpart (Fig.\,\ref{fig:appendix:sfr}). This is consistent with the fact that elliptical DP galaxies are situated in the main star forming sequence, whereas elliptical galaxies of the NBCS are associated with the quenched red sequence (see Sect.\,\ref{ssect:star:formation:agn}). 
Assuming a single stellar population, we find that elliptical galaxies with a DP have on average a 3 Gyr younger stellar population in comparison with those showing a single peak. Similarly, assuming an exponential declining star formation history, we detect two times larger time scales for DP elliptical galaxies.

\section{Additional material}
\begin{figure}[H]
\centering 
\includegraphics[width=0.45\textwidth]{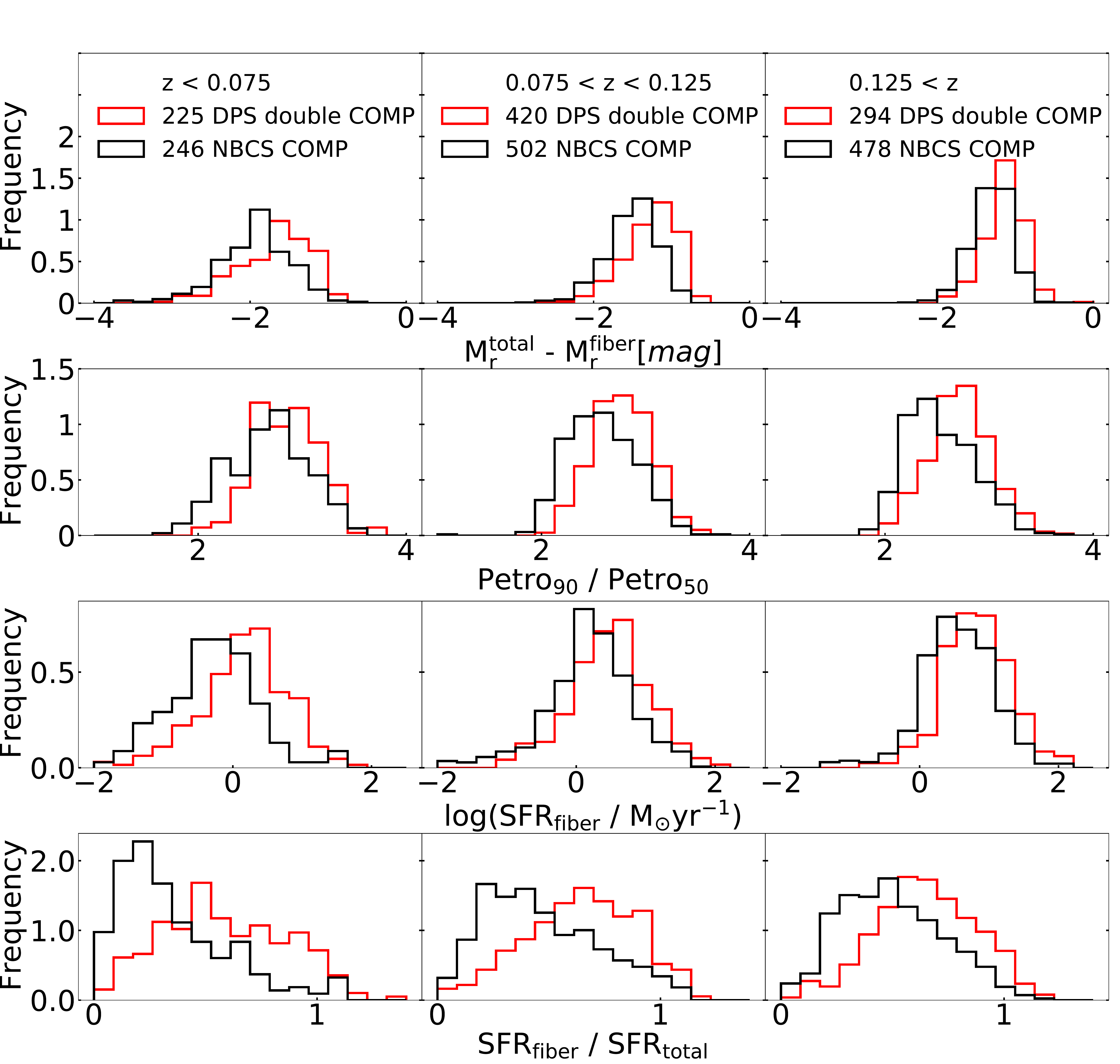}
\includegraphics[width=0.45\textwidth]{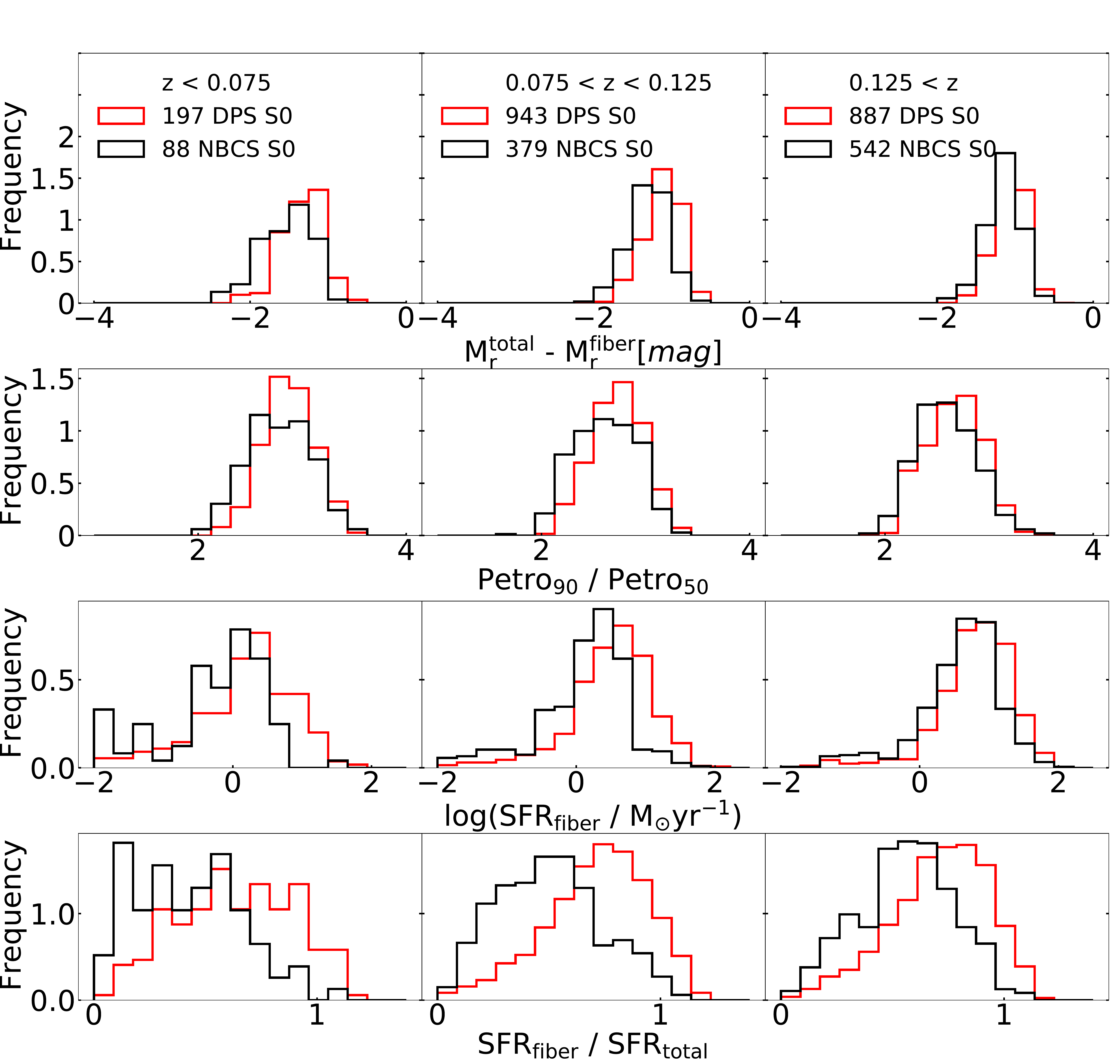}
\caption{Comparison of absolute $r$-band magnitude, Petrosian radii and (SFR$_{\rm fiber}$) and (SFR$_{\rm total}$) as in Fig.\,\ref{fig:fiber:sfr:sf}, but with DP galaxies classified as COMP in both peak components (red) and galaxies of the NBCS classified as COMP (black) in the upper panels and galaxies classified as S0s in the lower panel.}
\label{fig:fiber:sfr:comp}%
\end{figure}

\begin{figure*}[h]
  \centering 
   \textbf{Face-on LTG}\\
  \includegraphics[width=0.98\textwidth]{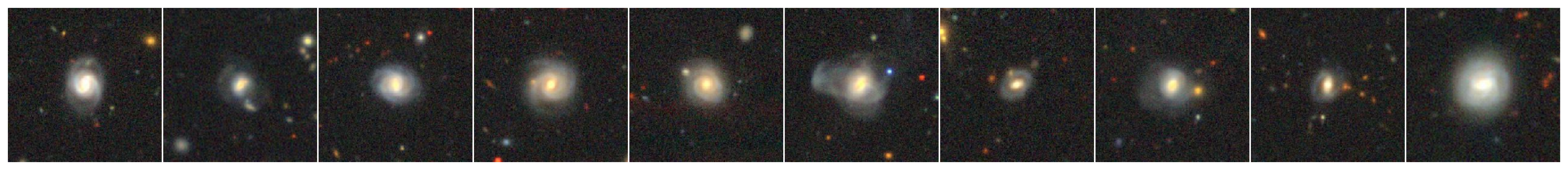}\\
   \vspace{-.1cm}
  \textbf{Intermediate inclination LTG}\\
 \includegraphics[width=0.98\textwidth]{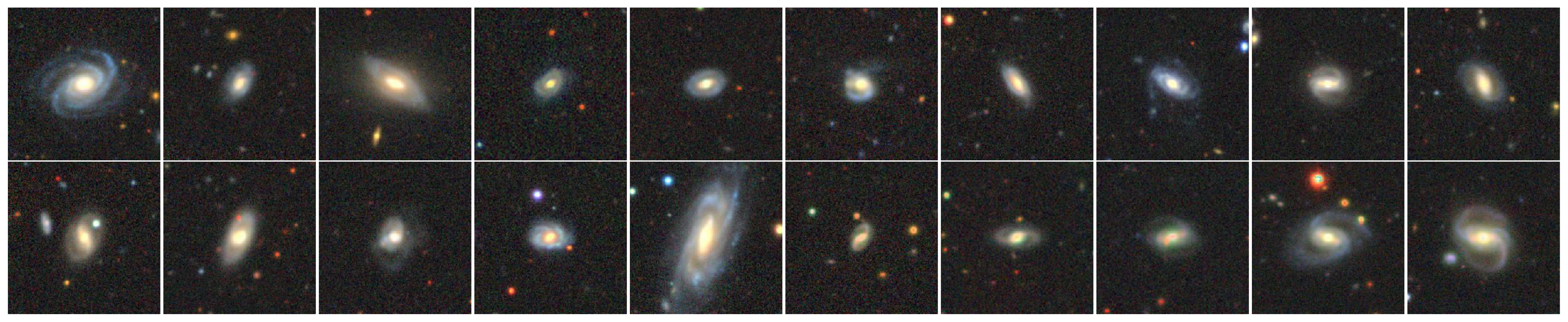}\\
 \vspace{-.1cm}
 \textbf{Edge-on LTG}\\
  \includegraphics[width=0.98\textwidth]{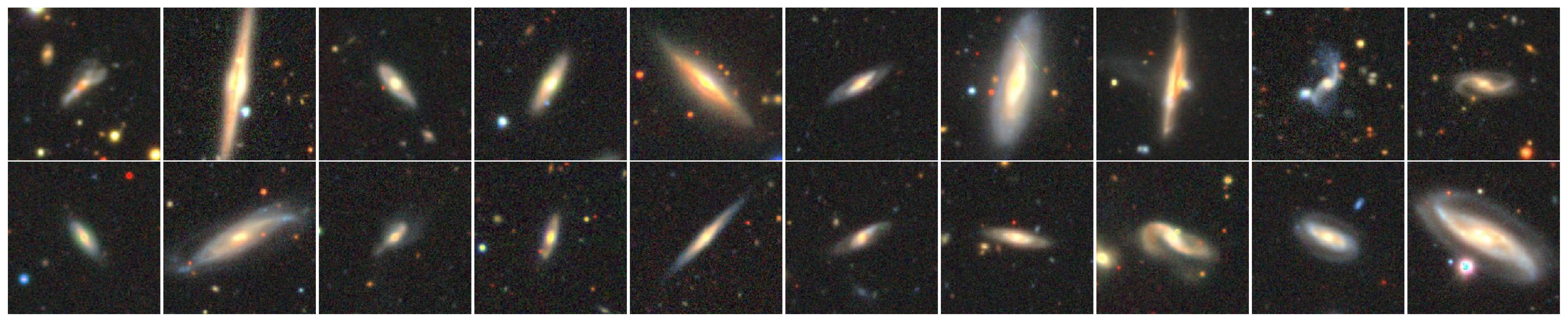}\\
   \vspace{-.1cm}
 \textbf{Elliptical}\\
  \includegraphics[width=0.98\textwidth]{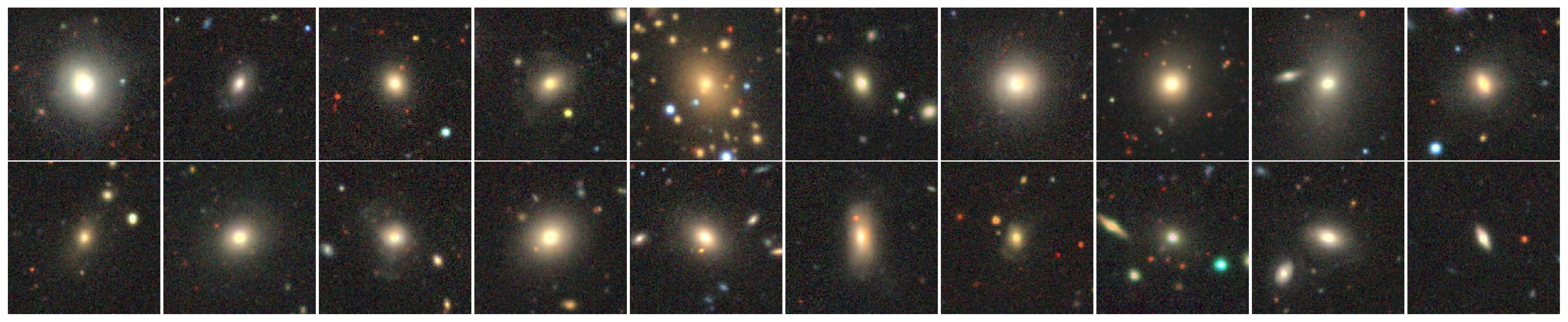}\\
   \vspace{-.1cm}
 \textbf{S0}\\
  \includegraphics[width=0.98\textwidth]{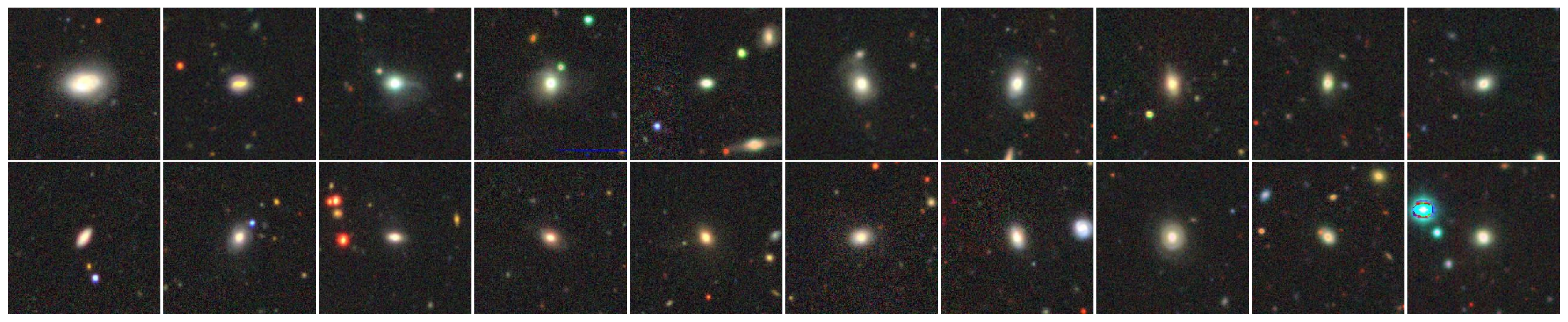}\\
  \vspace{-.1cm}
 \textbf{Merger}\\
  \includegraphics[width=0.98\textwidth]{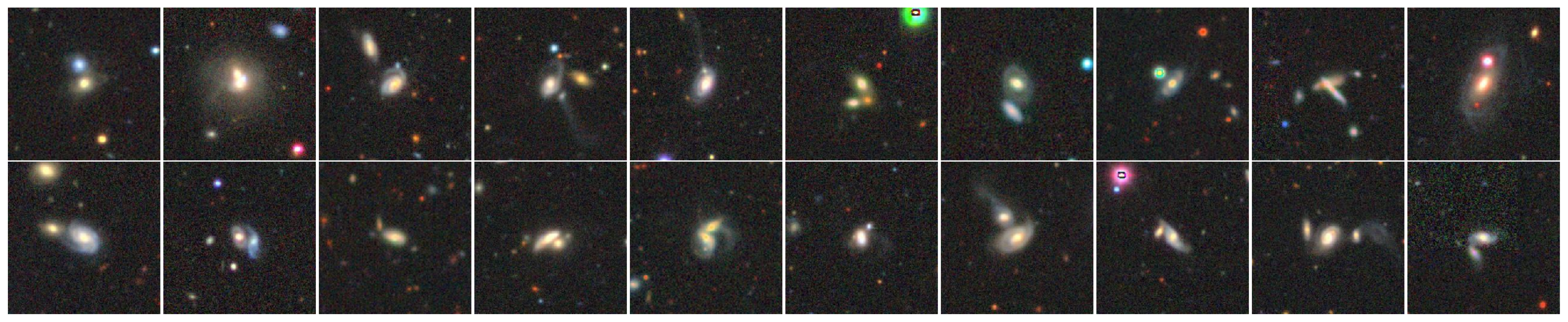}\\
  \caption{$62^{\prime\prime}\times 62^{\prime\prime}$ Legacy Survey snapshots \citep{2019AJ....157..168D} of 20 random galaxies of each morphological type, described in Sect.\,\ref{ssect:morph} (all 10 are shown for Face-on LTGs).}
  \label{fig:image:patern}
\end{figure*}

\vspace{-0.6cm}
\begin{table*}
\caption{Fraction of cross match for different BPT and morphological types}\label{table:bpt:morph1}
\vspace{-0.2cm}
\begin{minipage}{0.6\linewidth}
\small{
\begin{center}
\begin{tabular}{l r r r r r r r}
\hline\hline
BPT - class & LTG & Elliptical & S0 & Merger & Uncertain & Total\\
\hline
\multicolumn{7}{c}{DP- sample} \\
\hline
SF & 326(12.9\%)& 54(2.1\%)& 990(39.1\%)& 266(10.5\%)& 898(35.4\%) & 2534 \\
COMP & 431(20.0\%)& 69(3.2\%)& 715(33.2\%)& 208(9.7\%)& 730(33.9\%) & 2153 \\
AGN & 103(16.3\%)& 24(3.8\%)& 204(32.4\%)& 77(12.2\%)& 222(35.2\%) & 630 \\
LINER & 29(16.7\%)& 13(7.5\%)& 53(30.5\%)& 22(12.6\%)& 57(32.8\%) & 174 \\
Uncertain & 25(14.5\%)& 7(4.1\%)& 65(37.8\%)& 16(9.3\%)& 59(34.3\%) & 172 \\
\hline
\multicolumn{7}{c}{No-bias Control sample} \\
\hline
SF & 873(31.1\%)& 23(0.8\%)& 539(19.2\%)& 245(8.7\%)& 1131(40.2\%) & 2811 \\
COMP & 388(31.6\%)& 41(3.3\%)& 266(21.7\%)& 116(9.5\%)& 415(33.8\%) & 1226 \\
AGN & 177(25.8\%)& 29(4.2\%)& 129(18.8\%)& 80(11.6\%)& 272(39.6\%) & 687 \\
LINER & 76(24.7\%)& 38(12.3\%)& 56(18.2\%)& 35(11.4\%)& 103(33.4\%) & 308 \\
Uncertain & 25(26.0\%)& 3(3.1\%)& 19(19.8\%)& 11(11.5\%)& 38(39.6\%) & 96 \\
\hline
\multicolumn{7}{c}{Control sample} \\
\hline
SF & 18510(22.6\%)& 120(0.1\%)& 5328(6.5\%)& 5934(7.2\%)& 52173(63.6\%) & 82065 \\
COMP & 1123(23.8\%)& 76(1.6\%)& 1229(26.0\%)& 360(7.6\%)& 1933(40.9\%) & 4721 \\
AGN & 331(20.4\%)& 46(2.8\%)& 418(25.8\%)& 138(8.5\%)& 690(42.5\%) & 1623 \\
LINER & 122(22.8\%)& 56(10.4\%)& 106(19.8\%)& 49(9.1\%)& 203(37.9\%) & 536 \\
Uncertain & 71(15.2\%)& 4(0.9\%)& 64(13.7\%)& 43(9.2\%)& 285(61.0\%) & 467 \\
\hfill
\end{tabular}
\end{center}{}
}
\end{minipage}
\hspace{0.2\linewidth}
\begin{minipage}{0.2\linewidth}
{\justifying \small
{\noindent {\bf Notes:} This is a cross match of Tables\,\ref{table:bpt} and \ref{table:morph}. We present the fraction of each morphological type for all different subsets classified with the BPT diagram in Sect.\,\ref{ssect:bpt}. For the DPS we show the BPT-classification using the non-parametric fit. We present these fractions for the DPS, the NBCS and the CS. All rows add up to unity.}}
\end{minipage}
\end{table*}
\vspace{-1.0cm}
\begin{table*}
\caption{Fraction of cross match for different BPT and morphological types}\label{table:bpt:morph2}
\vspace{-0.2cm}
\begin{minipage}{0.6\linewidth}
\small{
\begin{center} 
\begin{tabular}{l r r r r r r } 
\hline\hline
BPT - class & LTG & Elliptical & S0 & Merger & Uncertain\\
\hline
\multicolumn{6}{c}{DP- sample} \\
\hline
SF & 326(35.7\%)& 54(32.3\%)& 990(48.8\%)& 266(45.2\%)& 898(45.7\%) \\
COMP & 431(47.2\%)& 69(41.3\%)& 715(35.3\%)& 208(35.3\%)& 730(37.1\%) \\
AGN & 103(11.3\%)& 24(14.4\%)& 204(10.1\%)& 77(13.1\%)& 222(11.3\%) \\
LINER & 29(3.2\%)& 13(7.8\%)& 53(2.6\%)& 22(3.7\%)& 57(2.9\%) \\
Uncertain & 25(2.7\%)& 7(4.2\%)& 65(3.2\%)& 16(2.7\%)& 59(3.0\%) \\
Total & 914 & 167 & 2027 & 589 & 1966  \\
\hline
\multicolumn{6}{c}{No-bias Control sample} \\
\hline
SF & 873(56.7\%)& 23(17.2\%)& 539(53.4\%)& 245(50.3\%)& 1131(57.7\%) \\
COMP & 388(25.2\%)& 41(30.6\%)& 266(26.4\%)& 116(23.8\%)& 415(21.2\%) \\
AGN & 177(11.5\%)& 29(21.6\%)& 129(12.8\%)& 80(16.4\%)& 272(13.9\%) \\
LINER & 76(4.9\%)& 38(28.4\%)& 56(5.6\%)& 35(7.2\%)& 103(5.3\%) \\
Uncertain & 25(1.6\%)& 3(2.2\%)& 19(1.9\%)& 11(2.3\%)& 38(1.9\%) \\
Total & 1539 & 134 & 1009 & 487 & 1959  \\
\hline
\multicolumn{6}{c}{Control sample} \\
\hline
SF & 18510(91.8\%)& 120(39.7\%)& 5328(74.6\%)& 5934(91.0\%)& 52173(94.4\%) \\
COMP & 1123(5.6\%)& 76(25.2\%)& 1229(17.2\%)& 360(5.5\%)& 1933(3.5\%) \\
AGN & 331(1.6\%)& 46(15.2\%)& 418(5.9\%)& 138(2.1\%)& 690(1.2\%) \\
LINER & 122(0.6\%)& 56(18.5\%)& 106(1.5\%)& 49(0.8\%)& 203(0.4\%) \\
Uncertain & 71(0.4\%)& 4(1.3\%)& 64(0.9\%)& 43(0.7\%)& 285(0.5\%) \\
Total & 20157 & 302 & 7145 & 6524 & 55284  \\
\hline
\hline
\end{tabular}
\end{center}{}
}
\end{minipage}
\hspace{0.2\linewidth}
\begin{minipage}{0.2\linewidth}
{\justifying \small
{\noindent {\bf Notes:} This is a cross match of Tables\,\ref{table:bpt} and \ref{table:morph}. We present the fraction of each subset classified with the BPT diagram in Sect.\,\ref{ssect:bpt} for all different morphological type specified in Sect.\,\ref{ssect:morph}. For the DPS we show the BPT-classification using the non-parametric fit. We present these fractions for the DPS, the CS and the NBCS. All columns add up to unity.}}
\end{minipage}
\end{table*}
\vspace{-0.6cm}
\begin{table*}
\caption{Galaxy environment}\label{table:environment}
\vspace{-0.3cm}
\begin{minipage}{0.6\linewidth}
\tiny{
\begin{center}
\begin{tabular}{r r r r r r  | r r r r r }
\hline\hline
\hline
& LTG & S0 & Elliptical & Merger & Uncertain & LTG & S0 & Elliptical & Merger & Uncertain \\
\hline
\multicolumn{4}{l}{\citet{2007ApJ...671..153Y}}  \hfill DPS & & & & & NBCS &  & \\
\hline
isolated & 59.7 \% & 68.6 \% & 56.3 \% & 54.0 \% & 63.4 \% & 65.7 \% & 67.6 \% & 47.8 \% & 58.1 \% & 67.9 \% \\
poor group & 24.3 \% & 16.6 \% & 29.3 \% & 25.0 \% & 16.0 \% & 21.0 \% & 15.0 \% & 25.4 \% & 25.9 \% & 13.3 \% \\
rich group & 5.0 \% & 2.4 \% & 6.0 \% & 3.6 \% & 2.8 \% & 4.6 \% & 3.9 \% & 11.2 \% & 3.3 \% & 3.4 \% \\
cluster & 6.1 \% & 2.4 \% & 3.0 \% & 4.2 \% & 3.3 \% & 4.5 \% & 2.2 \% & 10.4 \% & 4.5 \% & 3.4 \% \\
unknown & 4.8 \% & 10.1 \% & 5.4 \% & 13.2 \% & 14.5 \% & 4.2 \% & 11.4 \% & 5.2 \% & 8.2 \% & 12.0 \% \\
\hline
\multicolumn{4}{l}{\citet{2007ApJ...671..153Y} (${\rm z} < 0.11$)}  \hfill DPS & & & & & NBCS &  & \\
\hline
isolated & 54.8 \% & 65.2 \% & 50.9 \% & 43.6 \% & 62.5 \% & 62.8 \% & 66.9 \% & 39.1 \% & 48.3 \% & 65.1 \% \\
poor group & 28.1 \% & 23.0 \% & 34.3 \% & 33.3 \% & 20.2 \% & 22.4 \% & 20.1 \% & 28.7 \% & 32.8 \% & 15.1 \% \\
rich group & 5.1 \% & 2.8 \% & 8.3 \% & 6.7 \% & 3.6 \% & 5.8 \% & 6.1 \% & 14.9 \% & 5.0 \% & 6.0 \% \\
cluster & 7.7 \% & 4.6 \% & 3.7 \% & 9.3 \% & 6.8 \% & 6.2 \% & 4.7 \% & 13.8 \% & 9.5 \% & 6.6 \% \\
unknown & 4.4 \% & 4.4 \% & 2.8 \% & 7.1 \% & 7.0 \% & 2.9 \% & 2.3 \% & 3.4 \% & 4.5 \% & 7.2 \% \\
\hline
\multicolumn{4}{l}{\citet{2016A&A...596A..14S} (${\rm z} < 0.11$)}  \hfill DPS & & & & & NBCS &  & \\
\hline
isolated & 38.2 \% & 51.1 \% & 35.2 \% & 36.4 \% & 44.0 \% & 43.5 \% & 48.8 \% & 27.6 \% & 33.3 \% & 45.7 \% \\
poor group & 37.0 \% & 29.2 \% & 38.0 \% & 31.1 \% & 33.0 \% & 32.8 \% & 31.7 \% & 29.9 \% & 38.8 \% & 31.1 \% \\
rich group & 12.7 \% & 9.5 \% & 11.1 \% & 13.8 \% & 9.5 \% & 11.1 \% & 10.5 \% & 20.7 \% & 11.4 \% & 10.6 \% \\
cluster & 10.7 \% & 6.7 \% & 12.0 \% & 8.9 \% & 10.0 \% & 11.0 \% & 8.1 \% & 20.7 \% & 11.9 \% & 9.8 \% \\
unknown & 1.5 \% & 3.4 \% & 3.7 \% & 9.8 \% & 3.6 \% & 1.5 \% & 0.9 \% & 1.1 \% & 4.5 \% & 2.8 \% \\
\hline
\end{tabular}
\end{center}
}
\end{minipage}
\hspace{0.25\linewidth}
\begin{minipage}{0.15\linewidth}
{\justifying \small
{\noindent {\bf Notes:} Fraction of galaxies in different environments from \citet{2007ApJ...671..153Y} and \citet{2016A&A...596A..14S}}}
\end{minipage}
\end{table*}

\end{document}